\documentclass[twocolumn,twocolappendix]{aastex631}

\hypersetup{linkcolor=blue,citecolor=blue,filecolor=cyan,urlcolor=magenta}

\usepackage{multirow}
\usepackage{rotating}
\usepackage[percent]{overpic}

\newcommand{\Lx}{$L_{\rm X}$}
\newcommand{\lum}{erg s$^{-1}$}
\newcommand{\flux}{erg s$^{-1}$ cm$^{-2}$}
\newcommand{\XMM}{{\em XMM-Newton}}
\newcommand{\Chandra}{{\em Chandra}}
\newcommand{\Swift}{{\em Swift}}
\newcommand{\Msun}{$M_{\odot}$}
\newcommand{\Mjup}{$M_{\rm Jup}$}
\newcommand{\Mearth}{$M_{\oplus}$}
\newcommand{\Lsun}{$L_{\odot}$}
\newcommand{\Lbol}{$L_{\rm bol}$}
\newcommand{\Teff}{$T_{\rm eff}$}
\newcommand{\mpsini}{$M_{\rm p} \sin i$}

\graphicspath{{./}{figures/}}

\begin{document}

\title[X-ray Emission of Nearby Stars]{X-ray Emission of Nearby Low-mass and Sun-like Stars with Directly Imageable Habitable Zones}

\correspondingauthor{Breanna A. Binder}
\email{babinder@cpp.edu}

\author[0000-0002-4955-0471]{Breanna A. Binder}
\affiliation{Department of Physics and Astronomy, California State Polytechnic University, Pomona, CA, USA}

\author[0000-0002-1046-025X]{Sarah Peacock}
\affiliation{University of Maryland, Baltimore County, Baltimore, MD, 21250, USA}
\affiliation{NASA Goddard Space Flight Center, Greenbelt, MD 20771, USA}

\author[0000-0002-2949-2163]{Edward W. Schwieterman}
\affiliation{Department of Earth and Planetary Sciences, University of California, Riverside, CA, USA}
\affiliation{Blue Marble Space Institute of Science, Seattle, WA, USA}

\author[0000-0002-0569-1643]{Margaret C. Turnbull}
\affiliation{SETI Institute, Carl Sagan Center for the Study of Life in the Universe, Off-Site: Madison, WI 53713, USA}

\author{Azariel Y. Virgen}
\affiliation{Department of Physics and Astronomy, California State Polytechnic University, Pomona, CA, USA}

\author[0000-0002-7084-0529]{Stephen R. Kane}
\affiliation{Department of Earth and Planetary Sciences, University of California, Riverside, CA, USA}

\author[0000-0003-0944-2334]{Alison Farrish}
\affiliation{NASA Goddard Space Flight Center, Greenbelt, MD 20771, USA}

\author[0000-0001-6398-8755]{Katherine Garcia-Sage}
\affiliation{Heliophysics Division, NASA Goddard Space Flight Center, Greenbelt, MD, USA}

\begin{abstract}
Stellar X-ray and UV radiation can significantly affect the survival, composition, and long-term evolution of the atmospheres of planets in or near their host star's habitable zone (HZ). Especially interesting are planetary systems in the solar neighborhood that may host temperate and potentially habitable surface conditions, which may be analyzed by future ground and space-based direct-imaging surveys for signatures of habitability and life. To advance our understanding of the radiation environment in these systems, we leverage $\sim$3 Msec of \XMM\ and \Chandra\ observations in order to measure three fundamental stellar properties at X-ray energies for 57 nearby FGKM stellar systems: the shape of the stellar X-ray spectrum, the luminosity, and the timescales over which the stars vary (e.g., due to flares). These systems possess HZs that will be directly imageable to next-generation telescopes such as the {\em Habitable Worlds Observatory} and ground-based Extremely Large Telescopes (ELTs). We identify 29 stellar systems with \Lx/\Lbol\ ratios similar to (or less than) that of the Sun; any potential planets in the habitable zones of these stars therefore reside in present day X-ray radiation environments similar to (or less hostile than) modern Earth, though a broader set of these targets could host habitable planets. An additional 19 stellar systems have been observed with the \Swift\ X-ray Telescope; in total, only $\sim$30\% of potential direct imaging target stars has been observed with \XMM, \Chandra, or \Swift. The data products from this work (X-ray light curves and spectra) are available via a public Zenodo repository (doi: 10.5281/zenodo.11490574).
\end{abstract}


\section{Introduction} \label{sec:intro}

In the coming decades, ground- and space-based direct imaging will afford the best potential for detecting and spectrally characterizing terrestrial planets in the habitable zones of our nearest neighbors \citep{Turnbull+21, kopparapu2018,Li+21,Kane+18, Currie2023}. While the {\em James Webb Space Telescope} (JWST) has and will continue to characterize some rocky exoplanets \citep[e.g.,][]{Greene2023, Zieba2023, Lincowski2023, Lustig-Yaeger2023, Moran2023}, its ability to spectrally examine temperate terrestrial---and therefore potentially habitable---exoplanets is limited primarily to target systems with transiting planets and late M dwarf host stars. This is due to the lack of coronagraphic instrumentation on JWST to image rocky planets around more massive, Sun-like stars, and because the characterization of terrestrial planets through transit observations is most favorable for M dwarf hosts, which have a combination of favorable transit probabilities, planet-star size ratios, and planet-star separations that allow for the probing of low atmospheric altitudes when accounting for the effects of refraction \citep{GarciaMunoz2012, Betremieux2013, Betremieux2014}. Moreover, some key candidate biosignature molecules like O$_{2}$ cannot be detected directly at Earth-like abundances with JWST via transit spectroscopy even for the most favorable targets \citep[e.g., ][]{Lustig-Yaeger2019,Pidhorodetska2020,Meadows2023}, but could be more easily detected by reflected light observations of planets in nearby systems \citep{LUVOIR2019,Gaudi2020}.

The occurrence rates derived from Kepler indicate that terrestrial planets are likely present within the habitable zones (HZs) of a large fraction of nearby stars \citep[e.g.,][]{Bryson+21,Kunimoto+20,Datillo+23,Fulton+17,Burke+15,Kane+14}. The Astro2020 Decadal Survey has identified the discovery and characterization of these worlds as the highest priority for NASA's investments \citep{NAP26141}, and the {\em Habitable Worlds Observatory}\footnote{\url{https://habitableworldsobservatory.org/home}} will build on the LUVOIR and {\em Habitable Exoplanets Observatory} concepts \citep{LUVOIR2019,Gaudi2020} to realize this goal. Due to detection biases, the vast majority of nearby temperate rocky planets amenable for direct imaging have not yet been discovered, but will be revealed by HWO in its survey phase \citep[e.g.,][]{LUVOIR2019,Gaudi2020}, precursor science with extreme precision radial velocity (EPRV) measurements \citep{Crass2021,Morgan2021}, or direct imaging with upcoming 30-m class telescopes \citep{Fujii2018, Currie2023}.

A detailed understanding of the target stars themselves is a strong prerequisite for maximizing the science yield of exoplanet imaging missions, including extracting the statistical features of the overall planet population \citep[e.g., ][]{VanEylen+14} as well as characterizing the climates, atmospheric composition and stability \citep{Linsky+24,Louca+23}, and potential biosignatures of individual planets \citep{Rugheimer+15, schwieterman+2018}. Evaluating planetary habitability, specifically, must consider myriad factors beyond circumstellar location including the planet’s bulk composition \citep{Hinkel+18,Marounina+20, unterborn+2023} and stellar X-ray and extreme ultraviolet (XUV; 1-912 \AA; corresponding to an energy range of 0.01-12 keV) emission \citep{France+16,GarciaSage+17,Luger2015,richey+2019}. Planets orbiting in or near their star’s HZ can experience high levels of XUV radiation, which directly influences atmospheric survival, composition, and long-term evolution \citep{Peacock2020,Johnstone2021}. In some cases, planetary atmospheres and surface volatiles can be entirely driven away by XUV irradiation \citep[e.g., ][]{Sanz-Forcada2011,Fromont2024,Luger2015}. Further, the atmospheric states of planets orbiting M dwarf stars may be profoundly different than for those orbiting Sun-like FGK stars due their to extended super-luminous pre-main sequence phase \citep{Baraffe+15} during which XUV radiation could ablate away planetary atmospheres or lead to atmospheric oxidation via massive H escape \citep{GarciaSage+17,Luger2015,krissansen+2022}. Outgassing could replenish these atmospheres \citep{Kite+20,Swain+21,unterborn2022,Kane+20} though the retention and photochemistry of the atmosphere would continue to be controlled by the host star spectrum and activity \citep{roettenbacher2017,France+20,France+16,Loyd+18a,Loyd+18b, Amaral+2022}. 

While X-ray detections have been made for some HWO target stars \citep{Harada+24} and more detailed studies of the X-ray properties of low-mass nearby stars have been performed \citep{Brown+23,France+20,Loyd+18b}, no comprehensive and uniform X-ray analysis of potential direct imaging target stars has yet been done. In this paper, we present a comprehensive analysis of archival X-ray observations for all FGKM stellar systems with HZs that will be directly-imageable with upcoming telescopes such as HWO and ground-based Extremely Large Telescopes (ELTs). A follow-up paper yielding panchromatic XUV--IR spectra (including modeled EUV wavelengths) for each one of these systems is forthcoming (\textit{Peacock et al., in prep}). We describe our sample and the archival data in Sections~\ref{sec:sample} and \ref{sec:data}. Our analysis procedure is detailed in Section~\ref{sec:analysis}. We extract light curves for stars detected at high significance and identify periods of count rate variability (such as flaring events or epochs of elevated count rates). We then extract X-ray spectra for stars, separating variable- from non-variable epochs, and fit the resulting spectra with one-, two-, or three-component thermal plasma models. Finally, the average best-fit spectral models are used to convert observed count rates (or count rate upper limits, in the case of non-X-ray-detected stars) to luminosities. In Section~\ref{sec:discussion}, we discuss the evolution of the \Lx/\Lbol\ ratio with effective temperature and stellar age, and present an extensive appendix describing the X-ray observations and a summary of the known stellar physical parameters and exoplanet systems from the literature for each stellar system in our sample.

\section{The Sample}\label{sec:sample}

To identify the highest priority targets for ``deep dive'' investigations of nearby stars and their planetary systems, we have used the predicted imaging and spectroscopy capability of (1) NASA’s next generation space-based Great Observatories such as HWO, and (2) ground-based ELTs. To first order, the performance of these facilities is described in terms of two specifications: the inner working angle (IWA) — the smallest planet-star angular separation that can be imaged (expressed in milliarcseconds, or mas), and the minimum planet to star flux ratio (or fractional planet brightness, FPB) — where starlight suppression technology and postprocessing algorithms can separate the planet signal from stellar photons and systematic noise. Intriguingly (and somewhat vexingly), for HZ planets the IWA requirement favors stars that are more luminous and therefore have wider HZs, while the limiting flux ratio favors target stars that are less luminous and therefore outshine their HZ planets by a smaller factor \citep{Turnbull+12}. A third specification, planet limiting magnitude ($V_{\rm lim}$), determines which of these targets planets would be bright enough to obtain spectra in a reasonable amount of observing time -- and this requirement simply favors the nearest stars.

To form our preliminary target list, for space-based missions we chose stars for which the HZ (0.95 - 1.7 AU scaled by $\sqrt{L_{\rm bol}}$) falls at least partially outside an inner working angle of 58 mas, and for which Earth-sized planets of albedo $\sim$0.3 would be brighter than $V_{\rm lim} < 30$, with a fractional planet brightness (FPB) $> 4\times10^{-11}$. For ground-based observatories, we examined the 25 stars within 10 pc having the most favorable FPB whose HZs are at least partially visible outside an IWA of $\sim$30 mas, consistent with the literature \citep{brandl2018status,Quanz2015}. This results in a well-defined list of 175 target stars whose HZs fall at least partially within the detection space of upcoming exoplanet direct imaging observatories. We compared our target list with the potential target stars for HWO \citep{Mamajek+24}, who used slightly different cut-off requirements for $V_{\rm lim}$, FPB and IWA. Our list contains more M-dwarfs (more favorable targets for ELTs) while the \citet{Mamajek+24} list includes more F stars (more favorable for space-based observations). \citet{Mamajek+24} also excludes stars in close binaries or higher-order multiple systems, while we produce a more inclusive X-ray catalog of nearby stars, assuming technical challenges posed by binaries may eventually be surmounted. There are 229 unique stars when combining the two lists, which we summarize in Table~\ref{tab:combined_targets}. 

\begin{deluxetable}{cccccc}[!htbp]
\tablecaption{Potential Target Stars for Future HZ Imaging Surveys}
\label{tab:combined_targets}
\setlength{\tabcolsep}{1.5pt}
    \tablehead{
    \colhead{HD} & \colhead{Common} & \colhead{Distance} & \colhead{Spectral} & \colhead{From} & \colhead{Potential} \\
    ID & Name   &  (pc)    &  Type   & This Work?    & HWO Target\tablenotemark{a} 
    }
    \startdata
    166 & V439 And     & 13.77 & G8V   & \checkmark     & \checkmark \\
    693 & 6 Cet 	   & 18.89 &	F8V	&		& \checkmark \\
    739 & $\theta$ Scl & 21.72 &	F5V	&		& \checkmark \\
    1326 & GX And		& 3.57 &	M1.5V	&	\checkmark	&  \\
    1581 & $\zeta$ Tuc	& 8.61 &	F9.5V	&	\checkmark	& \checkmark \\
    \vdots & \vdots     & \vdots    & \vdots & \vdots & \vdots \\
    \enddata
    \tablecomments{$^a$From \citet{Mamajek+24}. Select entries are shown to illustrate the table form and content. The full machine-readable table (229 rows) is available online from the journal. }
\end{deluxetable}
\vspace{-1cm}

\begin{figure}[!htbp]
    \centering
    \includegraphics[width=1\linewidth,clip=true,trim=8cm 3cm 8cm 3cm]{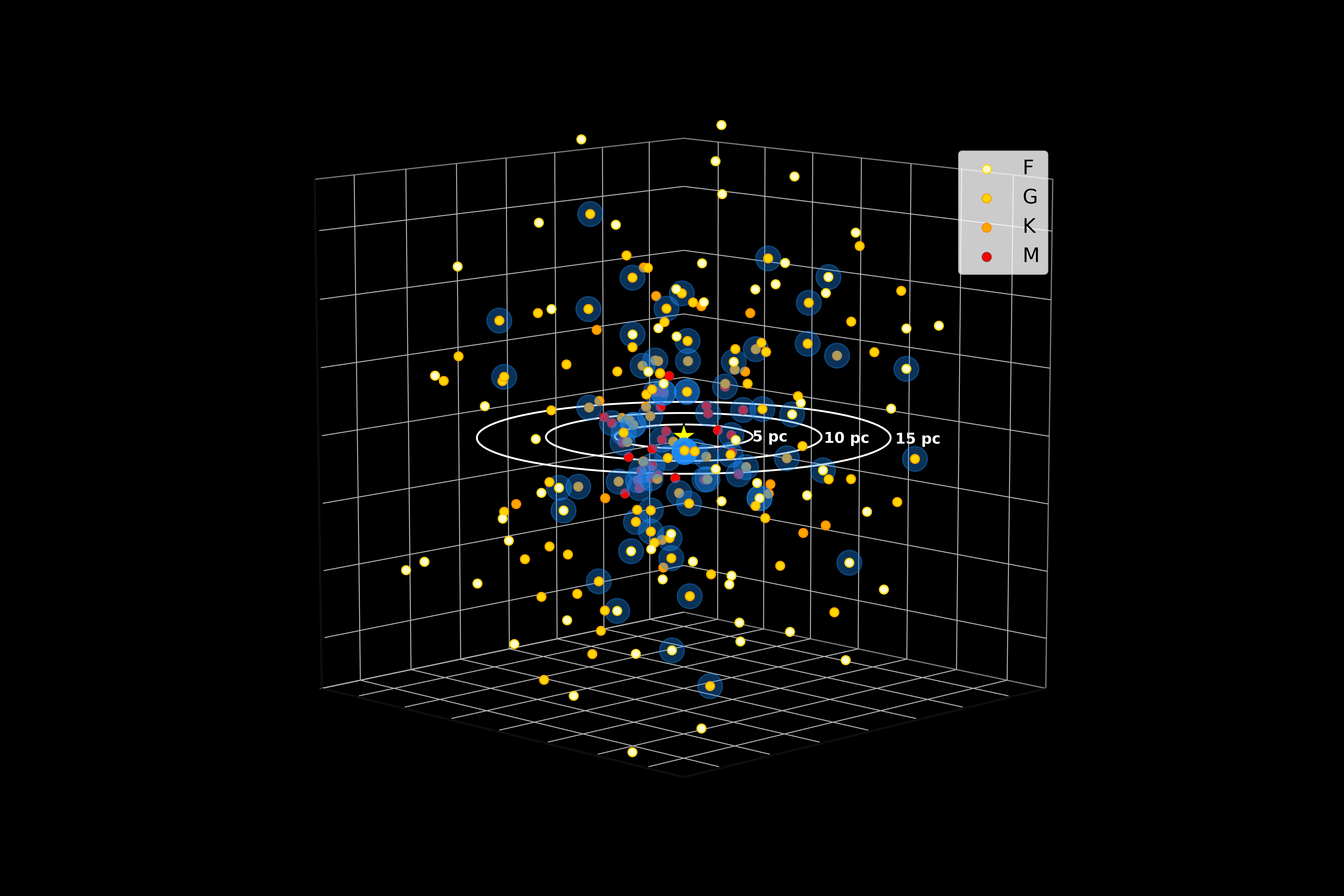}
    \caption{Map of nearby potential direct imaging target stars, color-coded by spectral type. Stars with blue halos have available X-ray observations. The yellow star at the center of the diagram represents the Sun, with white concentric rings showing distances of 5 pc, 10 pc, and 15 pc. An animated version of this figure (8 seconds in duration) is available from the online journal showing the 3D distribution through different elevation and azimuthal angles (colors and symbols are the same as the static figure).}
    \label{fig:nearby_stars_3d}
\end{figure}

Using this input list of 229 viable targets for direct imaging campaigns, we searched the  archives of the {\em XMM-Newton Observatory} and the {\em Chandra X-ray Observatory} for publicly available data. Many M stars (and some K stars) in this initial target list were previously studied as part of the Measurements of the Ultraviolet Spectral Characteristics of Low-mass Exoplanetary systems (MUSCLES) and Mega-MUSCLES {\em Hubble Space Telescope Treasury} programs, which characterized the X-ray/UV emission from K and M exoplanet host stars. We incorporate these results, presented in \citet{Brown+23}, into our work here (see Section~\ref{sec:discussion}). We retrieve 192 \XMM\ and \Chandra\ observations for 57 stellar systems containing Sun-like and low-mass stars from the X-ray archives, amounting to $\sim$3 Msec of observing time (not including the MUSCLES/Mega-MUSCLES observations). An additional 19 stellar systems have been observed by \Swift. Figure~\ref{fig:nearby_stars_3d} shows a map of the 3D spatial distribution of these nearby stars and indicates which stars have X-ray observations available.

There are three M-dwarfs highlighted in \citet{Mamajek+24} as strong potential HWO targets (Lacaille 8760, Lacaille 9352, and Lalande 21185). Both Lacaille 8760 and Lacaille 9352 were observed by \Swift\ and are included in this work. A detailed study of the panchromatic spectrum and habitability potential of Lalande 21185 is in preparation ({\it Schwieterman et al., in prep}) and is not included in the sample here. Roughly half of the target stars contained in our study are F- and G-type stars. Table~\ref{tab:stellar_properties} provides a summary of the stellar physical parameters for all stars included in our study, including the effective temperature (\Teff), spectral type \citep[from][]{Gray+06}, stellar mass ($M$), bolometric luminosity (\Lbol), and approximate age (in Gyr). All data were compiled using the NASA Exoplanet Archive\footnote{See \url{https://exoplanetarchive.ipac.caltech.edu/}}. Right ascension (R.A.), declination (Dec.), and distances were taken from Gaia DR3 \citep{GaiaDR3}. More detailed discussion of each stellar system is provided in Appendix~\ref{appendix:individual_stars}.

\section{X-ray Data} \label{sec:data}
We retrieved all publicly available imaging observations for our target stars from the \XMM\footnote{See \url{https://nxsa.esac.esa.int/nxsa-web/\#home}} and \Chandra\footnote{See \url{https://cda.harvard.edu/chaser/}} archives. Table~\ref{tab:observation_log} summarizes the number of observations and the approximate exposure times utilized in this work. Below, we provide a more detail description of how the observations were processed and the methods used to extract light curves and spectra for each facility.

\subsection{XMM-Newton}\label{sec:data_xmm}
We retrieved \XMM\ observations from the \XMM\ Science Archive and reprocessed the data from the raw \texttt{evt1} files using SAS 18\footnote{\url{https://www.cosmos.esa.int/web/xmm-newton/sas}}. The European Photon Imaging Camera (EPIC) on board \XMM\ carries a set of three X-ray CCD arrays: one array uses pn-CCDs (referred to as PN) and two utilize metal oxide semi-conductor CCDs (referred to as MOS1 and MOS2). The PN data were reprocessed with \texttt{epproc} and the MOS data were reprocessed using \texttt{emproc} and filtered on an energy range of 0.2-15 keV. Due to the brightness of many of the stars, pile-up (when two or more photons fall on a pixel in less than the readout time) is a significant concern. The PATTERN parameter records the number and pattern of CCD pixels that are triggered for a given event; we use the recommended PATTERN$=$0 to mitigate pile-up effects\footnote{See the \XMM\ ABC Guide at \url{https://heasarc.gsfc.nasa.gov/docs/xmm/sl/epic/image/sas_cl.html}}. Background light curves were extracted using \texttt{evselect} and inspected for background flares, and good time intervals (GTIs) were generated to reject observation times exhibiting strong background flaring events. 

We inspected the reprocessed, cleaned images for evidence of an X-ray source at the location of the target star. If the target star was detected at $\gtrsim$500 counts in the PN image, we extracted source light curves (binned to 100 s) to search for count-rate variability. We used \texttt{evselect} to extract both PN and MOS1/2 spectra for stars with $\gtrsim$2000 net counts, \texttt{rmfgen} to generate redistribution matrix files (RMFs), and \texttt{arfgen} to generate ancillary response functions (ARFs) for all available observations. All spectra were binned to contain at least 25 counts per bin.

The majority ($\sim$83\%) of the observations analyzed in this work are from the \XMM\ archive, containing $\sim$75\% of the total exposure time. However, with a spatial resolution of $\sim$15$^{\prime\prime}$, only stellar binaries with angular separations larger than $\sim$30$^{\prime\prime}$ can be resolved with the PN camera. Figure~\ref{fig:XMM_vs_Chandra_imaging} (left panel) shows a 10 ks \XMM/PN image of the binary system GL 412, where the two components (and a third, unidentified X-ray source) are resolved.

\startlongtable
\centerwidetable
\begin{deluxetable*}{ccccccccccc}
\tablecaption{Properties of Sample Stars}
\label{tab:stellar_properties}
    \tabletypesize{\scriptsize}
    \setlength{\tabcolsep}{0.5pt}
    \tablehead{
    \colhead{Star} & \colhead{Alternate} & \colhead{R.A.} & \colhead{Dec.} & \colhead{Spectral} & \colhead{Distance} & \colhead{Mass} & \colhead{$T_{\rm eff}$} & \colhead{\Lbol} & \colhead{Age$^*$} & \colhead{} \\ \cline{3-4}
    Name & Name(s)   &  \multicolumn{2}{c}{(J2000)}     &  Type$^*$   & (pc)    & (\Msun) & (K)      & (\Lsun) & (Gyr)   & References
    }
    \colnumbers
    \startdata
    $\eta$ Crv & HIP 61174 & 12:32:03.75 & -16:11:46.55 & F2V & 17.96$\pm$0.17 & 1.48$^{+0.29}_{-0.19}$ & 6854$^{+124}_{-99}$ & 4.62$^{+0.22}_{-0.25}$ & 1.5$^{+0.2}_{-0.4}$ & 1,2,50 \\
    $\xi$ Oph & HIP 84893 & 17:21:00.68 & -21:06:49.81 & F2V & 17.06$\pm$0.10 & 1.44$^{+0.30}_{-0.20}$ & 6756$^{+110}_{-127}$ & 4.13$^{+0.17}_{-0.18}$ & \nodata & 1,2 \\
    $\upsilon$ And A & HD 9826 & 01:36:47.60 & 41:24:13.54 & F8V & 13.41$\pm$0.06 & 1.29$\pm$0.04 & 6157$\pm$112 & 3.35$^{+0.10}_{-0.14}$ & 3.12 & 1,2,29 \\
    $\iota$ Hor & HR 810, HD 17051      & 02:42:34.03 & -50:47:57.54 & F9V      & 17.32$\pm$0.02 & 1.17$^{+0.17}_{-0.15}$ & 6146$^{+106}_{-184}$ & 1.74$^{+0.04}_{-0.06}$ & 0.47; 2.72 & 1-4,9 \\
    $\nu$ Phe   & HIP 5862      &  01:15:12.13 & -45:31:51.14   & F9V & 15.17$\pm$0.04 & 1.15$^{+0.18}_{-0.14}$ & 6119$^{+130}_{-154}$ & 2.00$\pm$0.06 & 2.88 & 1,2,3,9 \\
    $\gamma$ Pav & HIP 105858 & 21:26:26.81 & -65:21:45.50 &   F9V & 9.27$\pm$0.04 & 1.15$^{+0.18}_{-0.14}$ & 6109$^{+110}_{-176}$ & 1.47$\pm$0.05 & 1,7.25 & 1,3,44,45 \\
    $\beta$ Vir & HIP 57757 & 11:50:42.51 & 01:45:48.67 & F9V & 11.12$^{+0.07}_{-0.06}$ & 1.13$^{+0.16}_{-0.14}$ & 6071$^{+120}_{-90}$ & 3.77$^{+0.11}_{-0.14}$ & 2.96 & 1,2 \\
    GJ 1095 & HIP 35136 & 07:15:50.19 & +47:14:20.89 & F9V & 16.86$^{+0.19}_{-0.20}$ & 1.12 & 5866 & 1.54$\pm$0.06 & 1, 7 & 1,2,10 \\
    LHS 237 & GJ 288, HIP 37853 & 07:45:34.63 & -34:09:53.86 & F9V & 15.20$\pm$0.13  & 1.11 & 5744 & 1.51$\pm$0.06 & \nodata & 1,2 \\
    $\iota$ Per & HIP 14632 & 03:09:06.10 & +49:36:46.35 & F9.5V & 10.51$\pm$0.03 & 1.08$^{+0.18}_{-0.10}$ & 5938$^{+141}_{-169}$ & 2.23$^{+0.04}_{-0.10}$ & 4.20$\pm$0.48 & 1,2,9 \\
    $\beta$ Hyi & HIP 2021 & 00:25:55.80 & -77:15:10.06 & G0V & 7.47$\pm$0.03 & 1.14$^{+0.01}_{-0.02}$ & 5839 & 3.70$\pm$0.05 & 6.32 & 1,2,9 \\
    $\beta$ Com & HIP 64394             & 13:11:51.43 & +27:52:55.58 & G0V      & 9.18$\pm$0.03 & 1.10$^{+0.14}_{-0.13}$ & 5969$^{+188}_{-100}$ & 1.41$^{+0.05}_{-0.04}$ &  $<$1.12, 1.7 & 1,2,9 \\
    LHS 208     & HD 39091, $\pi$ Men   & 05:37:11.89 & -80:27:52.05 & G0V      & 18.27$\pm$0.02 & 1.094$\pm$0.039 & 6037$\pm$45 & 1.44$\pm$0.02 & 2.98$^{+1.40}_{-1.30}$ & 5 \\
    $\rho$ CrB & GJ 9537, HIP 78459  & 16:01:02.41 & +33:18:00.28 & G0V & 17.47$\pm$0.02 & 1.05$^{+0.12}_{-0.15}$ & 5833$^{+141}_{-124}$ & 1.81$^{+0.06}_{-0.04}$ & 11.04 & 1,2,9 \\
    GL 672 & HIP 84862 & 17:20:39.74 & +32:27:47.22 & G0V & 14.54$\pm$0.02 & 1.05$^{+0.16}_{-0.10}$ & 5817$^{+120}_{-84}$ & 1.29$^{+0.04}_{-0.03}$ & 12.04 & 1,2,9 \\
    $\chi^1$ Ori & HIP 27913, GJ 222    & 05:54:22.76 & +20:16:33.07 & G0V      & 8.66$\pm$0.08  & 1.028$^{+0.03}_{-0.028}$ & 5822 & 1.17$\pm$0.06 & 0.35; 4.32 & 1,2,6,9 \\
    GL 788 & HIP 100017 & 20:17:32.60 & +66:51:18.04 & G0V & 17.57$\pm$0.52 & 1.02$\pm$0.03 & 5886 &  1.18$\pm$0.10 & 3.76$^{+1.88}_{-1.92}$ & 2,9 \\
    47 UMa      & GJ 407, HD 95128 & 10:59:27.53 & +40:25:49.80 & G0V      & 13.80$\pm$0.03 & 1.005$\pm$0.047          & 5829$\pm$95           & 1.58$\pm$0.04 & 6.48$^{+1.44}_{-1.04}$ & 9,29 \\
    GL 620.1A   & HD 147513             & 16:24:01.39 & -39:11:34.68 & G1V      & 12.903$\pm$0.02 & 1.06$^{+0.12}_{-0.14}$ & 5873$^{+110}_{-103}$ & 1.032$^{+0.003}_{-0.004}$ & 2; $<$1; 0.4 & 1-4,9,19 \\ 
    GL 311      & $\pi^1$ UMa, HD 72905  & 08:39:11.64 & +65:01:16.67 & G1.5V    & 14.27$^{+0.14}_{-0.15}$ & 1.01 & 5920 & 1.03$\pm$0.05 & 0.2 & 2,6,8 \\
    $\alpha$ Cen A & GJ 559A, LHS 50    & 14:39:36.49 & -60:50:02.37 & G2V      & 1.347$\pm$0.003 & 1.06$^{+0.05}_{-0.04}$ & 5801 & 1.61$\pm$0.07 & 7.84$^{+1.08}_{-1.28}$ & 2,9 \\
    44 Boo A & $\iota$ Boo, HIP 73695 A & 15:03:46.74 & +47:39:15.89 & G2V & 12.8$\pm$0.2 & 1.04$\pm$0.1 & 5877 & 1.552 & 1.5 & 46,47,48 \\
    18 Sco      & GJ 616, HIP 79672     & 16:15:37.52 & -08:22:17.91 & G2V      & 14.13$\pm$0.02 & 1.04$^{+0.12}_{-0.14}$ & 5791$^{+117}_{-93}$ & 1.09$^{+0.04}_{-0.02}$ & 5.84$^{+1.88}_{-1.96}$ & 1,9,10 \\
    51 Peg & HIP 113357, GJ 882 & 22:57:28.22 & +20:46:08.78 & G2V & 15.46$\pm$0.03 & 1.03$^{+0.17}_{-0.09}$ & 5758$^{+102}_{-120}$ & 1.37$^{+0.02}_{-0.05}$ & 6.76 & 1,2 \\
    $\zeta^1$ Ret & LHS 171, HIP 15330 & 03:17:49.26 & -62:34:20.76 & G2V & 12.12$\pm$0.08 & 0.96$^{+0.03}_{-0.06}$ & 5699 & 0.80$\pm$0.05 & 1.56$\pm$0.44 & 2,9 \\
    HD 136352   & GJ 582, $\nu^2$ Lupi  & 15:21:45.55 & -48:19:07.88 & G2V      & 14.68$\pm0.02$ & 0.87$\pm$0.04    & 5664$\pm$61 & 1.04$\pm$0.06 & 12.3$^{+1.2}_{-2.9}$ & 11,12 \\ 
    GL 327 & HIP 43726 & 08:54:17.50 & -05:26:03.56 & G3V & 17.42$^{+0.44}_{-0.47}$ & 1.09$\pm$0.04 & 5790 & 1.09$^{+0.05}_{-0.04}$ & 1.32 & 2,9 \\
    $\mu$ Ara   & HD 160691             & 17:44:08.68 & -51:50:05.65 & G3IV/V   & 15.60$\pm$0.03  & 1.03$^{+0.15}_{-0.11}$ & 5772$^{+119}_{-84}$ & 1.90$^{+0.04}_{-0.06}$ & 5.7$\pm$0.6 & 1,2,7 \\
    $\kappa^1$ Cet & HD 20630           & 03:19:21.98 & +03:22:14.22 & G5V      & 9.14$\pm$0.03  & 1.02$^{+0.16}_{-0.10}$ & 5712$^{+161}_{-124}$ & 0.86$^{+0.02}_{-0.03}$ & 0.35; 0.7; 2.2 & 1,2,6,20,21 \\
    HD 140901  & GL 599 & 15:47:28.54 & -37:55:02.15 & G7IV/V & 15.25$\pm$0.01 & 1.06$\pm$0.05 & 5587$^{+153}_{-125}$ & 1.22$\pm$0.03 & 3.2$\pm$2.8 & 1,49 \\
    GJ 777A & LHS 3510, HD 190360 & 20:03:38.25 & +29:53:40.08 & G7IV/V & 16.01$\pm$0.02 & 0.99$\pm$0.08 & 5552$\pm$11 & 1.20$\pm$0.06 & 13.4 & 1,2,9 \\
    $\xi$ Boo A & 37 Boo, HD 131156 & 14:51:23.53 & +19:06:00.97 & G7V & 6.71$\pm$0.16 & 0.90$\pm$0.04 & 5551$\pm$21 & 0.55$\pm$0.04 & $<$4 & 38 \\
    $\delta$ Pav & GJ 780               & 20:08:46.81 & -66:11:13.53 & G8IV     & 6.09$\pm$0.01   & 0.99$^{+0.14}_{-0.10}$ & 5590$^{+156}_{-140}$ & 1.25$^{+0.02}_{-0.04}$ & 6.2 & 1,2\\ 
    LHS 2156 & HIP 47080 & 09:35:38.55 & +35:48:32.33 & G8IV/V & 11.20$\pm$0.02 & 0.96$^{+0.16}_{-0.09}$ & 5499$^{+156}_{-136}$ & 0.80$^{+0.02}_{-0.04}$ & \nodata & 1,2 \\
    82 Eri      & GJ 139, HD 20794      & 03:20:00.08 & -43:03:59.59 & G8V      & 6.003$\pm$0.009  & 0.94$\pm$0.12  & 5413$^{+91}_{-118}$ & 0.65$\pm$0.01 & 5.76; $>12$ & 1,2,9,13 \\ 
    $\tau$ Ceti & HD 10700, GJ 71   & 01:44:02.17 & -15:56:01.25 & G8.5V   & 3.603$\pm$0.007      & 0.92$^{+0.12}_{-0.10}$ & 5333$^{+124}_{-75}$ & 0.51$\pm$0.01 & $>$12 & 1,2,9 \\
    44 Boo Ba & & 15:03:46.74 & +47:39:15.89 & K0V & 12.8$\pm$0.2 & 0.98 & 5300 & 0.51 & \nodata & 46,47,48 \\
    55 Cnc A & $\rho^1$ Cnc, HIP 43587 & 08:52:35.22 & +28:19:47.22 & K0IV/V & 12.59$\pm$0.01 & 0.90$^{+0.14}_{-0.09}$ & 5252$^{+123}_{-172}$ & 0.64$\pm$0.02 & 9.5 & 1,2 \\
    70 Oph A    & LHS 458, HIP 88601A   & 18:05:27.47 & +02:29:42.81 & K0V      & 5.15$\pm$0.03  & 0.89$\pm$0.02  & 5300$\pm$50 & 0.65$\pm$0.02 & 6.2$\pm$1.0 & 1,22 \\ 
    40 Eri A    & GJ 166A, HD 26965     & 04:15:13.91 & -07:40:05.08 & K0.5V    & 5.04$\pm$0.01  & 0.85$^{+0.11}_{-0.10}$ & 5092$^{+140}_{-149}$ & 0.43$\pm$0.01 & 6.9$\pm$4.7 & 1,14,23 \\ 
    $\delta$ Eri & HIP 17378 & 03:43:14.80 & -09:45:36.30 & K1III/IV & 9.04$\pm$0.07 & 1.19 & 5095 & 3.37$^{+0.07}_{-0.06}$ & 6.28 & 2 \\
    GL 451A     & Groombridge 1830  & 11:53:04.16 & +37:41:34.18 & K1IV     & 9.16$^{+0.16}_{-0.17}$   & 0.66 & 4950 & 0.23$\pm$0.04 & 5.0$\pm$0.3 & 2,6 \\
    GL 117      & HIP 13402, HD 17925    & 02:52:32.56 & -12:46:14.00 & K1.5V    & 10.355$\pm$0.005 & 0.89$^{+0.09}_{-0.13}$ & 5225$^{+91}_{-136}$ & 0.40$\pm$0.01 & 1.5;$<$1.2;0.1 & 1,2,6,9 \\ 
    $\alpha$ Cen B & GJ 59 B, LHS 51   & 14:39:35.06 & -60:50:15.10 & K2IV      & 1.347$\pm$0.03 & 0.87$\pm$0.07 & 5234$\pm$63 & 0.52$\pm$0.05 & 8, $>$11 & 2,9,30 \\
    GL 783 A     & HD 191408A, LHS 486   & 20:11:12.54 & -36:06:29.58 & K2.5V    & 6.05$\pm$0.03  & 0.74 & 4922 & 0.28$\pm$0.10 & 7.05$\pm$0.65 & 1,2,6,25 \\ 
    GL 183      & HD 32147, HIP 23311   & 05:00:49.59 & -05:45:30.96 & K3V      & 8.845$\pm$0.003  & 0.81$^{+0.11}_{-0.09}$ & 4931$^{+133}_{-134}$ & 0.29$\pm$0.01 & 2 & 1,3,6 \\ 
    GL 892 & HD 219134 & 23:13:21.06 & +57:10:10.80 & K3V & 6.531$\pm$0.004 & 0.79$\pm$0.03 & 4817$\pm$62 & 0.29$\pm$0.01 & 11.0$\pm$2.2,12.46 & 1,2,36 \\
    GJ 667 A    & HIP 84709             &  17:18:56.53 & -34:59:25.40       & K3V     & 6.97$\pm$0.83 & 0.59          & 4810          & 0.13$^{+0.17}_{-0.09}$ & \nodata  & 2,3,31 \\
    LHS 1875 & GL 250, HIP 32984 & 06:52:17.47 & -05:10:25.42 & K3.5V & 8.747$\pm$0.004 & 0.76$^{+0.09}_{-0.10}$ & 4719$^{+157}_{-129}$ & 0.22$^{+0.02}_{-0.01}$ & 3.1,$<$0.48 & 1,2,9 \\
    GL 570 A & HIP 73184 &  14:57:29.18 & -21:25:23.31 & K4V & 5.881$\pm$0.003 & 0.75$^{+0.08}_{-0.10}$ & 4681$^{+165}_{-126}$ & 0.22$^{+0.01}_{-0.02}$ & $<$0.6;3 & 1,2,9 \\
    $\xi$ Boo B &  & 14:51:23.19 & +19:06:03.97& K4V & 6.71$\pm$0.16 & 0.66$\pm$0.07 & 4350$\pm$150 & 0.10$\pm$0.02 & $<$4 & 38 \\
    44 Boo Bb & & 15:03:46.74 & +47:39:15.89 & K4V & 12.8$\pm$0.2 & 0.55 & 5035 & 0.24 & \nodata & 46,47,48 \\
    GJ 667 B    & WDS J17190-3459 B     &   \nodata & \nodata   & K4V     & \nodata       & \nodata       & \nodata     & \nodata     & \nodata & 31 \\
    70 Oph B    & LHS 459   & 18:05:27.82 & +02:29:39.12 & K5V & 5.15$\pm$0.03 & 0.73$\pm$0.01 & 4390$\pm$200 & 0.15$\pm$0.02 & 6.2$\pm$1.0 & 1,22 \\
    61 Cygni A  &  HIP 104214           & 21:06:59.64 & +38:45:49.90 & K5V      & 3.497$\pm$0.001   & 0.67$^{+0.08}_{-0.09}$  & 4304$^{+161}_{-128}$ & 0.15$\pm$0.01 & 6 & 1,2,26  \\ 
    61 Cygni B  & HIP 104217             & 21:07:00.88 & +38:45:21.86 & K7V      & 3.497$\pm$0.001   & 0.61$^{+0.08}_{-0.09}$ & 3949$^{+156}_{-137}$ & 0.104$^{+0.012}_{-0.009}$ & 6 & 1,2,26 \\ 
    GL 412A    & HIP 54211             & 11:05:22.09 & +43:31:51.41 & M1V      & 4.83$\pm$0.03   & 0.38 & 3688 & 0.020$^{+0.090}_{-0.014}$ & 3 & 2,15,28 \\
    GL 570 B & (HIP 73182) & 14:57:27.70 & -21:25:07.94 & M1.5V & 5.86$\pm$0.06 & 0.55$\pm$0.05 & 3345 & 0.13$\pm$0.09 & \nodata & 1,2,37 \\
    GJ 832  & LHS 3685, HD 204961   & 21:33:33.90 & -49:00:45.47 & M1.5V & 4.964$\pm$0.001 & 0.44$\pm$0.02 & 3601 & 0.03$^{+0.09}_{-0.02}$ & 9.24 & 1,2,4 \\
    GJ 667 C    &                        &  17:19:00.30 & -34:59:52.06       & M1.5V  & 7.244$\pm$0.005 & 0.33$\pm$0.02 & 3350$\pm$50 & 0.0137$\pm$0.0009 & $>$2 & 31 \\
    Kapteyn's Star & GJ 191             & 05:11:50.38 & -45:02:37.73 & M2V    & 3.9331$\pm$0.0004  & 0.28$\pm$0.02 & 3742$\pm$157 & 0.016$\pm$0.004 & 11.5$^{+0.5}_{-1.5}$ & 1,2,16 \\ 
    Wolf 1055 & HD 180617, GL 752 A & 19:16:55.26 & +05:10:08.04 & M2.5V & 5.912$\pm$0.02 & 0.48$\pm$0.02 & 3534$\pm$51 & 0.032$\pm$0.001 & \nodata & 40 \\
    AD Leo   & GJ 388 & 10:19:35.70 & 19:52:11.30 & M3V & 4.964$\pm$0.001 & 0.423$\pm$0.012 & 3477$\pm$23 & 0.0236$\pm$0.0001 & 0.4$\pm$0.1 & 39 \\
    GL 570 C & HIP 73182 & 14:57:27.70 & -21:25:07.94 & M3V & 5.86$\pm$0.06 & 0.39$\pm$0.03 & \nodata & 0.018 & \nodata & 37 \\
    Wolf 1061       & GJ 628, LHS 419 & 16:30:17.96 & -12:40:04.26 & M3.5V & 4.306$\pm$0.001 & 0.30$\pm$0.02 & 3309$\pm$157 & 0.011$^{+0.003}_{-0.002}$ & \nodata & 1,2 \\
    Luyten's Star & GJ 273 & 07:27:25.11 & +05:12:33.78 & M3.5V & 5.92$\pm$0.02 & 0.29$\pm$0.02 & 3382$\pm$49 & 0.0087$^{+0.0063}_{-0.0065}$ & \nodata & 1,2,43 \\
    55 Cnc B &   & 08:52:40.28 & +28:18:54.91 & M4.5V & 12.48$\pm$0.02 & 0.26$\pm$0.02 & 3187$\pm$157 & 0.008$\pm$0.002 & \nodata & 1,35 \\
    GL 783B     & HD 191408B, LHS 487   & 20:11:12.80 & -36:06:32.17 & M4.5V & 6.05$\pm$0.03 & 0.24 & \nodata & 0.0009 & 7.05$\pm$0.65 & 25 \\
    GJ 777B & LHS 3509 & 20:03:37.57 & +29:53:45.49& M4.5V & 15.97$\pm$0.01 & 0.21$\pm$0.02 & 3099$\pm$157 & 0.0047$\pm$0.0012 & \nodata & 1,34 \\
    40 Eri C    & LHS 25      & 04:15:19.12 & -07:40:15.25 & M4.5V & 5.04$\pm$0.01 & 0.204$\pm$0.006 & 3100 & 0.008  & 6.9$\pm$4.7 & 1,14,23,24 \\ 
    $\upsilon$ And B &  & 01:36:50.16 & 41:23:25.98 & M4.5V & 13.47$\pm$0.02 & 0.19$\pm$0.02 & 3159$\pm$157 & 0.0044$^{+0.0012}_{-0.0011}$ & \nodata & 29 \\ 
    Proxima Centauri & GJ 551           & 14:29:34.16 & -62:40:33.87 & M5.5V    & 1.3012$\pm$0.0003   & 0.13$\pm$0.02 & 2900$\pm$100 & 0.0016$^{+0.0004}_{-0.0006}$ & 4.85 & 1,17,18 \\
    GL 412B     & WX UMa, LHS 39    & 11:05:24.50 & +43:31:33.27 & M6.6V & 4.83$\pm$0.03 & 0.095 & 2700 & 0.00095 & 3 & 27,28 \\
    VB 10 & GL 752 B & 19:16:57.61 & +05:09:01.59 & M8V & 5.912$\pm$0.02 & 0.0881$^{+0.0026}_{-0.0024}$ & 2508$^{+63}_{-60}$ & 0.000499±0.000004 & 1 & 41,42 \\
    \enddata
    \tablerefs{1: \citet{TESSinput}, 2: \href{https://nexsci.caltech.edu/missions/EXEP/EXEPstarlist.html}{ExoCat-1}, \citet{Turnbull15}, 3: \citet{Gray+06}, 4: \citet{Sanz-Forcada+10}, 5: \citet{Huang+18}, 6: \citet{Mamajek+08}, 7: \citet{Benedict+22}, 8: \citet{GaiaEDR3}, 9: \citet{Takeda+07}, 10: \citet{Gray+03}, 11: \citet{Delrez+21}, 12: \citet{Udry+19}, 13: \citet{Pepe+11}, 14: \citet{Ma+18}, 15: \citet{Mann+15}, 16: \citet{Anglada-Escude+14}, 17: \citet{Faria+22}, 18: \citet{Kervella+03}, 19: \citet{Ghezzi+10}, 20: \citet{Dorren+94}, 21: \citet{Gudel+97}, 22: \citet{Eggenberger+08}, 23: \citet{Mason+17}, 24: \citet{Johnson+83}, 25: \citet{Turnbull+03}, 26: \citet{Kervella+08}, 27: \citet{Casagrande+08}, 28: \citet{Mann+15}, 29: \citet{Rosenthal+21}, 30: \citet{Santos+13}, 31: \citet{Anglada-Escude+13}, 32: \citet{Bidelman85}, 33: \citet{Feng+19}, 34: \citet{Hawley+96}, 35: \citet{Alonso-Floriano+15}, 36: \citet{Gillon+17}, 37: \citet{Mariotti+90}, 38: \citet{Fernandes+98}, 39: \citet{Kossakowski+22}, 40: \citet{Burt+21}, 41: \citet{Pravdo+09}, 42: \citet{Pineda+21}, 43: \citet{Astudillo-Defru+17}, 44: \citet{Holmberg+09}, 45: \citet{Mosser+08}, 46: \citet{Zirm11}, 47: \citet{Latkovic+21}, 48: \citet{Zasche+09}, 49: \citet{Philipot+23}, 50: \citet{Nordstrom+04} \\
    $^*$Significant disagreements in literature are discussed in Appendix~\ref{appendix:individual_stars}.}
\end{deluxetable*}

\begin{deluxetable}{cccccc}[!htbp]
    \caption{Summary of X-ray Observations}\label{tab:observation_log}
    \tabletypesize{\footnotesize}
    \setlength{\tabcolsep}{1.5pt}
    \tablehead{
                      & \multicolumn{2}{c}{\XMM} && \multicolumn{2}{c}{\Chandra} \\ \cline{2-3} \cline{5-6}
    \colhead{Star} & \# Obs. & Exp. Time (ks)  && \# Obs. & Exp. Time (ks) 
    }
    \startdata
    $\eta$ Crv   & 0 & \nodata && 4 & 38.5 \\
    $\xi$ Oph    & 0 & \nodata && 1 & 19.8 \\
    $\upsilon$ And & 1 & 5.3    && 4 & 58.9 \\
    $\iota$ Hor  & 32 & 211.3   && 0 & \nodata    \\
    $\nu$ Phe    & 1 & 16.0     && 0 & \nodata \\
    $\gamma$ Pav & 1 & 15.9     && 0 & \nodata \\
    $\beta$ Vir  & 1 & 41.1     && 0 & \nodata \\
    GJ 1095      & 0 & \nodata  && 1 & 96.2  \\
    LHS 237      & 1  & 15.6    && 0 & \nodata  \\
    $\iota$ Per  & 0 & \nodata  && 1 & 4.9  \\
    \vdots      & \vdots & \vdots && \vdots & \vdots \\
    \hline
    Total       & 151 & 2235.7 && 41 & 815.0 \\
    \enddata
    \tablecomments{Select entries are shown to illustrate the table form and content. The full machine-readable table (57 rows) is available online from the journal.}
\end{deluxetable}

\vspace{-1.5cm}

\subsection{Chandra}\label{sec:data_chandra}
We retrieved \Chandra\ ACIS imaging observations from the \Chandra\ archive\footnote{\url{https://cda.harvard.edu/chaser/}} and reprocessed all \texttt{evt1} data using the CIAO v4.15 \citep{Fruscione+06} task \texttt{chandra\_repro} and standard reduction procedures. We ran the point source detection algorithm \texttt{wavdetect} on each individual exposure to generate a list of X-ray source positions and error ellipses. The major and minor axes of these error ellipses were increased by a factor of five, and all X-ray sources were masked so that a background light curve could be extracted for each observation. We inspected these background light curves for background flaring events. We created good time intervals (GTIs) using \texttt{lc\_clean} and filtered all event data on these GTIs. Finally, we restricted the energy range of our resulting ``clean'' \texttt{evt2} files to 0.5-7 keV.

\begin{figure*}
    \centering
    \begin{tabular}{cc}
    \includegraphics[width=0.37\linewidth]{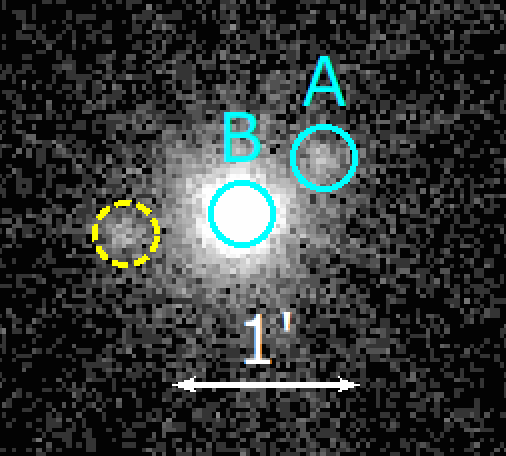} &
    \includegraphics[width=0.58\linewidth]{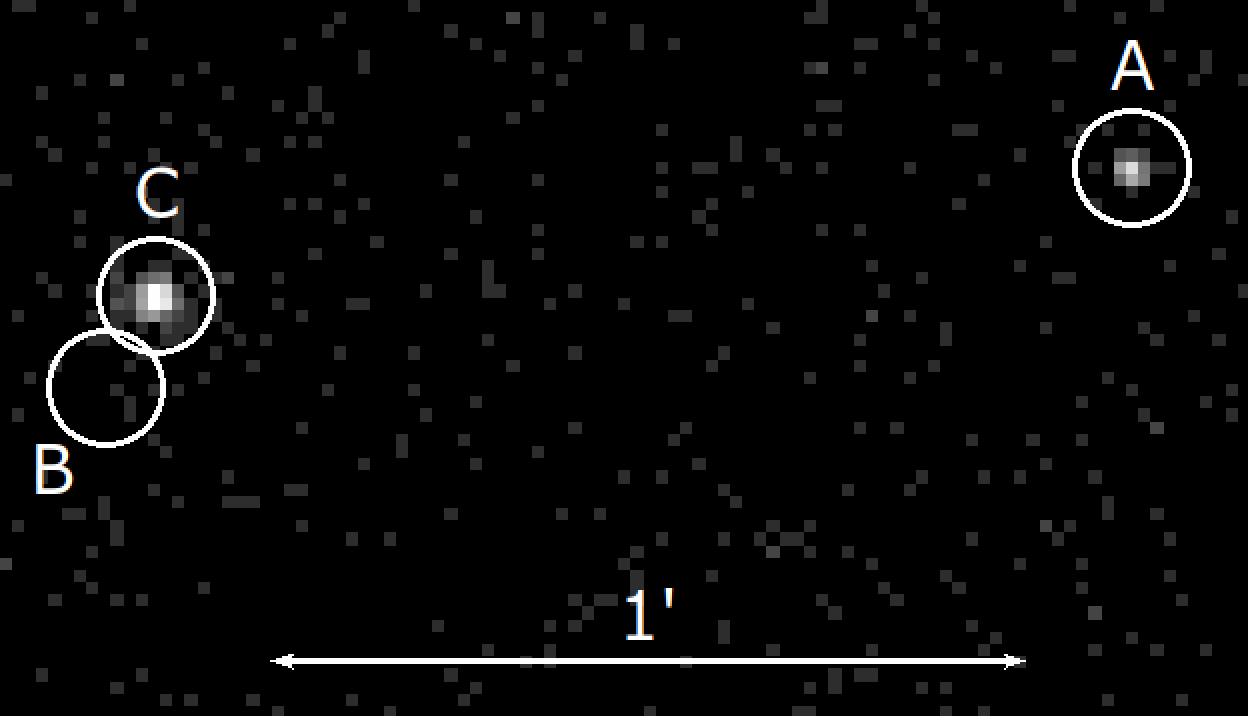} \\
    \end{tabular}
    \caption{Comparison of the spatial resolution of \XMM\ ({\it left}) and \Chandra\ ({\it right}); a 1$^{\prime}$ scale bar is shown in both images. {\it Left}: The GL 412 binary star system with the locations of GL 412A (M1V) and GL 412B/WX Ma (M6.6V) shown by cyan circles. The circles have radii of 10$^{\prime\prime}$ and are separated by $\sim$30$^{\prime\prime}$. A third, unidentified X-ray source (dashed yellow circle) is located $\sim$37$^{\prime\prime}$ from GL 412B. {\it Right}: The 40 Eri triple star system with the locations of 40 Eri A (K0V), 40 Eri B (DA white dwarf), and 40 Eri C  (M4.5V) circled in white. The separation between the B and C components is $\sim$7$^{\prime\prime}$. X-ray sources are found coincident with 40 Eri A and C, while 40 Eri B is undetected.}
    \label{fig:XMM_vs_Chandra_imaging}
\end{figure*}

We re-ran \texttt{wavdetect} on the cleaned image to identify significant X-ray sources and obtain first-pass estimates of their source properties. Due to its exquisite spatial resolution (0.496$^{\prime\prime}$ pixel size) and a typical on-axis PSF size less than 1$^{\prime\prime}$ for the ACIS instruments, \Chandra\ is the instrument of choice for resolving crowded fields. For stellar systems with multiple components and angular separations of less than a few arcseconds, \Chandra\ is currently the only telescope that can be used to definitively associate X-ray emission with a given stellar component. Figure~\ref{fig:XMM_vs_Chandra_imaging} (right panel) illustrates the spatial resolving power of \Chandra\ with a 5 ks exposure of the triple star system 40 Eri, where the brightest X-ray source is clearly associated with the position of 40 Eri C, a fainter X-ray source is found at the location of 40 Eri A, and 40 Eri B is undetected.

If the target star was detected in the cleaned \texttt{evt2} image with $\gtrsim$50 net counts, we extracted source light curves (binned to 100 s) to search for evidence of X-ray variability. For sources with $\gtrsim$500 net counts, the CIAO tool \texttt{specextract} was used to extract spectra and generate RMFs and ARFs. All spectra were binned to contain at least 10 counts per bin. 

\subsection{Swift}\label{sec:data_swift}
The X-Ray Telescope (XRT) onboard the Neil Gehrels \Swift\ Observatory additionally observed some nearby stars. \Swift\ is optimized to quickly slew to observe transient events, and typically obtains short ``snapshot'' X-ray observations ($\sim$2 ks or less) with XRT of relatively bright X-ray sources. These observations are useful for providing flux estimates (or flux upper limits) for stars of interest, but often cannot provide the longer-duration light curves or higher signal-to-noise spectra that can be obtained from \XMM\ or \Chandra. We therefore use the High Energy Astrophysics Science Archive Research Center (HEASARC)\footnote{\url{https://heasarc.gsfc.nasa.gov/cgi-bin/W3Browse/swift.pl}} to determine which stars were observed by \Swift\ but not \XMM\ or \Chandra, and we use the \Swift\ data products generator\footnote{\url{https://www.swift.ac.uk/user_objects/index.php}} to create stacked images of these additional stars and measure count rates (or count rate upper limits). The 19 stars that were observed only by \Swift/XRT, but not by \XMM\ or \Chandra, are not considered part of our primary stellar sample, but we still provide updated provide X-ray luminosity estimates in Appendix~\ref{appendix:swift}.

\section{Analysis Procedure}\label{sec:analysis}
We first inspected all X-ray observations for evidence of X-ray emission at the location of the target stars. For stars that were detected, we first extract light curves (binned to 100 s) to search for rapid variability within the observation. For brighter stars (i.e., those with $\gtrsim$2000 net counts with \XMM\ or $\gtrsim$500 net counts with \Chandra), changes in observed count rate are readily apparent by visual inspection; for fainter stars, a more quantitative assessment of variability is needed in addition to visual inspection. 

\subsection{Assessing Count Rate Variability}
We use the Anderson-Darling test, which is a non-parametric statistical test that is more sensitive to short duration variations than the more commonly used Kolmogorov-Smirnov (KS) test \citep{Feigelson+22}. The Anderson-Darling statistic, $A^2$, is computed as \citep{Rahman+04}:

\begin{equation}
    A^2 = -N-\frac{1}{N} \sum_{i=1}^N [(2i-1) \text{ln}(t_i)+(2N+1-2i)\text{ln}(1-t_i)],
\end{equation}

\noindent where $N$ is the number of photons observed and $t_i$ is the arrival time of the $i$-th bin, scaled to the interval (0,1). We use \texttt{scipy}'s \texttt{anderson} routine to compute $A^2$ and the critical values for the 5\% and 1\% significance levels when compared to a normal distribution. An $A^2$ value that is larger than a critical value for a given significance level is evidence of variability; the typical 5\% critical value is $\sim$0.75 and the typical 1\% critical value is $\sim$1.05 for the observations considered here. We additionally compare the observed light curve to a constant count rate and compute a standard reduced $\chi^2_r$ ($\chi^2$ divided by the degrees of freedom) statistic.

Variability in observed count rate is, to first order, driven by a change in X-ray flux incident on the detector. In our case, this change in flux is due to changes in the intrinsic X-ray luminosity of the target star. However, changes in the underlying spectra of the star can additionally impart more subtle changes on the light curve, as the sensitivity of \XMM\ and \Chandra\ to X-ray photons is energy-dependent. We therefore ``flag'' segments of each light curve according to the type of variability observed, hereafter referred to as the VarFlag, so that the spectral properties can be measured and an accurate count rate-to-luminosity conversion factor can be established as a function of time during the observation. The VarFlags are defined as follows:

\begin{itemize}
    \item Q: quiescent periods, when the count rate is low and constant within the uncertainties
    \item F: strong flaring events, where the count rate is above $\sim$80\% the quiescent maximum count rate
    \item D: decaying periods following flares or descending count rates that follow highly elevated count rates; these periods are roughly defined as immediately following an obvious flare where the count rate is below 80\% the maximum flare count rate but still above $\sim$150\% of the quiescent count rate
    \item E: elevated count rate periods, when count rates are approximately 50\% higher than the quiescent count rate but not obviously associated with a strong flaring event
    \item R: rising count rate periods that precede a strong flare
\end{itemize}

Dividing a single observation into a series of sub-exposures is a subjective process: we wish to define sub-exposures that are short enough to capture unique variations seen in the light curve (such as rapid flaring events, some lasting only 200-300 s) but long enough so that a sufficient number of X-ray counts are available for adequate spectral modeling (the $\sim$2000 and $\sim$500 count limits for \XMM\ and \Chandra, respectively). Our main goal is to accurately measure the X-ray luminosity of the star as a function of time.

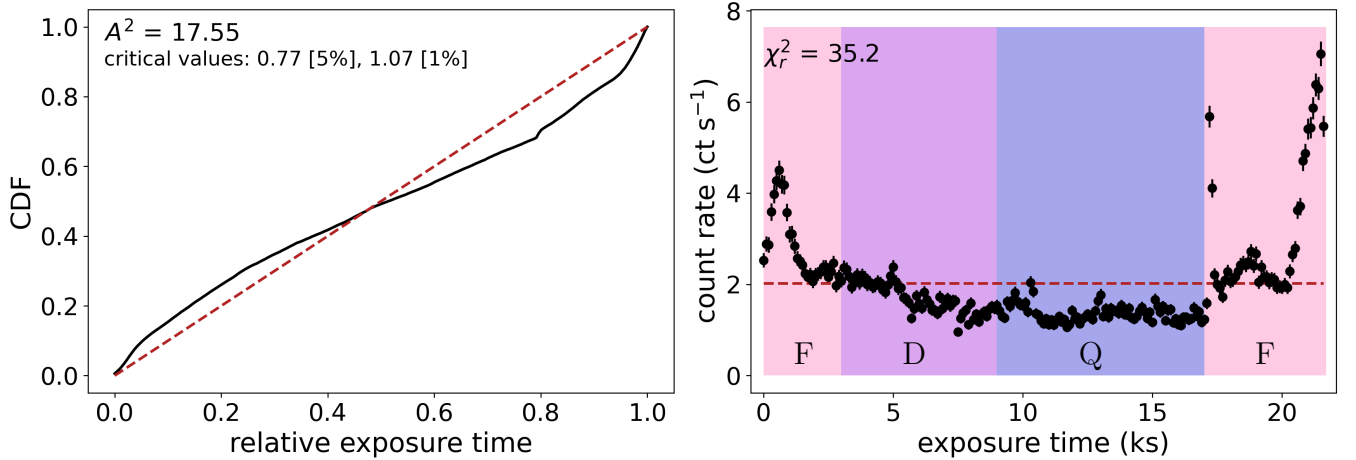
\begin{figure*}
\centering
    \begin{overpic}[width=1\linewidth]{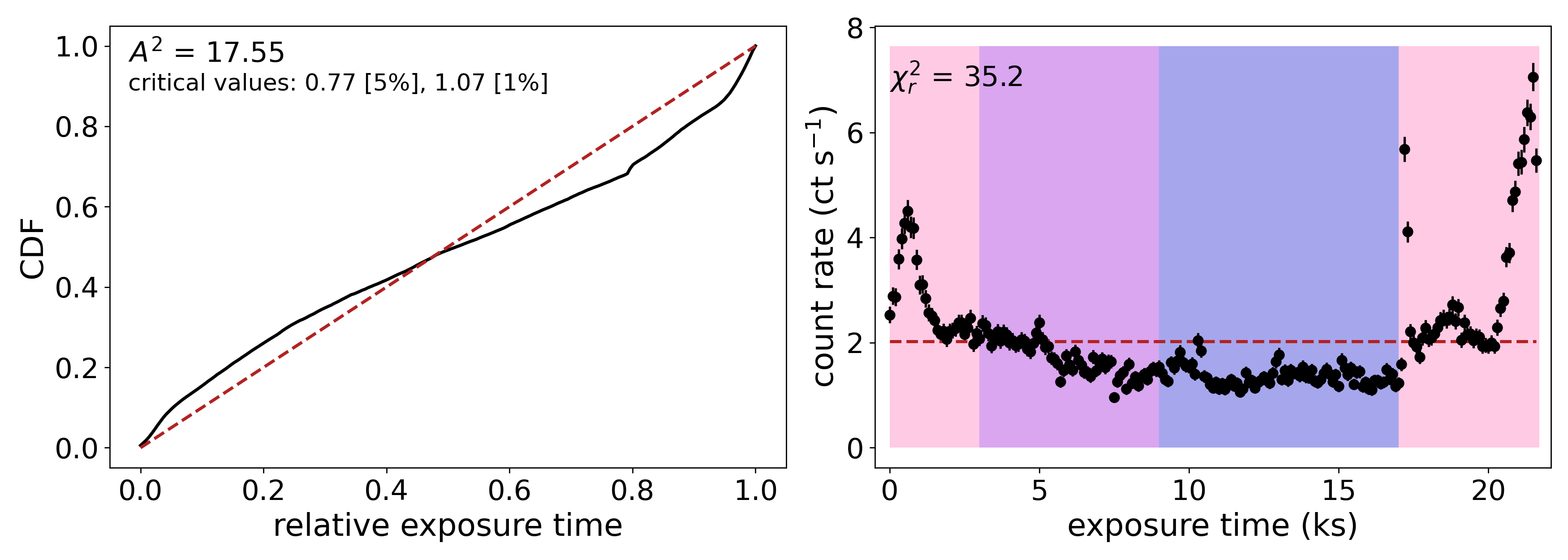}
        \put (59,8) {\large F}
        \put (67,8) {\large D}
        \put (80,8) {\large Q}
        \put (93,8) {\large F}
    \end{overpic}
    \caption{{\it Left}: Photon arrival time CDF of Proxima Centauri (\XMM\ observation ID 0801880501, black) compared to a constant count rate (dark red, dashed). The $A^2$ statistic and critical values are shown in the upper left corner. {\it Right}: The light curve data (black circles) compared to a constant count rate (dark red, dashed), with sections of the light curve color-coded by VarFlag. Quiescent periods are shown in dark blue-violet (``Q''), flaring periods in pink ``F''), and descending count rates in purple (``D'').}
    \label{fig:ProxCen_lc}
\end{figure*}

Figure~\ref{fig:ProxCen_lc} shows an example light curve of Proxima Centauri (\XMM\ observation number 0801880501) and a cumulative distribution function (CDF) of photon arrival times compared to a constant count rate. The $A^2$ statistic, critical values, and $\chi^2_r$ are shown, and the light curve has been color-coded according to the VarFlag assigned to each segment of the light curve. Table~\ref{tab:variability_metrics} summarizes the variability metrics for every X-ray observation of every star available in the archives (the full machine-readable table is available online from the journal). While some stars in our sample have been the targets of extensive observing campaigns by \XMM\ and/or \Chandra\ for many years, observational data is quite sparse for other stars. For stars with many observations, long-term periodic X-ray variability has been detected and is likely the result of coronal activity cycles \citep[e.g., $\iota$ Hor, the $\alpha$ Cen system;][]{Sanz-Forcada+19,Ayres+23}. Similar long-term X-ray variability cycles may exist for sparsely observed stars in our sample, but will not be detectable in single snapshot observations. Additional observations are required to constrain the long-term coronal activity cycle and/or flaring activity for many stars in the present sample.

\begin{deluxetable}{clcccc}
    \caption{X-ray Variability Metrics}\label{tab:variability_metrics}
    \setlength{\tabcolsep}{2pt}
    \tablehead{
        &               & Exp. & Net & & \\
     Star &   Observation ID & Time (ks) &  Counts & $\chi^2_r$ & $A^2$ 
    }
    \startdata
    $\iota$ Hor &    XMM/0673610201 & 5.3  & 4230$\pm$70 & 1.5 & 0.49  \\
    $\beta$ Vir &   XMM/0044740201 & 41.1 & 18450$\pm$160 & 3.1 & 0.73 \\
    $\upsilon$ And A & XMM/0722030101 & 5.3 & 940$\pm$40 & 2.1 & 0.36 \\
    44 Boo       & XMM/0100650101 & 18.8 & 287110$\pm$550 & 3.1 & 0.64 \\
    GL 620.1A    & XMM/0822070201 & 7.8 & 12220$\pm$110 & 1.9 & 0.41  \\
    $\kappa^1$ Cet & Chandra/22327  & 14.0 & 6600$\pm$80 & 1.7 & 3.26 \\ 
    Wolf 1061    & Chandra/20163  & 38.3 & 250$\pm$40 & 0.9 & 29.78 \\
    Prox Cen     & XMM/0801880501 & 20.0 & 66790$\pm$260  & 35.2 & 17.55  \\
    GL 412B       & XMM/0742230101 & 10.3 & 32030$\pm$210 & 243.1 & 3.43 \\
    VB 10    & XMM/0504010101 & 24.2 & 660$\pm$110 & 1.1 & 11.01 \\
    \vdots & \vdots & \vdots & \vdots & \vdots & \vdots \\
    \enddata
    \tablecomments{Select entries are shown to illustrate the table form and content. The full machine-readable table (200 rows) is available online from the journal. Net counts are reported for 0.2-15 keV for \XMM\ observations and 0.5-7 keV for \Chandra\ observations.}
\end{deluxetable}

\subsection{Spectral Modeling}
All X-ray spectral modeling was performed with XSPEC v12.11.1 \citep{Arnaud96} using standard $\chi^2$ statistics. Depending on the number of counts and the overall complexity of the spectra, we used one-, two-, or three-temperature {\sc apec} models to parameterize the X-ray spectra. Since all of the stars are located very close to the Sun (within $\sim$20 pc), the intervening hydrogen column density is expected to be low; we adopt a fixed absorbing column $N_{\rm H}$ (using the {\sc tbabs} model) of $10^{19}$ cm$^{-2}$ for all stars \citep[a similar assumption was made by][]{Brown+23}. We assume abundances from \citet{Wilms+00} for both the plasma emission and interstellar absorption models. The plasma temperature and normalizations were left as free parameters during the fit. The normalization of each {\sc apec} component is defined as (neglecting cosmological terms):

\begin{equation}\label{eq:vem}
\mathcal{N}=\frac{10^{-14}}{4\pi d^2} \int n_e n_{\rm H} dV,
\end{equation}

\noindent where $d$ is the distance to the source, $n_e$ and $n_{\rm H}$ are the electron and hydrogen column densities (in units of cm$^{-3}$), respectively, and $dV$ is the volume element (in units of cm$^3$). The integral $\int n_e n_{\rm H} dV$ is the volume emission measure (VEM). Once the best-fit model is obtained, we use the {\sc cflux} convolution model to measure the flux on each plasma component. 

While the spectra of some stars required only one or two {\sc apec} components to achieve a statistically acceptable fit, most spectra required a third {\sc apec} component. We initialized all spectral models with the default (solar) abundance. For some stars (e.g., $\iota$ Hor, $\kappa^1$ Cet) the best fit model was insensitive to the exact choice of abundance; for these stars the abundance was kept fixed at the solar value. If a statistically acceptable fit was not obtained with solar abundances, we allowed the abundance to be a free parameter of the fit, subject to the constraint that the abundance be the same for all plasma components.

\begin{figure}
    \centering
    \includegraphics[width=1\linewidth]{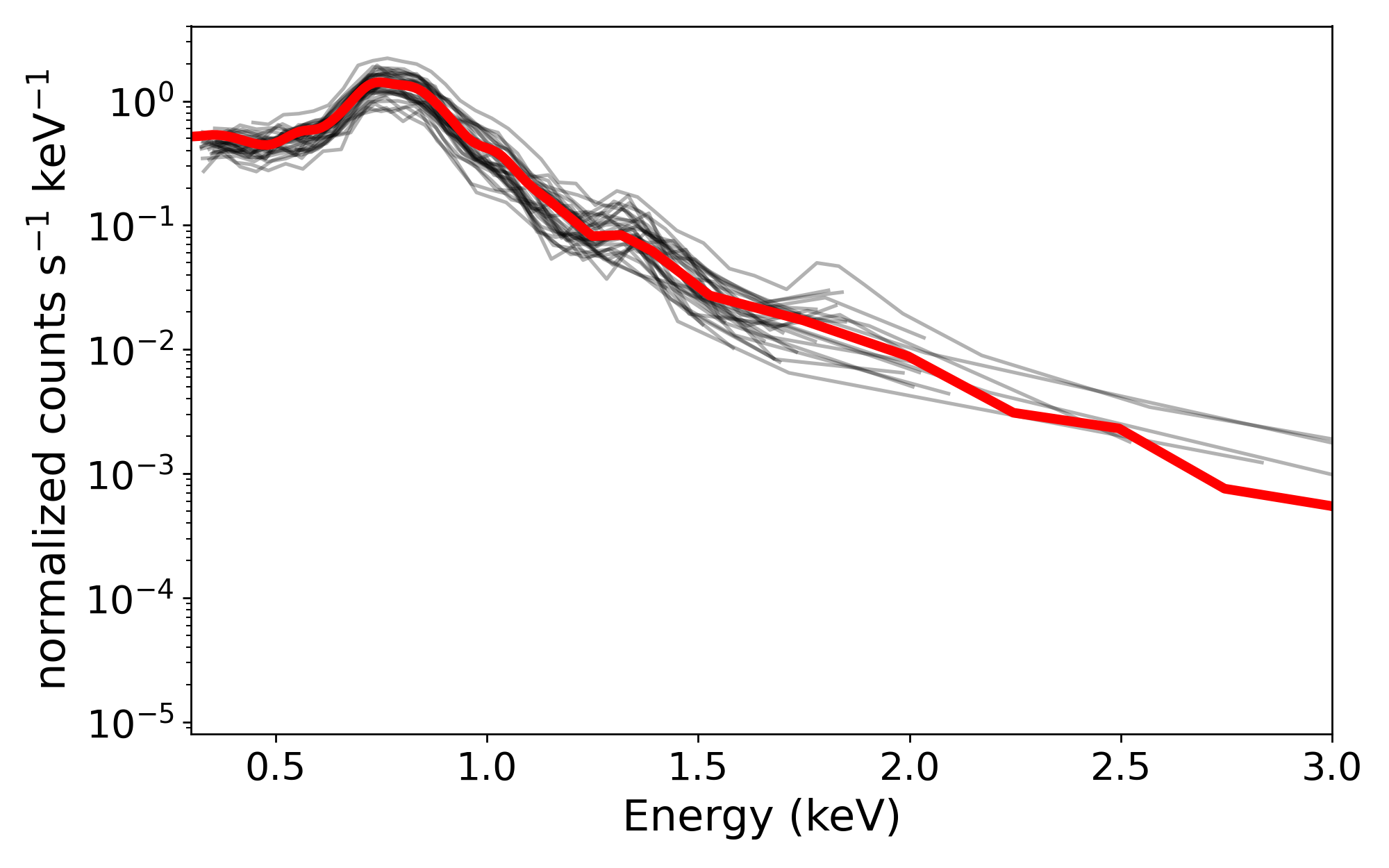} \\
    \caption{Best-fit spectra for all 38 \XMM\ observations of $\iota$ Hor (black) with the average best-fit spectrum superimposed (red).}
    \label{fig:iotaHor_average_spectrum}
\end{figure}

For stars that exhibited count rate variability, spectra were extracted and modeled for each sub-exposure. Table~\ref{table:all_spectral_fits} provides full best-fit parameters for all stars with enough counts to perform spectral modeling for all observations and sub-exposures (the full machine-readable table is available online from the journal). For stars with multiple observations or multiple sub-exposures with the same VarFlag, we present the average best-fit model for a given VarFlag in Table~\ref{table:spectral_modeling}. For uniformity, we refer to the coolest component of a three-component model as ``APEC \#1," the hottest component as ``APEC \#3," and the intermediate-temperature component as ``APEC \#2." The two-temperature thermal plasma models have their individual components matched to the three-temperature component with the most similar temperature.  Figure~\ref{fig:iotaHor_average_spectrum} shows an example of the averaged spectrum for $\iota$ Hor, which did not exhibit significant count rate variability in any of the publicly available 38 \XMM\ observations and was therefore assigned a VarFlag of Q (quiescent). The best-fit models for the individual exposures are shown in black, with the average best-fit spectrum superimposed in red.

\begin{deluxetable*}{cccccccccccccc}
\tablecaption{All Best-Fit Parameters from X-ray Spectral Modeling}
\label{table:all_spectral_fits}
    \setlength{\tabcolsep}{0.4pt}
    \tablehead{
        &&& \multicolumn{2}{c}{APEC \#1} && \multicolumn{2}{c}{APEC \#2} && \multicolumn{2}{c}{APEC \#3} \\ \cline{4-5} \cline{7-8} \cline{10-11}
        Star & ObsID & VarFlag & $kT_1$ & $\mathcal{N}_1$ && $kT_2$ & $\mathcal{N}_2$ && $kT_3$ & $\mathcal{N}_3$ & log$L_{\rm X}$ & abund & $\left<\chi^2/{\rm dof} \right>$ \\
        && & (keV) & ($10^{-4}$) && (keV) & (10$^{-4}$)  && (keV) & (10$^{-4}$) & [erg s$^{-1}$] & & 
    }
    \startdata
    $\eta$ Crv & 14474 & Q & 0.16$^{+0.07}_{-0.14}$ & 3.3$^{+10.0}_{-1.6}$ && 0.50$\pm$0.08 & 4.6$^{+5.0}_{-2.9}$ && \nodata & \nodata & 28.43$\pm$0.06 & 0.32$^{+0.56}_{-0.16}$ & 1.14 \\
    $\beta$ Vir & 0044740201 & Q & 0.09$\pm$0.01 & 4.7$^{+1.2}_{-0.9}$ && 0.37$\pm$0.01 & 2.4$\pm$0.1 && \nodata & \nodata & 28.08$\pm$0.02 & 1 (fixed) & 2.48 \\
    $\chi^1$ Ori & 0111500101/0-10ks & E & 0.10$\pm$0.02 & 39.7$^{+55.7}_{-19.9}$      &&    0.49$\pm$0.03 & 52.9$^{+10.5}_{-9.4}$      &&      0.84$^{+0.10}_{-0.09}$ & 12.1$^{+7.8}_{-4.8}$  & 28.88$^{+0.03}_{-0.04}$ & 0.31$^{+0.07}_{-0.05}$ & 1.31 \\
    $\beta$ Com & 0148680101/0-5ks & Q & 0.07$\pm$0.02 & 23.5$^{+114.8}_{-15.1}$      &&    0.39$\pm$0.02 & 3.4$\pm$0.3      &&      \nodata & \nodata  & 28.11$\pm$0.02 & 1 (fixed) & 1.55 \\
    44 Boo & 0100650101/0-6ks & D & 0.25$\pm$0.02 & 23.8$^{+4.4}_{-6.6}$ && 0.50$^{+0.13}_{-0.12}$ & 9.8$^{+5.4}_{-3.6}$ && 0.91$^{+0.03}_{-0.02}$ & 275.3$^{+13.8}_{-15.7}$ & 28.67$\pm$0.01 & 0.09$\pm$0.01 & 1.45 \\
    GL 620.1A & 0822070301 & Q & 0.11$^{+0.02}_{-0.01}$ & 9.5$^{+5.3}_{-3.6}$      &&    0.46$\pm$0.03 & 1.4$^{+2.5}_{-2.3}$      &&      0.77$^{+0.05}_{-0.04}$ & 6.3$^{+1.9}_{-1.7}$  & 28.82$\pm$0.03 & 0.43$^{+0.09}_{-0.06}$ & 1.60 \\
    \vdots & \vdots & \vdots & \vdots & \vdots & & \vdots & \vdots & &\vdots & \vdots & \vdots & \vdots & \vdots \\
    \enddata
    \tablecomments{Select entries are shown to illustrate the table form and content. The full machine-readable table (360 rows) is available online from the journal.}
\end{deluxetable*}

\begin{figure*}
    \centering
    \includegraphics[width=1\linewidth,clip=true,trim=0cm 0cm 0cm 0cm]{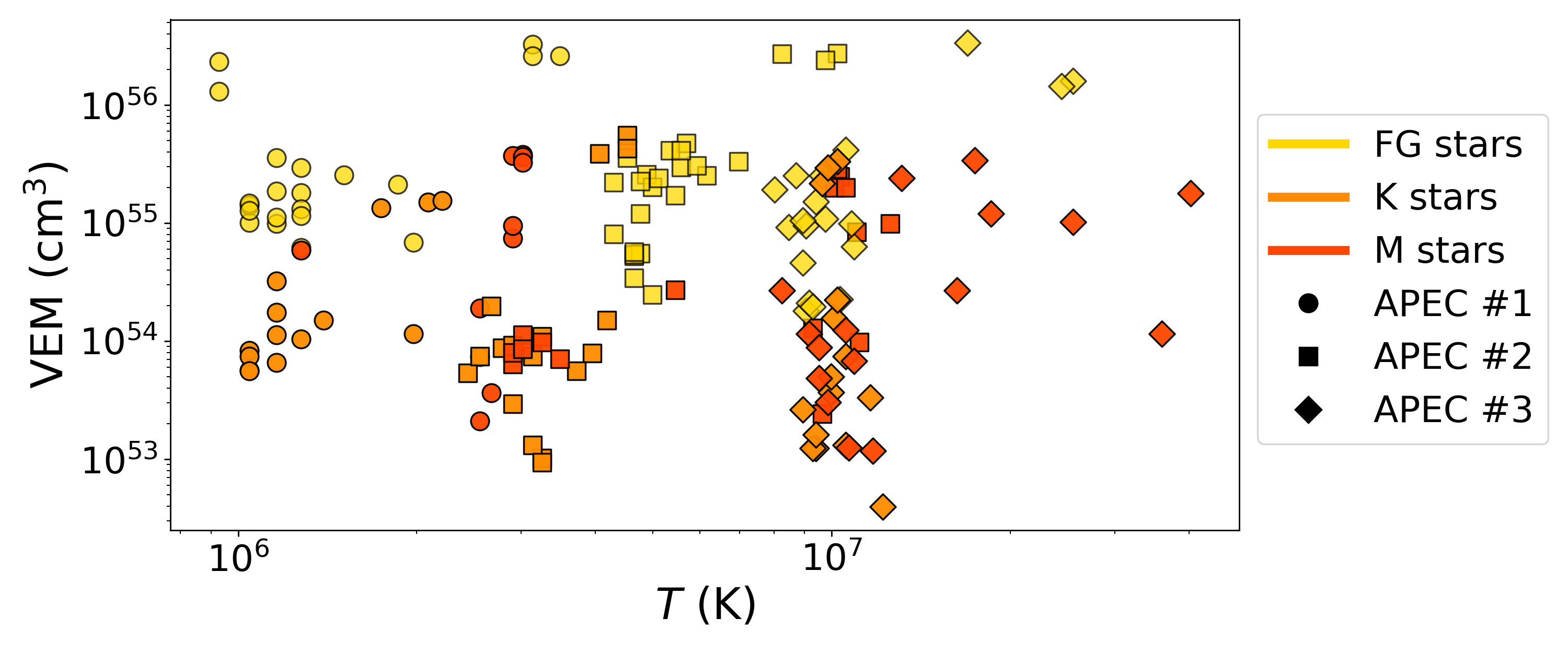}
    \caption{Best fit VEMs and temperatures for thermal plasma components from XSPEC modeling. VEMs and temperatures from APEC \#1 are shown in circles, those from APEC \#2 are shown in squares, and those from APEC \# are shown in diamonds. Values measured for F and G stars are shown in yellow, K stars in orange, and M stars in red. M stars exhibit the largest variations in both VEM and temperature across all three APEC components.}
    \label{fig:temp_vem}
\end{figure*}

Figure~\ref{fig:temp_vem} shows the VEM for each star (calculated from the best-fit normalization, defined in Equation~\ref{eq:vem} above) as a function of temperature for the coolest APEC component (circles), the intermediate APEC component (squares), and the hottest APEC component (diamonds) using the measurements summarized in Table~\ref{table:spectral_modeling}. The points are color-coded by stellar spectral type: F and G stars are shown in yellow, K stars in orange, and M stars in red. In general, the temperatures and VEMs of M stars exhibit the largest spread in coronal temperatures relative to more massive stars. The coronal temperatures of FG and K stars are similar for the coolest and hottest APEC components, although VEMs of FG stars are roughly an order of magnitude higher than those of K stars at similar temperatures. The VEMs do not significantly change within a given spectral type group, but the intermediate-temperature APEC component of FG stars is on average hotter (by $\sim$2 MK) than the intermediate APEC component of K stars.

The three-component thermal plasma model that generally describes the X-ray emission from F and G stars is described by plasma temperatures of $\sim$1.3 MK, 5.1 MK, and 9.4 MK, with corresponding (log) VEMs of 55.2 [cm$^3$], 55.4 [cm$^3$], and 55.0 [cm$^3$], respectively. For K stars, the temperatures become 1.2 MK, 3.3 MK, 10 MK, with (log) VEM values of 54.1 [cm$^3$], 53.9 [cm$^3$], and 53.6 [cm$^3$], respectively. For these hotter stars, periods of elevated count rates or flaring events different from the quiescent periods in that VEMs increased, while the plasma temperatures remained mostly constant. This result is in sharp contrast to the model fits for M stars, which showed more significant variations in both temperature and VEM during flaring events. 

\begin{deluxetable*}{ccccccccccccc}
\tablecaption{Average Best-Fit Parameters from X-ray Spectral Modeling}
\label{table:spectral_modeling}
    \setlength{\tabcolsep}{0.85pt}
    \tablehead{
        && \multicolumn{2}{c}{APEC \#1} && \multicolumn{2}{c}{APEC \#2} && \multicolumn{2}{c}{APEC \#3} \\ \cline{3-4} \cline{6-7} \cline{9-10}
        Star & VarFlag & $kT_1$ & $\mathcal{N}_1$ && $kT_2$ & $\mathcal{N}_2$ && $kT_3$ & $\mathcal{N}_3$ & log$L_{\rm X}$ & abund & $\left<\chi^2/{\rm dof} \right>$ \\
        & & (keV) & ($10^{-4}$) && (keV) & (10$^{-4}$)  && (keV) & (10$^{-4}$) & [erg s$^{-1}$] & & 
    }
    \startdata
    $\eta$ Crv      & Q & 0.20$^{+0.12}_{-0.09}$ & 2.3$^{+4.3}_{-1.1}$ && 0.55$^{+0.09}_{-0.07}$ & 3.1$^{+2.2}_{-1.7}$ && \nodata & \nodata & 28.41$^{+0.05}_{-0.10}$ & 0.27$^{+0.61}_{-0.18}$ & 1.31 \\
    $\upsilon$ And A & Q & 0.29$\pm$0.03 & 0.8$\pm$0.1 && \nodata & \nodata && \nodata & \nodata & 27.57$\pm$0.03 & 1 (fixed) & 0.91 \\
    $\iota$ Hor     & Q & 0.09$\pm$0.03 & 7.7$\pm$1.1                     && 0.42$\pm$0.08 & 2.9$\pm$0.2           && 0.71$\pm$0.14 & 1.6$\pm$0.2  & 28.75$^{+0.10}_{-0.14}$ & 1 (fixed) & 1.27 \\
    $\beta$ Vir & Q & 0.09$\pm$0.01 & 4.7$^{+1.2}_{-0.9}$ && 0.37$\pm$0.01 & 2.4$\pm$0.1 && \nodata & \nodata & 28.08$\pm$0.02 & 1 (fixed) & 2.48 \\
    $\chi^1$ Ori    & Q & 0.08$\pm$0.03 & 258$\pm$228                     && 0.42$\pm$0.07 & 28.6$\pm$9.2          && 0.69$\pm$0.06 & 21.3$\pm$8.2 & 28.94$^{+0.07}_{-0.10}$ & 0.20$\pm$0.12 & 1.55 \\
    \vdots    & \vdots & \vdots & \vdots && \vdots & \vdots && \vdots & \vdots & \vdots & \vdots & \vdots \\
    \enddata
        \tablecomments{Select entries are shown to illustrate the table form and content. The full machine-readable table (76 rows) is available online from the journal.}

\end{deluxetable*}

\subsection{X-ray Luminosity}
In addition to insights about the temperatures and VEMs of coronal structures associated with our target stars, the best-fit X-ray spectral models provide us with a count rate-to-luminosity conversion factor that mitigates against instrumental effects from the extracted light curves and gives us a clearer view of the luminosity evolution of the target star. Furthermore, we can ``unfold'' our X-ray spectra (i.e., de-convolve the observed spectrum and the energy-dependent detector response function) to assess the intrinsic energy emission from the target star as a function of energy (or wavelength). 

\begin{figure*}
\centering
    \begin{tabular}{c}
       \includegraphics[width=1\linewidth]{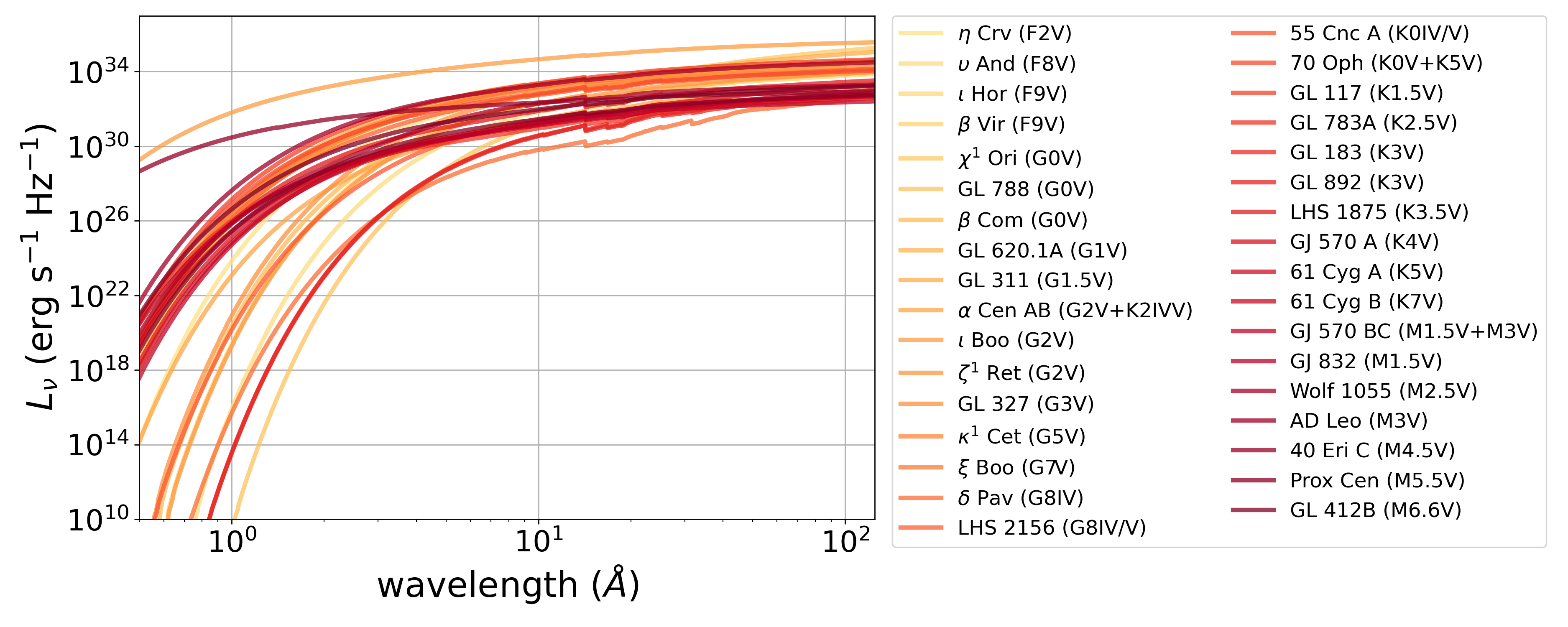}  \\
    \end{tabular}
    \caption{Average continuum quiescent spectra for each target star (0.5-150 \AA, corresponding to an energy range of $\sim$25-0.08 keV). Spectra have been corrected for minor foreground absorption. Stars are color-coded by spectral type, with earlier-type stars shown in yellow and later-type stars shown in dark red.}
    \label{fig:average_unabs_spectra}
\end{figure*}

\begin{figure}
    \centering
    \includegraphics[width=1\linewidth,clip=true,trim=0cm 0cm 0cm 0cm]{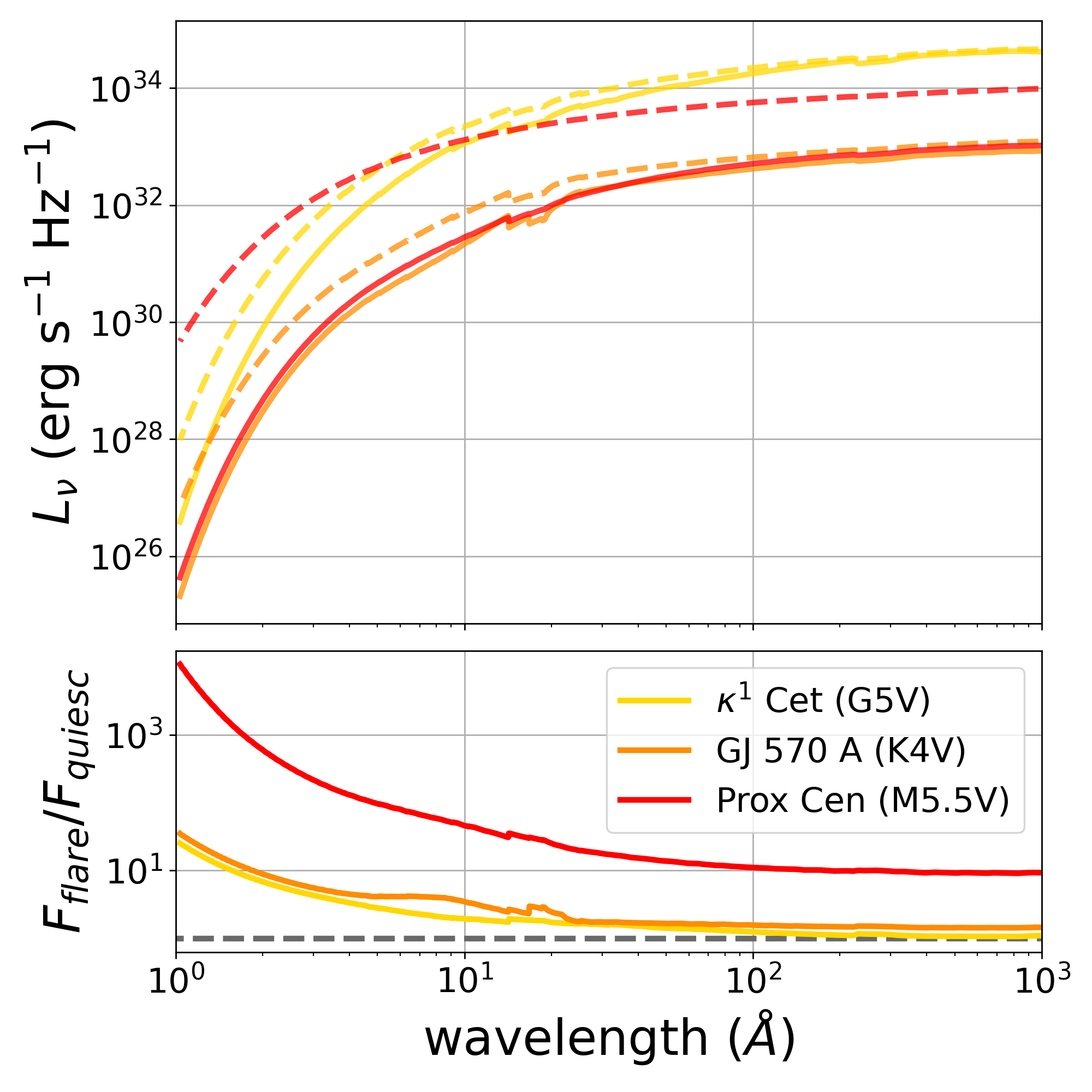}
    \caption{{\it Top}: Quiescent (solid lines) and flaring (dashed lines) continuum spectra of three representative stars from our sample ($\kappa^1$ Cet in yellow, GJ 570 A in orange, and Prox Cen in red). {\it Bottom}: The ratio of flaring spectrum to quiescent spectrum, showing a pronounced enhancement in X-ray emission at $<$10 \AA\ during flaring periods for all three stars. The flaring spectrum of Prox Cen remains an order of magnitude brighter than the quiescent spectrum even at longer wavelengths. }
    \label{fig:FQ_extend}
\end{figure}

Figure~\ref{fig:average_unabs_spectra} shows the continuum luminosity of each star (in units of \Lsun) as a function of wavelength during quiescent times. All spectra have been corrected for minor foreground absorption, and we use the XSPEC command \texttt{fakeit} to extend our best-fit model to a wavelength range of $\sim$1-200 \AA. For nearly all stars, flaring events are driven by (sometimes dramatic) enhancements in X-ray luminosity at wavelengths $\lesssim$10 \AA\ (energies $\gtrsim$1.2 keV). Figure~\ref{fig:FQ_extend} shows quiescent and flaring spectra for three representative stars ($\kappa^1$ Cet, GJ 570A, and Prox Cen) from 1-1000 \AA\ (top panel), as well as the ratio of the flaring and quiescent spectra (bottom panel). For both $\kappa^1$ Cet (G5V) and GJ 570A (K4V), the flux at $\sim$1 \AA\ ($\sim$12 keV) during flaring times is a factor of $\sim$30 larger than during quiescent times, but the fluxes at longer wavelengths ($\gtrsim$100 \AA; energies $\lesssim$0.12 keV) are similar. In the case of Prox Cen (M5.5V), however, the flaring spectrum is brighter than the quiescent spectrum by at least an order of magnitude at all wavelengths, with short-wavelength ($\lesssim$ 3 \AA) emission enhanced by more than a factor of 10$^3$. This finding is unlikely to be the result of the reduced soft energy responses of \Chandra\ or \XMM, as the loss of effective area due to contamination build-up occurs most significantly at energies below 1 keV \citep[e.g.,][]{Grant+24}. Nevertheless, we use detector response functions specific to the orbital cycle in which the individual observations were obtained to de-convolve the detector response as accurately as possible. Plots of individual stellar X-ray spectra, color-coded by VarFlag, are discussed in Appendix~\ref{appendix:individual_stars}.

A question that our analysis can elucidate is: For what fraction of time is a star observed to be emitting at a given X-ray luminosity? While this is related to the flaring frequency and the typical peak luminosity and duration of flares, our goal in this work is not to directly characterize flaring events or the physical mechanisms that cause them. Rather, we wish to predict the X-ray environments in which HZ planets reside. 

\begin{figure}
    \centering
    \includegraphics[width=1\linewidth]{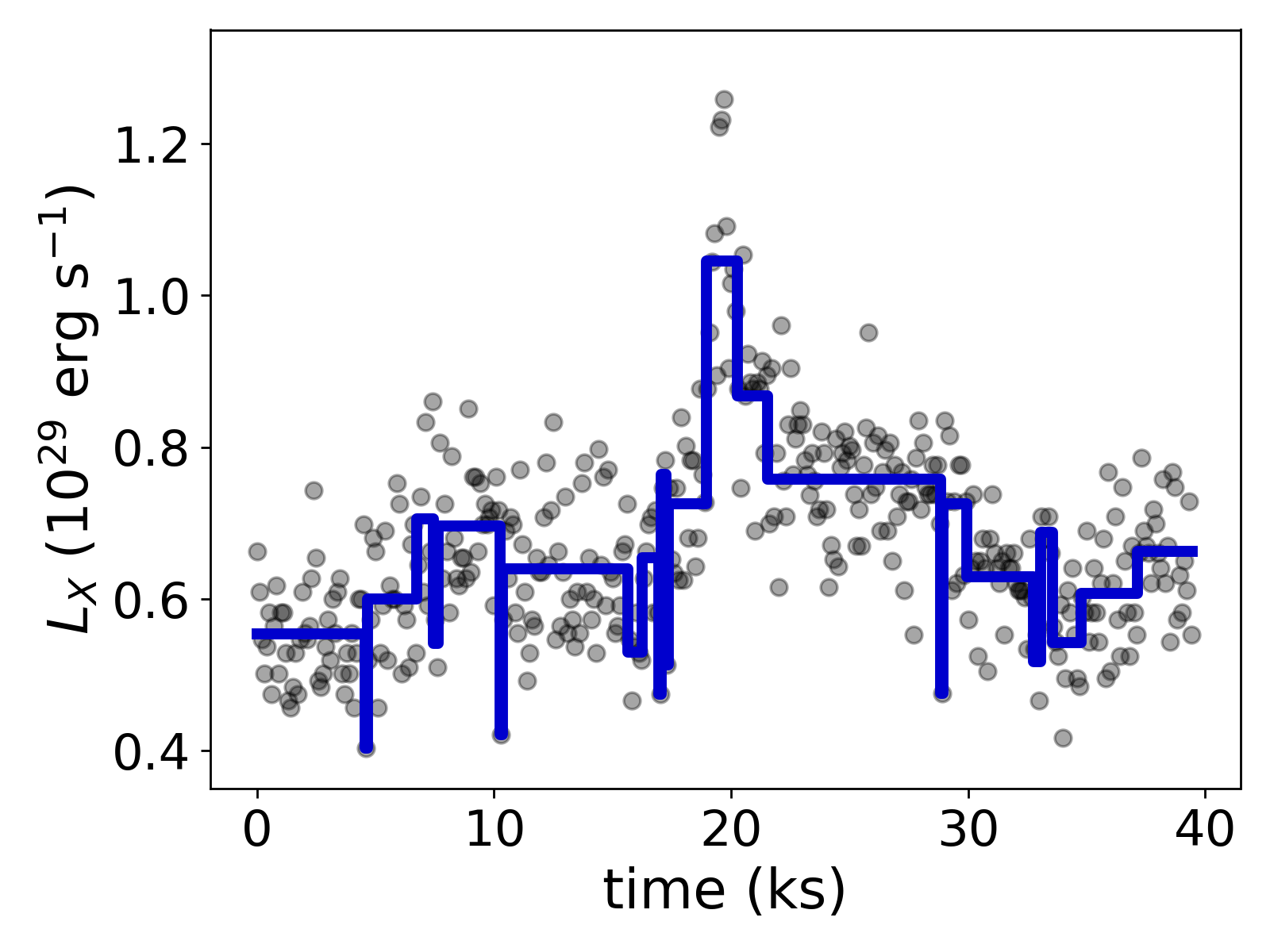}
    \caption{The light curve of $\kappa^1$ Cet from \XMM\ observation 0111410101. The MOS1 count rates were converted to luminosities (gray circles) using the best-fit spectral models from three sub-exposures (from 0-14 ks, 14-25 ks, and 25-39 ks). The Bayesian blocked light curve is superimposed in blue.}
    \label{fig:bayesian_block}
\end{figure}

To do this, we use our best fit spectral models to first convert the observed count rates to luminosities. For stars that are too faint to have their spectra directly modeled, we use the spectral model of the star with the closest spectral type and physical parameters to convert the observed count rates into luminosities. For stars that are detected but too faint for reliable light curve extraction, we simply estimate their time-averaged X-ray luminosity; for un-detected stars we calculate 3$\sigma$ luminosity upper limits. We then apply Astropy's \citep{astropy13,astropy18} \texttt{bayesian\_block} algorithm to the light curves; the ``fitness'' parameter is used to define the measurement uncertainties. The Bayesian block algorithm \citep{Scargle98,Scargle+13} is a segmentation technique, which aims to optimally split (in this case) time series data into ``blocks'' such that each block is statistically different from its neighboring block. An example Bayesian blocked light curve for $\kappa^1$ Cet is shown in Figure~\ref{fig:bayesian_block}.

\begin{figure}
    \includegraphics[width=1\linewidth]{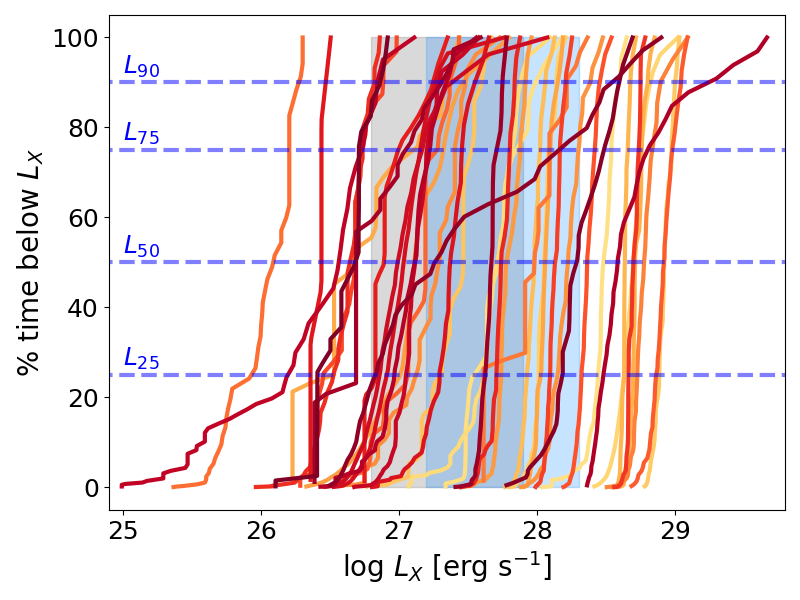} \\
    \includegraphics[width=1\linewidth]{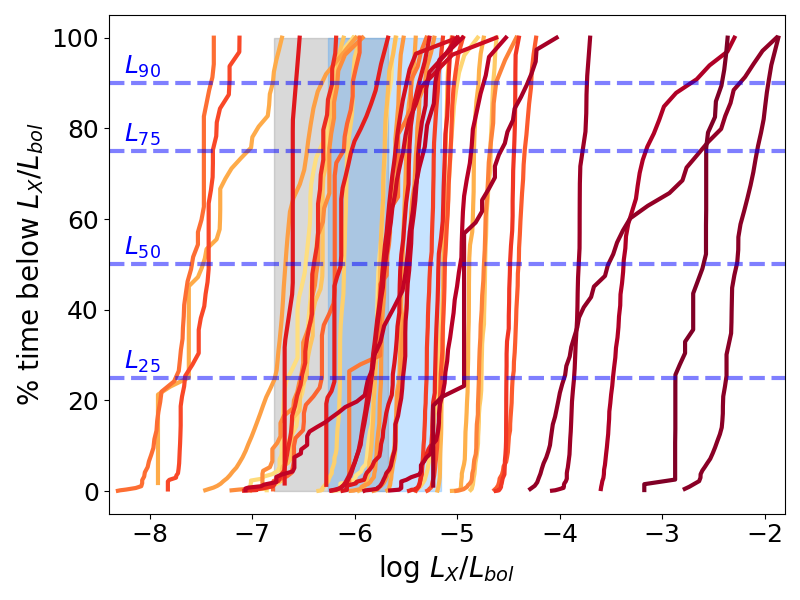} \\
    \caption{{\it Top}: The fraction of time that a star is observed to be fainter than a specific luminosity. {\it Bottom}: The fraction of time that a star is observed to be below a \Lx/\Lbol\ value. Stars are color-coded by spectral type, with earlier-type stars shown in yellow and later-type stars shown in dark red. The 25th, 50th, 75th, and 90th percentile luminosities and \Lx/\Lbol\ ratios are indicated by blue dashed lines (values for each star summarized in Table~\ref{tab:luminosities}). The shaded gray region indicates typical ranges for the current Sun, and the blue shaded regions indicate approximate ranges for the early Sun ($\sim$4 Gyr ago; see text).} 
    \label{fig:time_vs_Lx}
\end{figure}

The output of the \texttt{bayesian\_block} routine is a list of time bins and the luminosity the star was emitting at during each bin. From these lists, we measure the fraction of the time that each star was observed to be fainter than a given \Lx\ or \Lx/\Lbol\ value. The results are shown in Figure~\ref{fig:time_vs_Lx}. For comparison, \Lx\ of the modern-day Sun is 6.31$\times10^{26}$ \lum\ (log\Lx/\Lbol\ = -6.78) when quiet and 8.04$\times10^{27}$ \lum\ (log\Lx/\Lbol\ = -5.68) when active \citep{Linsky+20}, while $\sim$4 Gyr ago the \Lbol\ of the Sun was $\sim$75\% the modern value and its \Lx\ was a factor of $\sim$2-3 higher \citep{Obridko+20}. On this plot, a perfectly non-variable star would be a straight vertical line, while stars that experience significant changes in X-ray luminosity (i.e., Proxima Centauri) will show a pronounced horizontal extent. We measure the luminosities at which the stars spend less than 25\%, 50\%, 70\%, and 90\% of the observed time (which we refer to as $L_{25}$, $L_{50}$, $L_{75}$, and $L_{90}$, respectively). These values are summarized in Table~\ref{tab:luminosities}; we also include the time-averaged luminosities (with uncertainties) of faintly detected stars and the 3$\sigma$ upper limits for non-detected stars in the $L_{50}$ column. 

\begin{table*}[!htbp]
\centerwidetable
\footnotesize
\setlength{\tabcolsep}{2.5pt}
    \caption{X-ray Luminosities and \Lx/\Lbol\ Ratios (or 3$\sigma$ Upper Limits) for All Sample Stars}\label{tab:luminosities}
    \begin{tabular}{ccccccccc}
    \hline \hline
             & log$L_{25}$    & log$L_{50}$    & log$L_{75}$ & log$L_{90}$ & log($L_{25}$/\Lbol) & log($L_{50}$/\Lbol) & log($L_{75}$/\Lbol) & log($L_{90}$/\Lbol) \\
        Star & [erg s$^{-1}$] & [erg s$^{-1}$] & [erg s$^{-1}$] & [erg s$^{-1}$] & & & &  \\
    \hline
    $\eta$ Crv      & 28.19             & 28.39         & 28.47     & 28.59 & -6.06     & -5.86     & -5.78     & -5.66 \\
    $\xi$ Oph   & 27.47 & 27.64 & 27.77 & 28.03 & -6.73 & -6.56 & -6.43 & -6.17 \\
    $\upsilon$ And A & 27.27 & 27.55 & 27.72 & 27.87 & -6.84 & -6.56 & -6.39 & -6.24 \\
    $\beta$ Vir & 27.90 & 27.99 & 28.05 & 28.15 & -6.26 & -6.17 & -6.11 & -6.01 \\
    $\beta$ Hyi     & 26.23 & 26.53 & 26.94 & 27.37 & -7.92 & -7.62 & -7.21 & -6.78 \\
    $\kappa^1$ Cet  & 28.63             & 28.71         & 28.78     & 28.88 & -4.89 & -4.81 & -4.74 & -4.67 \\
    HD 140901       & 27.62 & 27.92 & 28.05 & 28.23 & -6.05 & -5.75 & -5.62 & -5.44 \\
    $\delta$ Eri    & 26.39 & 26.53 & 26.69 & 26.89 & -7.72 & -7.58 & -7.42 & -7.22 \\
    GL 570 A        & 27.15             & 27.36         & 27.52     & 27.68 & -5.78 & -5.57 & -5.41 & -5.25 \\
    61 Cyg A        & 26.75             & 26.92         & 27.08     & 27.28 & -6.00 & -5.84 & -5.68 & -5.56 \\
    61 Cyg B        & 26.69             & 26.87         & 27.04     & 27.42 & -5.89 & -5.71 & -5.54 & -5.35 \\
    AD Leo          & 28.42             & 28.51         & 28.69     & 29.36 & -3.54 & -3.45 & -3.27 & -2.98 \\
    Proxima Centauri & 26.70            & 26.93         & 27.71     & 28.63 & -4.09 & -3.86 & -3.08 & -2.33 \\
    \vdots & \vdots & \vdots & \vdots & \vdots & \vdots & \vdots & \vdots & \vdots \\
    \hline \hline
    \end{tabular}
    \tablecomments{Select entries are shown to illustrate the table form and content. The full machine-readable table (63 rows) is available online from the journal.}
\end{table*}

\section{Discussion}\label{sec:discussion}
We compared the 25th and 90th percentile \Lx/\Lbol\ ratios for stars in our sample to the current minimum and maximum ratios for the Sun \citep{Linsky+20}, and find that eight stars that were detected in the X-ray observations at high significance exhibit \Lx/\Lbol\ ratios similar to that of the modern Sun: $\xi$ Oph (F2V), $\beta$ Vir (F9V), 47 UMa (G0V), $\beta$ Com (G0V), 55 Cnc A (G8V), GL 892 (K3V), GL 783 (K3V), and GL 183 (K3V). Three of these stars ($\beta$ Vir, $\beta$ Com, and GL 183) are younger than the Sun, while four are older (47 UMa, 55 Cnc A, GL 892, and GL 783) and one ($\xi$ Oph) does not have an age estimate available in the literature. Twelve stars have \Lx/\Lbol\ ratios consistent with those of the early Sun: $\eta$ Crv (F2V), $\beta$ Com, GL 311 (G1.5V), $\zeta^1$ Ret (G2V), GL 327 (G3V), HD 140901 (G7IV), the unresolved binary 70 Oph (K0V+K5V), LHS 1875 (K3V), GL 570A (K4V), 61 Cyg A (K5V) and B (K7V), and the unresolved GL 570 BC pair (M1.5V+M3V). We note that one star, $\beta$ Com, exhibits a range in \Lx/\Lbol\ that fits within the narrow overlap region between the modern and early Sun (log\Lx/\Lbol $\approx$ -6.4 to -5.7). An additional 11 stars were not detected or only marginally detected and have \Lx/\Lbol\ upper limits lower than the modern day minimum solar \Lx/\Lbol: $\nu$ Phe, LHS 237, LHS 208, $\iota$ Per, $\rho$ CrB, GL 672, 51 Peg, GJ 777A, 82 Eri, GL 451A, and Kapteyn's Star. For roughly a dozen stars in our sample, the observed \Lx\ (or \Lx\ upper limit) translates into X-ray surface fluxes below the $\sim$10$^4$ \flux\ minimum observed in the quiet Sun and other ``coronal hole'' stars \citep{Schmitt97}. A closer look at these very X-ray faint stars will be the subject of an upcoming study (Binder et al., {\it in preparation}). In Table~\ref{tab:SolComparison}, we summarize these X-ray-detected Solar-like stars (their spectral types and approximate ages are taken from Table~\ref{tab:stellar_properties}) and the relevant solar \Lx/\Lbol\ comparison (modern or early Sun). We additionally provide the ``tier'' each star was assigned in \citet{Mamajek+24}, indicating the priority which each star should be considered for an HWO direct imaging survey (with A indicating the highest priority group, and tiers B and C requiring further study).

\begin{table*}[]
    \centering
    \caption{Potential HWO Target Stars with Solar-Like \Lx/\Lbol\ Ratios}
    \begin{tabular}{ccccc}
    \hline \hline
        Star Name & Spectral Type & Age (Gyr) & Sun Comparison & HWO Tier  \\
        \hline
    $\eta$ Crv & F2V &  $<$2       & early         &  C \\
    $\xi$ Oph &  F2V &    unknown   &  modern      &    C \\
    $\nu$ Phe   & F9V & $\sim$3     & $<$modern min &  B \\
    $\beta$ Vir & F9V &  $\sim$3   &  modern      &    C \\
    LHS 237     & F9V  & unknown    & $<$modern min &  none \\
    $\iota$ Per & F9.5V   & $\sim$4   & $<$modern min &  A \\
    LHS 208     & G0V   & $\sim$3   & $<$modern min   &  B \\
    $\rho$ CrB  & G0V   & $\sim$11  & $<$modern min &  B \\
    GL 672      & G0V   & $\sim$12  & $<$modern min &  C \\
    47 UMa     & G0V &  $\sim$6.5  & modern       &   A \\
    $\beta$ Com & G0V &  $<$2      &  modern,early &   A \\
    GL 311     & G1.5V& $<$2       & early         &  B \\
    $\zeta^1$ Ret & G2V & $<$2     &  early        &   A \\
    51 Peg      & G2V   & $\sim$7   & $<$modern min & none  \\
    GL 327     & G3V &  $<$2       & early         &  C \\
    GJ 777A     & G7IV/V  & $\sim$13  & $<$modern min &  B \\
    HD 140901  & G7IV/V & $\sim$3    & early         &  C \\
    82 Eri      & G8V   & $\sim$6   & $<$modern min &  B \\
    70 Oph AB  & K0V+K5V &  $\sim$6   &  early     &  B \\
    55 Cnc A    & K0IV/V &  $\sim$9.5 &  modern       &   C \\
    GL 451A     & K1IV  & $\sim$5    & $<$modern min &  C \\
    GL 783     & K2.5V &  $\sim$7    & modern        &  B \\
    GL 892     & K3V &  $\sim$11   & modern        &  A \\
    GL 183     & K3V &   $\sim$2   &  modern       &   B \\
    LHS 1875   & K3.5V & $<$3       &  early         &  C \\
    GL 570A    & K4V & $<$3       &  early         &  A \\
    61 Cyg AB  & K5V+K7V & $\sim$6 & early      &  A \\
    GL 570 BC  & M1.5V+M3V & $<$3  &  early   & none \\
    Kapteyn's Star & M2V    & $\sim$11  & $<$modern min & none \\
    \hline \hline
    \end{tabular}
    \label{tab:SolComparison}
\end{table*}

Any HZ planets that may exist around the stars listed in Table~\ref{tab:SolComparison} are currently experiencing high energy radiation environments that are similar to or less hostile than the modern Earth. Out of the 229 unique stars we identified as potential targets for future direct imaging surveys, only $\sim$25\% had ever been imaged with either \XMM\ or \Chandra\ (this fraction increases to $\sim$30\% when \Swift\ imaging is included), and out of the 57 stellar systems with high quality X-ray observation we find that 28 systems ($\sim$50\%) have \Lx/\Lbol\ ratios similar to or less than that of either the modern or early Sun. These stars span the full range of spectral types and ages considered here, and represent all three tiers of the \citet{Mamajek+24} ranking system. It is therefore likely that there are an additional $\sim$100 potential direct imaging target stars with HZs with current X-ray radiation environments similar to or even less hostile than that of the Sun. We note that threshold \Lx/\Lbol ratios below that of the modern or early Sun could be a conservative condition for finding habitable planets. Although, the past evolution of the star would also have to be taken into account. Detailed coupled geological and atmospheric evolution analyses would be most informative for predicting the targets with the best chances of maintaining potentially habitable conditions \citep[e.g.,][]{krissansen+2022}. A systematic X-ray survey of these as-of-yet unobserved nearby stars is needed to inform the selection of the best targets for direct imaging surveys.

\begin{figure}
    \centering
    \includegraphics[width=1\linewidth,clip=true,trim=0cm 0cm 0cm 0cm]{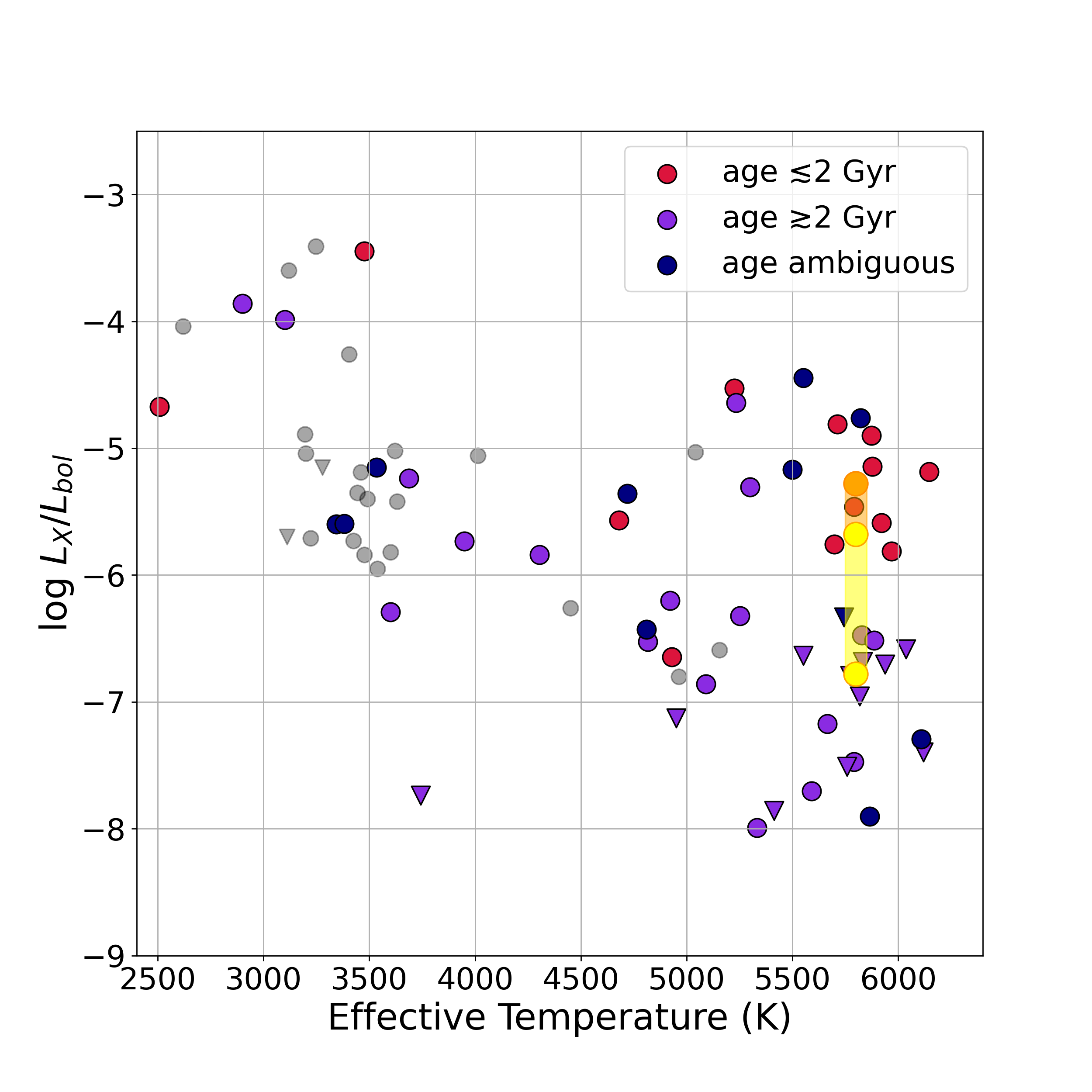}
    \caption{The \Lx/\Lbol\ ratio as a function of $T_{\rm eff}$. Stars in our sample with firm X-ray detections are shown in circles, while upper limits are shown for stars that were not X-ray detected (downward pointing triangles). Stars that are believed to be young ($\lesssim$2 Gyr) are shown in red, older stars ($\gtrsim$2 Gyr) are shown in purple, and stars without age estimates or with ambiguous/discrepant age estimates are shown in dark blue. The MUSCLES/Mega-MUSCLES sample \citep{Brown+23} is shown in gray. The yellow circles and shaded area indicate the quiescent and flaring \Lx/\Lbol\ ratios for the Sun \citep{Linsky+20}.}
    \label{fig:MUSCLES_compare}
\end{figure}

In general, M stars exhibit significantly more X-ray variability than earlier type stars and spend a significant amount of time at \Lx/\Lbol\ ratios dramatically larger than that of the Sun. We can supplement our sample of Sun-like and low mass stars with X-ray observations of an 23 additional M- and K-type stars from the MUSCLES/Mega-MUSCLES survey \citep{Brown+23}. \citet{Brown+23} found a major decrease in the X-ray-to-bolometric luminosity ratio (\Lx/\Lbol) with increasing effective temperature. We compute \Lx/\Lbol\ (or upper limits, in the case of X-ray non-detections) for all stars in our sample using $L_{50}$ from Table~\ref{tab:luminosities} and \Lbol\ from Table~\ref{tab:stellar_properties}. In Figure~\ref{fig:MUSCLES_compare}, we plot the (logarithmic) \Lx/\Lbol\ ratio as a function of $T_{\rm eff}$ (also from Table~\ref{tab:stellar_properties}) for our full sample (colored points), with the MUSCLES/Mega-MUSCLES sample shown in gray. Our sample extends this inverse relationship between \Lx/\Lbol\ and $T_{\rm eff}$ by $\sim$1000 K in $T_{\rm eff}$, with \Lx/\Lbol\ ratios that are about an order of magnitude lower than observed for K and M stars. 

There is a small clustering of stars with $T_{\rm eff}\sim5200-6200$ K with significantly higher \Lx/\Lbol\ ratios than the majority of comparable stars, likely due to young stellar ages. Chromospheric activity in solar-type stars is observed to decrease rapidly after $\sim$2-3 Gyr \citep{Pace+04,Zhang+19}, after which activity levels remain mostly constant. We indicate the approximate age of each star in Figure~\ref{fig:MUSCLES_compare}: young stars ($\lesssim$2 Gyr) are shown in red, older stars ($\gtrsim$2 Gyr) are shown in purple, and stars without age estimates or ambiguous ages are shown in dark blue. Most of these high-\Lx/\Lbol, high-$T_{\rm eff}$ stars are indeed believed to be consistent with the younger ages where each spectral type emits elevated high energy emission \citep{Yowell+23}. There are three stars with ambiguous ages that lie within this region of the diagram ($\chi^1$ Ori, 55 Cnc A, and GL 451A); the observed X-ray emission of these stars suggests that younger age estimates may be preferred for these stars.

\begin{figure}
    \centering
    \includegraphics[width=1\linewidth,clip=true,trim=0cm 0cm 0cm 0cm]{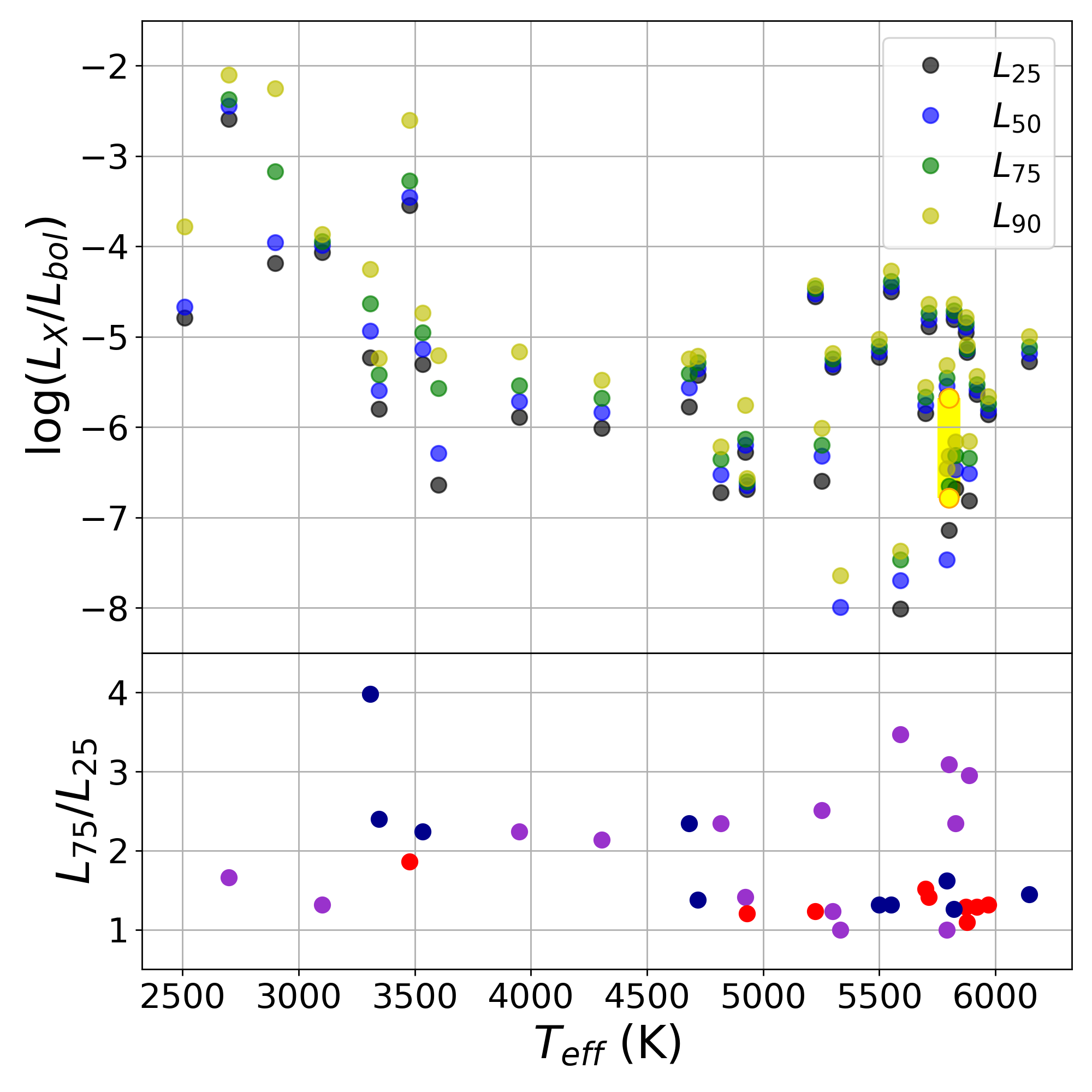}
    \caption{{\it Top}: the \Lx/\Lbol\ ratio as a function of $T_{\rm eff}$ assuming $L_{25}$ (black), $L_{50}$ (blue), $L_{75}$ (green), and $L_{90}$ (dark yellow). The yellow circles and shaded area indicate the quiescent and flaring \Lx/\Lbol\ ratios for the Sun \citep{Linsky+20}. {\it Bottom}: the ratio of $L_{75}/L_{25}$ as a function of $T_{\rm eff}$, color-coded by age (as in Figure~\ref{fig:MUSCLES_compare}: young stars shown in red, old stars shown in purple, and stars with ambiguous ages shown in dark blue).}
    \label{fig:LXLbol_IQR}
\end{figure}

In Figure~\ref{fig:LXLbol_IQR} we again show the \Lx/\Lbol\ ratio as a function of $T_{\rm eff}$ for stars in our sample, but we use $L_{25}$, $L_{50}$, $L_{75}$ and $L_{90}$ to compute the \Lx/\Lbol\ ratio. This allows us to assess the degree to which intrinsic X-ray variability contributes to the scatter in \Lx/\Lbol\ observed in Figure~\ref{fig:MUSCLES_compare}. The ratio of $L_{75}/L_{25}$ provides an estimate of how much a star varies ``normal'' about the median \Lx. We find that the variability of the youngest stars in our sample exhibit the smallest dynamic range, with $L_{75}/L_{25}\sim1.3$ for stars with $T_{\rm eff}\gtrsim$5000 K (for the coolest young star in the plot, AD Leo, $L_{75}/L_{25}\sim1.8$). The $L_{75}/L_{25}$ ratio for older stars shows more dispersion, varying by up to a factor of $\sim$3.5.

\vfill~
\section{Conclusions}
We have analyzed $\sim$3 Msec of high quality \XMM\ and \Chandra\ observations of 57 nearby stellar systems, as well as snapshot \Swift\ images of an additional 19 stellar systems, to provide a uniform catalog of the X-ray properties of potential target stars for future direct imaging surveys. All X-ray light curves and spectroscopic data is publicly available on Zenodo (DOI: 10.5281/zenodo.11490574). The X-ray environments in the HZs around F, G, and early K stars are more likely to be similar to those that the Earth has evolved in, and we identify 29 stars with current \Lx/\Lbol\ ratios similar to or less than that of the Sun. Our results are consistent with \citet{Yowell+23}, who found that the X-ray fluxes of late M stars can be a factor of $\sim$3-15 times larger than for K stars. In addition to high average \Lx/\Lbol\ ratios, M stars exhibit more frequent and dramatic flares than earlier-type stars. We note that, of the 229 candidate target stars for direct imaging, only $\sim$30\% have been observed by \XMM, \Chandra, or \Swift. 

\pagebreak

\begin{acknowledgments}
This work is supported by NASA Exoplanets Research Program (XRP) award \#80NSSC23K0039 (PI Turnbull). This research has made use of the NASA Exoplanet Archive, which is operated by the California Institute of Technology, under contract with the National Aeronautics and Space Administration under the Exoplanet Exploration Program. B.B. is grateful for the warm hospitality of the University of California, Santa Barbara Kavli Institute for Theoretical Physics, where the majority of this manuscript was written; this research was supported in part by grant no. NSF PHY-2309135 to the Kavli Institute for Theoretical Physics. S.P. acknowledges support from NASA under award number 80GSFC21M0002. E.S. and S.P. acknowledge support from the CHAMPs (Consortium on Habitability and Atmospheres of M-dwarf Planets) team, supported by the National Aeronautics and Space Administration (NASA) under grant nos. 80NSSC21K0905 and 80NSSC23K1399 issued through the Interdisciplinary Consortia for Astrobiology Research (ICAR) program. This work has made use of data from the European Space Agency (ESA) mission {\it Gaia} (\url{https://www.cosmos.esa.int/gaia}), processed by the {\it Gaia} Data Processing and Analysis Consortium (DPAC, \url{https://www.cosmos.esa.int/web/gaia/dpac/consortium}). Funding for the DPAC has been provided by national institutions, in particular the institutions participating in the {\it Gaia} Multilateral Agreement. This work made use of data supplied by the UK Swift Science Data Centre at the University of Leicester. We would like to thank Eric Mamajek for useful discussions related to this work, and the anonymous reviewer for helpful comments that improved our paper. 
\end{acknowledgments}

\facilities{CXO, XMM, Swift (XRT), Exoplanet Archive}

\software{astropy \citep{astropy13,astropy18,astropy22}
          }

\appendix
\section{Discussion of Individual Stars}\label{appendix:individual_stars}
In this appendix we provide additional context for each stellar system (e.g, when there is a notable discrepancy in the published literature with regards to a given physical parameter) and summarize the known or candidate exoplanets associated with each star. We then present the results of our X-ray analysis for each stellar system. Table~\ref{tab:match_stars} summarizes the observations of faint/non-detected stars, including if the star was detected in a given observation or not, the count rate (or upper limit on the count rate) and detection significance, and the ``match star'' that was used to convert count rates to luminosities.

\begin{figure*}
\centering
\begin{tabular}{ccc}
    \includegraphics[height=1.75in]{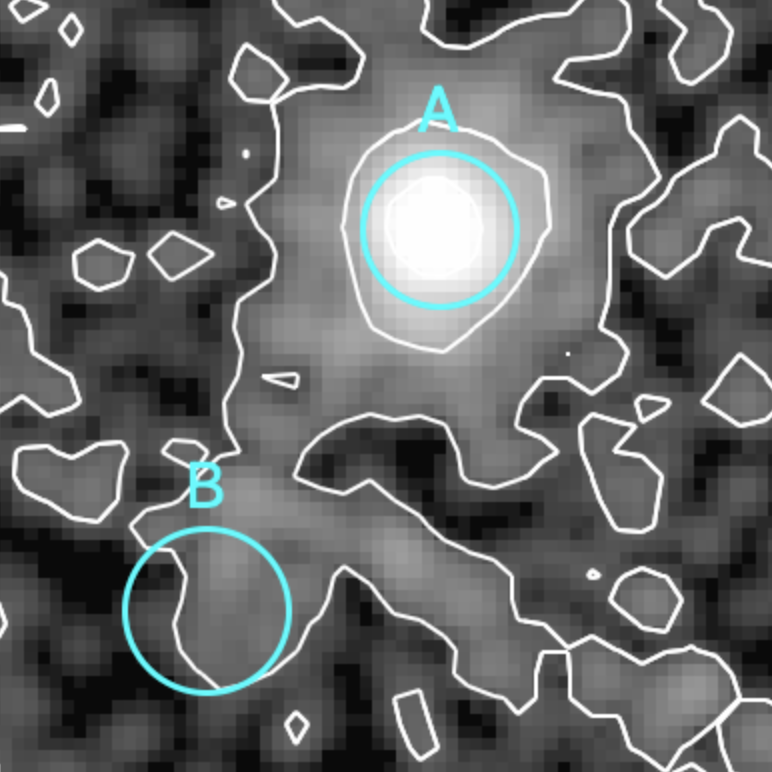} &
    \includegraphics[height=1.75in,clip=true,trim=1.5cm 0cm 0cm 0cm]{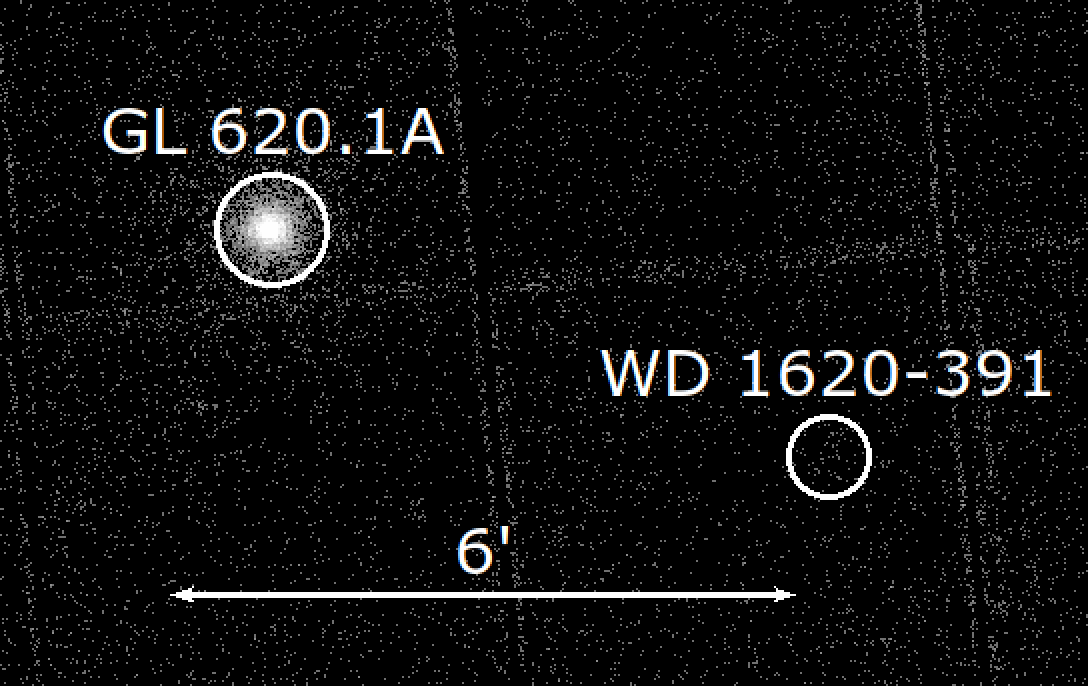} &
    \includegraphics[height=1.75in]{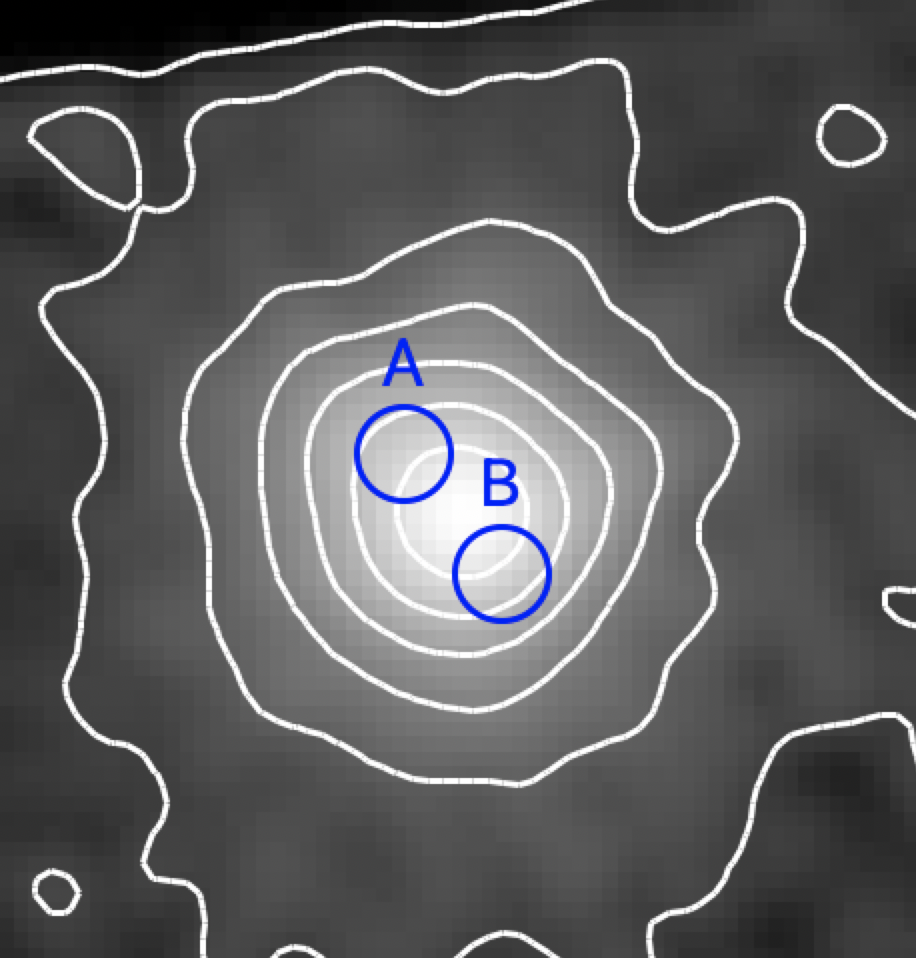} \\
    (a) & (b) & (c) \\
    
    \includegraphics[height=1.75in]{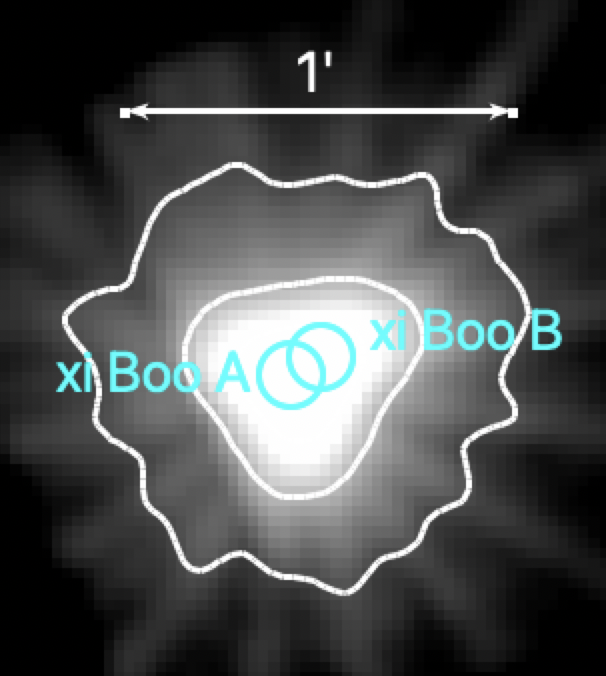} &
    \includegraphics[height=1.75in,clip=true,trim=3.5cm 0cm 3.5cm 0cm]{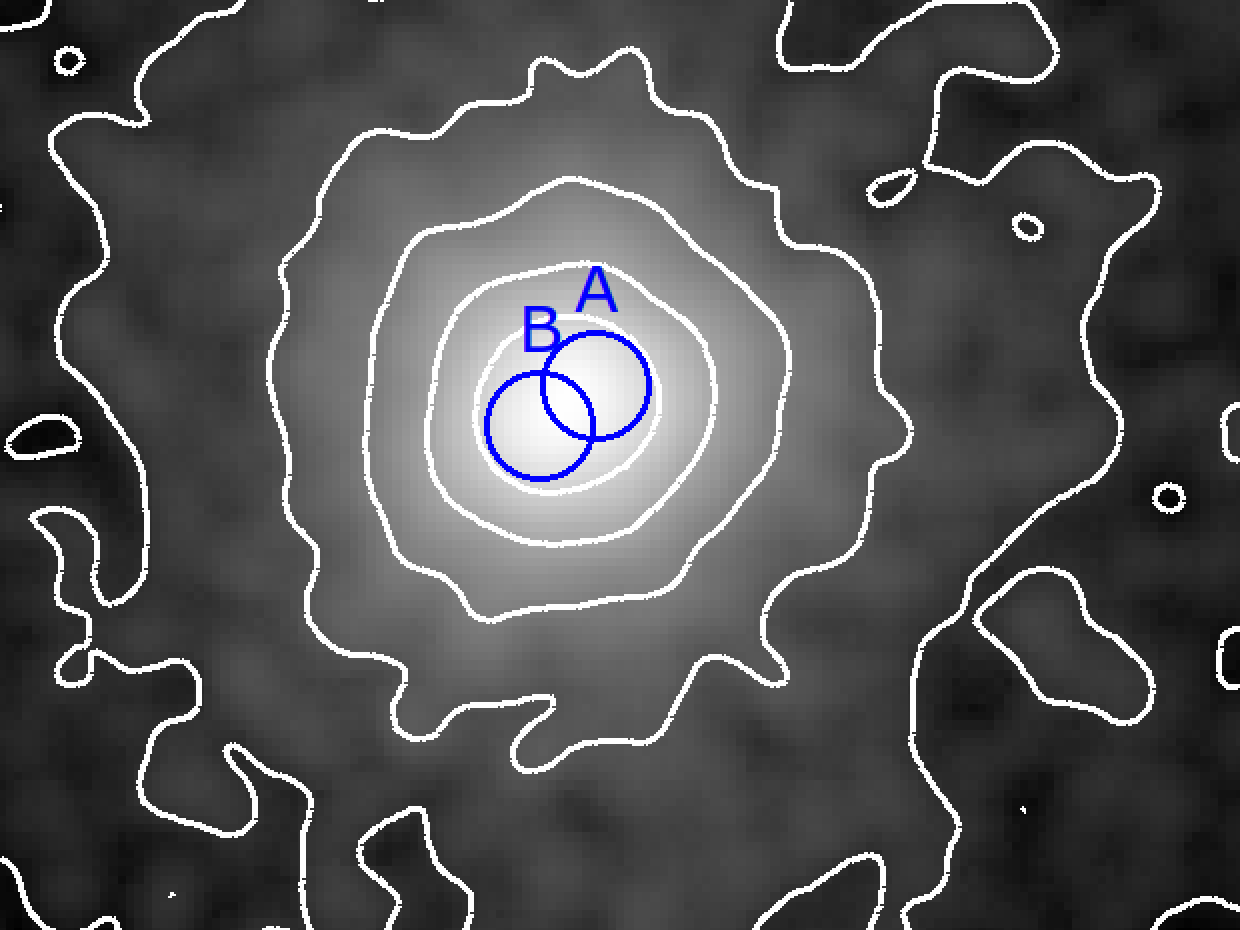} &
    \includegraphics[height=1.75in,clip=true,trim=2.5cm 0cm 2.5cm 0cm]{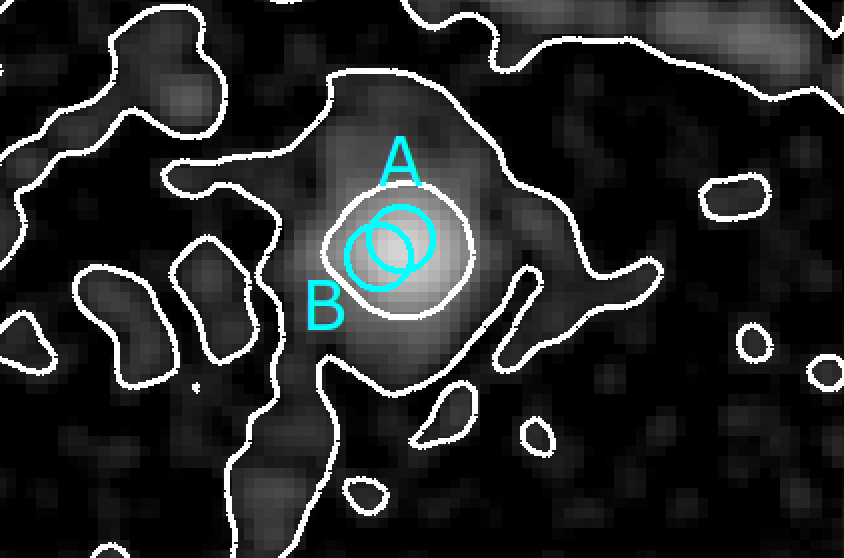} \\
    (d) & (e) & (f) \\

    \includegraphics[height=1.75in]{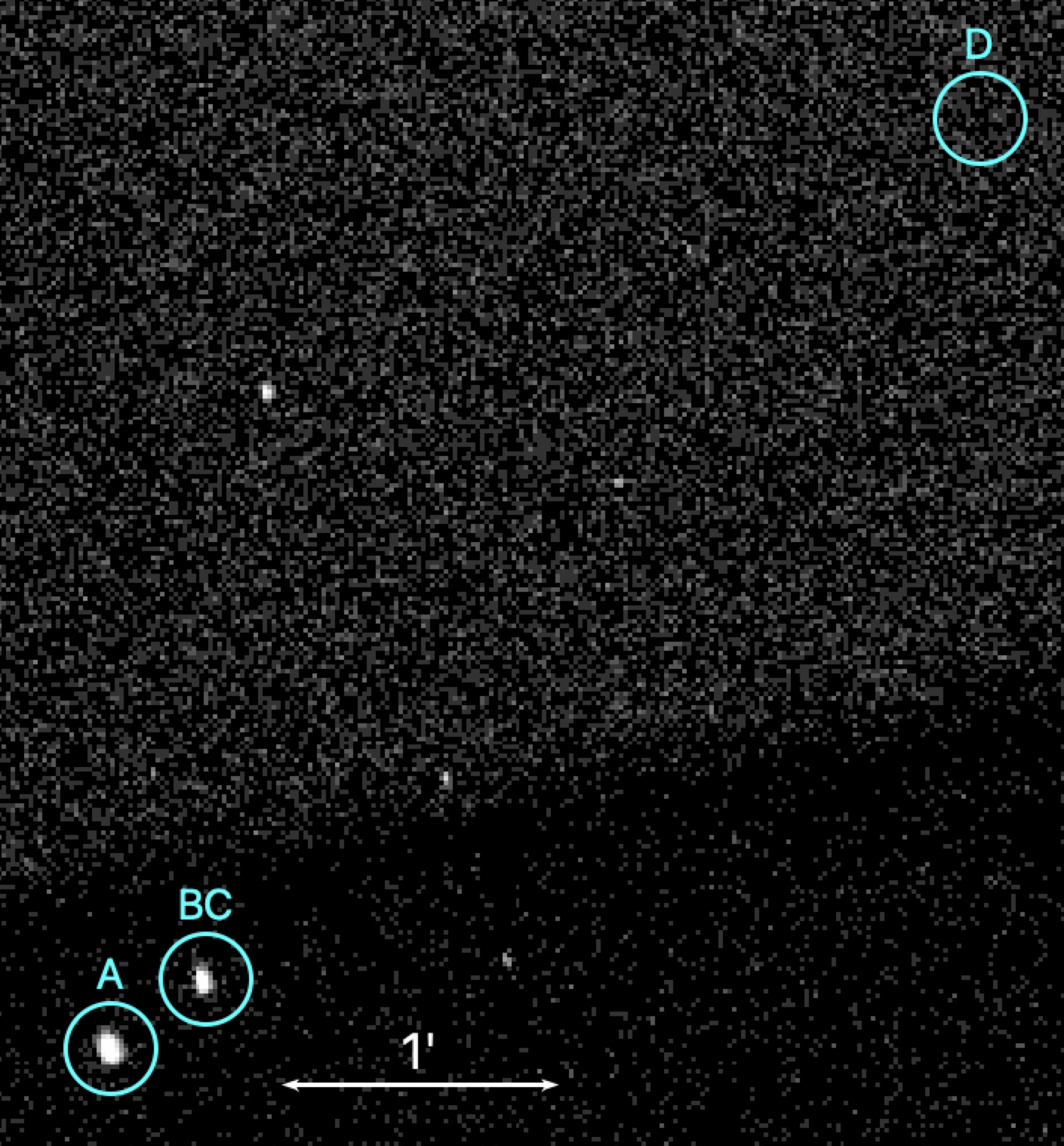} &
    \includegraphics[height=1.75in]{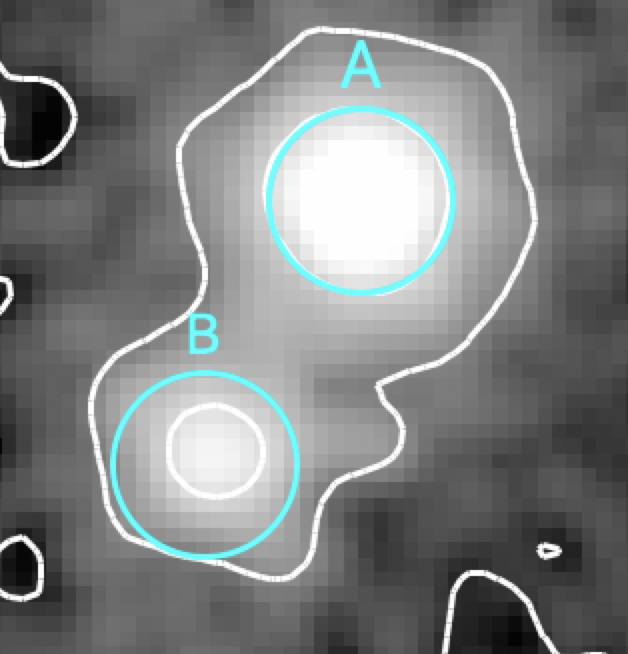} &
    \includegraphics[height=1.75in]{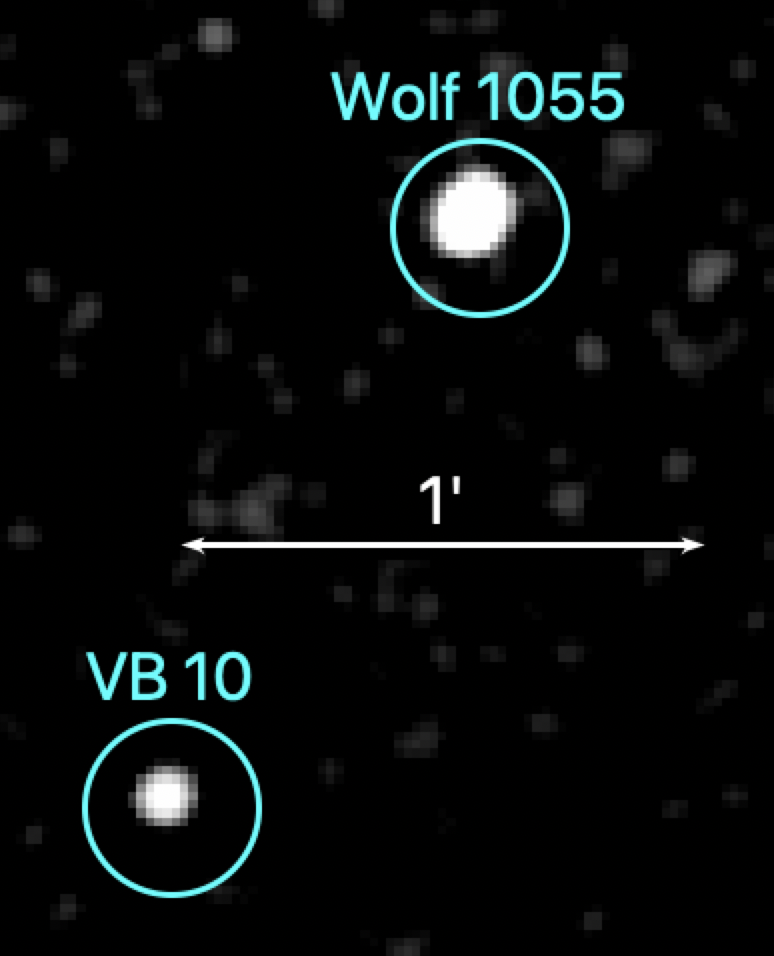} \\
    (g) & (h) & (i) \\
\end{tabular}
    \caption{Images of resolved and unresolved binary systems with \XMM\ and \Chandra. (a) \XMM/PN image of $\upsilon$ And. The locations of the F8V star (A) and the M4.5V star (B) are shown by cyan circles (both with radii 10$^{\prime\prime}$). The stars are separated by $\sim$0.9$^{\prime}$. (b) \XMM\ image of GL 620.1A and the white dwarf WD 1620-391. No X-rays are detected at the location of WD 1620-391 in either observation. (c) The $\alpha$ Cen binary system as seen by the PN camera on \XMM. The positions of the G2V star (A) and K1V star (B) are shown in blue circles (both with radii of 5$^{\prime\prime}$ and separated by $\sim$18$^{\prime\prime}$). (d) The $\xi$ Boo binary system as seen by the MOS2 camera on \XMM. The locations of the G8V star (A) and K4V star (B) are shown in cyan circles (both with radii of 5$^{\prime\prime}$). (e) The 70 Oph binary star system as seen by the PN camera on \XMM. The positions of the K0V star (A) and the K5V star (B) are shown in blue circles (both with radii of 5$^{\prime\prime}$ and separated by $\sim$3.7$^{\prime\prime}$). (f) The GL 783 binary star system as seen by the PN camera on \XMM. The positions of the K3V star (A) and the M4.5V star (B) are shown in cyan circles (both with radii of 5$^{\prime\prime}$ and separated by $\sim$2.6$^{\prime\prime}$. (g) \Chandra\ image of the GL 570 system. The primary GL 570A (K4V) is easily resolved from the binary red dwarf pair GL 570BC (M1.5V and M3V, respectively). Cyan circles have a radius of 10$^{\prime\prime}$. The brown dwarf GL 570D is not detected in X-rays. (h) The 61 Cyg binary system as seen by the PN camera on \XMM. The positions of the K5V star (A) and K7V star (B) are shown in cyan circles (both with a radius of 10$^{\prime\prime}$ and separated by $\sim$12$^{\prime\prime}$). (i) \Chandra\ image of Wolf 1055 and VB 10 (both circles have radii of 10$^{\prime\prime}$ and are separated by $\sim$1.2$^{\prime}$).  \\
    Some images have been smoothed for display purposes only; white contours indicate the intensity distribution of X-ray counts on the detector.}
    \label{fig:Xray_imaging_binaries}
\end{figure*}

\begin{figure*}
\figurenum{13}
\plotone{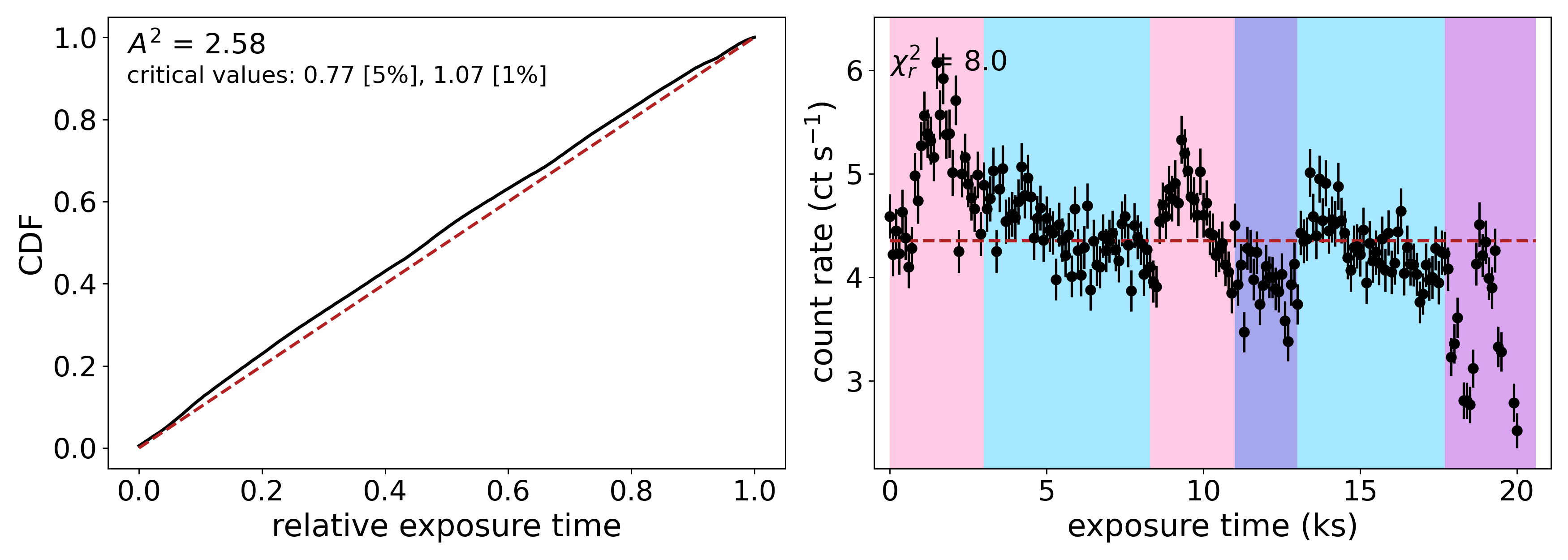}
\caption{{\it Left}: Photon arrival time CDF for $\alpha$ Cen (black) compared to a constant count rate (dark red, dashed). The $A^2$ statistic and critical values are shown in the upper-left corner. {\it Right}: The light curve data (black circles) compared to a constant count rate (dark red, dashed). Quiescent periods are shown in blue-violet, flaring periods in pink, elevated count rate periods in blue, and descending periods in purple. The complete figure set (193 images) is available in the online journal.}\label{fig:LightCurves-FigureSet}
\end{figure*}

\begin{figure}
\figurenum{14}
\plotone{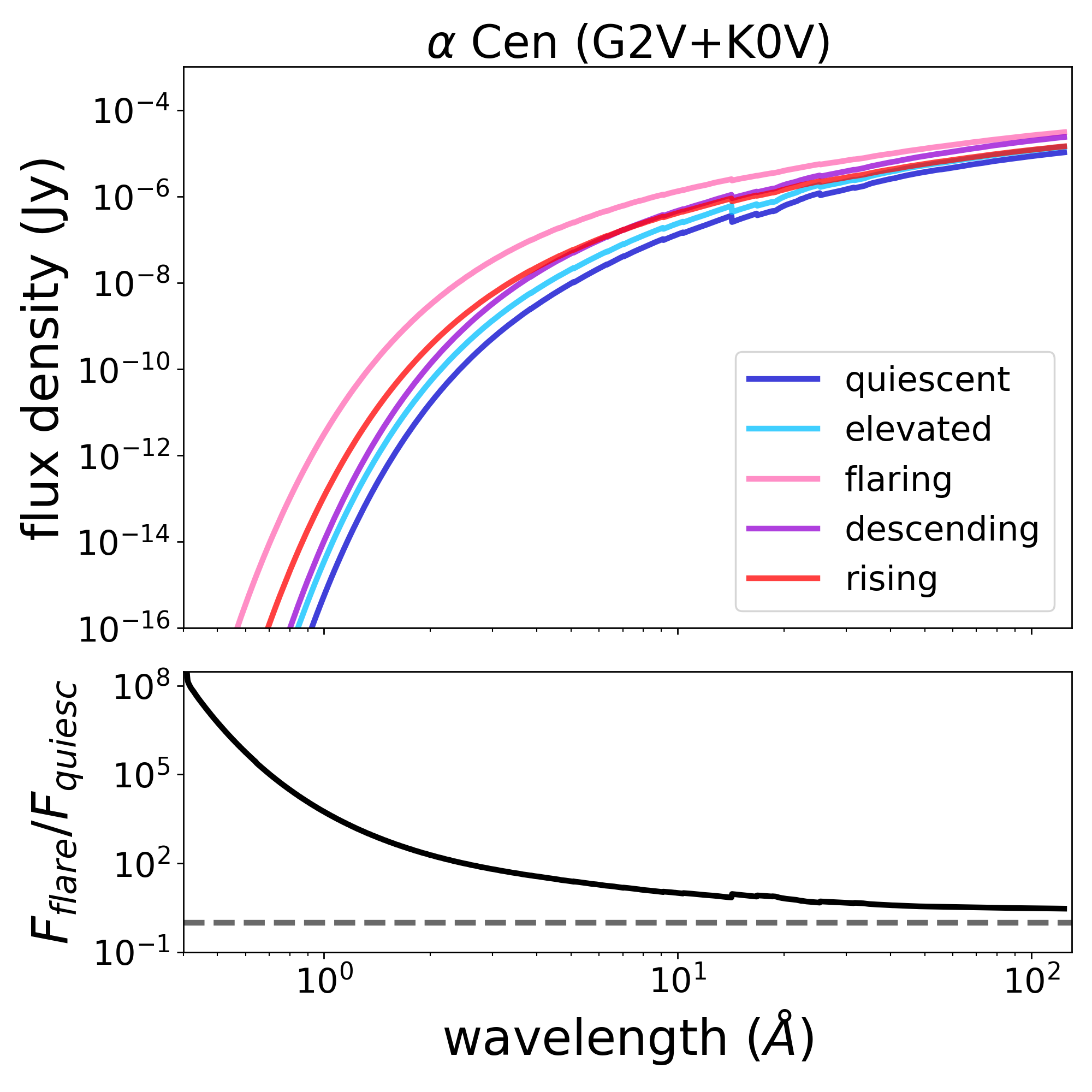}
\caption{{\it Top}: The best-fit continuum spectra for $\alpha$ Cen color-coded by VarFlag. Quiescent periods are shown in dark blue, elevated count rate periods in light blue, flaring periods in pink, periods of descending count rates in purple, and periods of rising count rates in red. {\it Bottom}: The ratio of the flaring-to-quiescent spectrum as a function of wavelength. The complete figure set (18 images) is available in the online journal.}\label{fig:Spectra-FigureSet}
\end{figure}

Figure~\ref{fig:Xray_imaging_binaries} shows X-ray images of resolved and unresolved binary star systems included in our study. Individual images are discussed further for each star below. Figure Set~\ref{fig:LightCurves-FigureSet} shows the light curves extracted for all stars from all observations in which they were detected, with sub-exposures colored by VarFlag. Figure Set~\ref{fig:Spectra-FigureSet} shows the average best-fit spectra for all sufficiently bright stars. For stars that exhibit flaring events, we also show the ratio of the flaring-to-quiescent spectra as a function of wavelength.

\begin{table*}[!ht]
\centering
\footnotesize
    \caption{Summary of Faint Star Observations}\label{tab:match_stars}
    \begin{tabular}{cccccc}
    \hline \hline
        Star     & Observation ID    & Detected?   & Count Rate (ct s$^{-1}$) & Significance ($\sigma$) & Match Star \\
    \hline
    $\xi$ Oph   & Chandra/27852 & Yes & (8.1$\pm$1.0)$\times10^{-3}$ & 8.0 & $\eta$ Crv \\
    $\nu$ Phe & XMM/0206540101 & No    & $<$4.3$\times10^{-3}$ & \nodata & $\iota$ Hor \\
    $\gamma$ Pav & XMM/0670380101 & Yes & (1.1$\pm$0.4)$\times10^{-2}$ & 2.8 & $\iota$ Hor \\
    GJ 1095     & Chandra/4199  & Yes   & (2.1$\pm$0.7)$\times10^{-4}$ & 2.9 & $\iota$ Hor \\
    LHS 237     & XMM/0840210501 & No   & $<$3.7$\times10^{-2}$     & \nodata & $\iota$ Hor \\
    LHS 208     & XMM/0865400201 & No   & $<$1.4$\times10^{-2}$ & \nodata & GL 620.1A \\
    LHS 208     & Chandra/22293  & No   & $<$5.7$\times10^{-4}$ & \nodata & GL 620.1A \\
    $\iota$ Per & Chandra/12338  & No   & $<$1.2$\times10^{-2}$ & \nodata & $\iota$ Hor \\
    $\rho$ CrB  & Chandra/12396  & No   & $<$3.8$\times10^{-3}$ & \nodata & $\iota$ Hor \\
    GL 672      & Chandra/12397  & No   & $<$2.1$\times10^{-3}$ & \nodata & GL 620.1A \\
    47 UMa      & XMM/0304203401 & Yes  & (3.9$\pm$0.3)$\times10^{-2}$ & 12.1 & GL 620.1A \\
    $\beta$ Hyi & Chandra/12337  & Yes  & (5.1$\pm$0.6)$\times10^{-2}$ & 12.5 & GL 620.1A \\
    $\beta$ Hyi & XMM/0006010401 & Yes  & (5.6$\pm$0.4)$\times10^{-2}$ & 14.0 & GL 620.1A \\
    18 Sco      & XMM/0303660101 & Yes  & (2.2$\pm$0.3)$\times10^{-2}$ & 7.3  & GL 311 \\
    18 Sco      & Chandra/12393  & Yes  & (5.5$\pm$1.1)$\times10^{-4}$ & 5.0  & GL 311 \\
    $\mu$ Ara   & XMM/0551021001 & No   & $<$1.6$\times10^{-2}$ & \nodata & GL 327 \\
    $\mu$ Ara   & XMM/0551023101 & No   & $<$2.2$\times10^{-2}$ & \nodata & GL 327 \\
    HD 136352   & XMM/0884680201 & Yes  & (4.1$\pm$1.5)$\times10^{-3}$ & 2.7 & $\kappa^1$ Cet \\
    51 Peg      & XMM/0551020901 & No   & $<$2.3$\times10^{-3}$ & \nodata & $\kappa^1$ Cet \\
    51 Peg      & Chandra/10825  & No   & $<$3.1$\times10^{-3}$ & \nodata & $\kappa^1$ Cet \\
    GJ 777A     & XMM/0304201101 & No   & $<$1.2$\times10^{-2}$ & \nodata & $\delta$ Pav \\
    GJ 777A     & XMM/0304202601 & No   & $<$7.8$\times10^{-3}$ & \nodata & $\delta$ Pav \\
    HD 140901   & Chandra/13769  & Yes  & (8.9$\pm$0.9)$\times10^{-3}$ & 11.0 & $\delta$ Pav \\
    82 Eri      & XMM/0670380601  & No    & $<$1.2$\times10^{-2}$ & \nodata & 55 Cnc A \\
    82 Eri      & Chandra/22292   & No    & $<$8.1$\times10^{-4}$ & \nodata & 55 Cnc A \\
    GL 451A     & Chandra/9931    & No    & $<$4.9$\times10^{-4}$ & \nodata & GL 183 \\
    $\tau$ Cet  & XMM/0670380501  & Yes   & (1.2$\pm$0.2)$\times10^{-2}$ & 7.0 & 55 Cnc A \\
    $\tau$ Cet  & Chandra/1886    & Yes   & (2.4$\pm$0.3)$\times10^{-3}$ & 8.0 & 55 Cnc A \\
    $\delta$ Eri & XMM/0205720101 & Yes   & (7.8$\pm$1.1)$\times10^{-3}$ & 7.3 & 55 Cnc A\\
    40 Eri A    & Chandra/13644   & Yes   & (1.4$\pm$0.2)$\times10^{-2}$ & 7.7 & GL 117 \\
    GL 412 A    & XMM/0742230101  & Yes   & (5.7$\pm$0.2)$\times10^{-2}$ & 28.5 & GJ 832 \\
    GJ 667 C    & Chandra/17317   & Yes   & (5.7$\pm$0.9)$\times10^{-3}$ & 6.3 & GJ 832 \\
    GJ 667 C    & Chandra/17318   & Yes   & (3.1$\pm$0.5)$\times10^{-4}$ & 6.2 & GJ 832 \\
    Kapteyn     & Chandra/merged  & No    & $<$1.2$\times10^{-4}$ & \nodata & Wolf 1055 \\
    Luyten      & Chandra/20164   & Yes   & (3.5$\pm$0.7)$\times10^{-3}$ & 5.0 & AD Leo \\
    $\upsilon$ And B & XMM/0722030101 & No & $<$3.2$\times10^{-2}$ & \nodata & 40 Eri C \\
    VB 10       & XMM/0504010101  & Yes   & (2.7$\pm$0.5)$\times10^{-2}$ & 5.4 & GL 412B \\
    VB 10       & Chandra/616     & Yes   & (1.2$\pm$0.8)$\times10^{-3}$ & 1.5 & GL 412B \\
    VB 10       & Chandra/7428    & Yes   & (1.6$\pm$0.4)$\times10^{-3}$ & 4.0 & GL 412B \\
    \hline \hline
    \end{tabular}
\end{table*}

\subsection{$\eta$ Crv}
There are no significant discrepancies in the stellar physical parameters of $\eta$ Crv. Although there are no known or candidate exoplanets, the system is known to host two debris disks: one at a distance of $\sim$160-180 AU, and a hotter disk at $\sim$3-7 AU \citep{Wyatt+05,Wyatt+07,Smith+08}.

$\eta$ Crv was observed four times with \Chandra. All observations were taken with ACIS-S in 1/4 subarray mode for $\sim10$ ks. In all observations, $\eta$ Crv was detected with $\sim$1000 net counts (0.5-7 keV), but no significant variability was observed in the X-ray light curves (see Figure Set~\ref{fig:LightCurves-FigureSet}). The spectrum of $\eta$ Crv is well-described by a two-component thermal plasma model, with no significant changes in the temperatures or normalizations across the different observations.
\subsection{$\xi$ Oph}
There are no significant discrepancies in the stellar physical parameters of $\xi$ Oph, and the star does not host any no known or candidate exoplanets. $\xi$ Oph was observed once with \Chandra/ACIS-S for 19.8 ks and robustly detected with $\sim$160 counts (8$\sigma$ significance). There are hints of low-level variability in the X-ray emission of $\xi$ Oph, but the relatively low number of net counts prohibits spectral modeling.
\subsection{$\upsilon$ And}
$\upsilon$ And A is an F8V dwarf in a presumed binary system with a proper motion companion $\upsilon$ And B, an M4.5V dwarf (also referred to as $\upsilon$ And D). The two stars have an angular separation of $\sim$0.9$^{\prime}$ (corresponding to $\sim$750 AU). There are no significant discrepancies in the physical parameters of either star. $\upsilon$ And A hosts three confirmed exoplanets ($\upsilon$ And b, c, and d) which were discovered via RV measurements. The masses of $\upsilon$ And b, c, and d are \mpsini\ $\sim$0.67 $M_{\rm Jup}$, $\sim$0.67 $M_{\rm Jup}$, $\sim$2 $M_{\rm Jup}$, and $\sim$4 $M_{\rm Jup}$, respectively, with orbital periods (orbital distances) of $\sim$4.6 days (0.06 AU), $\sim$241 days (0.82 AU), and $\sim$1282 days (2.5 AU), respectively \citep[all planetary data taken from][]{Rosenthal+21}.

The $\upsilon$ And system was observed four times with \Chandra/ACIS-S in 1/8 subarray mode (for $\sim$15 ks per observation) and once with \XMM\ for 5.3 ks. $\upsilon$ And A was detected in all five observations with sufficient counts to extract light curves (see Figure Set~\ref{fig:LightCurves-FigureSet}) and with enough counts in \Chandra\ observations 10976 and 10977 to enable spectral modeling. There is no evidence for strong X-ray variability in any observation, and both \Chandra\ spectra were consistent with a single-temperature thermal plasma model with a temperature of $\sim$0.3 keV ($\sim$3.4 MK; see Table~\ref{table:spectral_modeling}). $\upsilon$ And B was out of the field of view of the \Chandra\ observations. However, the star was within the field of view of the \XMM\ PN detector but fell just on the gap between two CCDs and was not detected (see Figure~\ref{fig:Xray_imaging_binaries}). 
\subsection{$\iota$ Hor}
$\iota$ Hor is characterized either as a late F-type \citep[F9V;][]{Gray+06} or early G-type \citep[G0V;][]{Turnbull15} star. There is additional disagreement on the age of the system: \citet{Turnbull15} quotes an age of 2.72 Gyr, while \citet{Sanz-Forcada+10} suggest a much younger age of $\sim$0.47 Gyr \citep[see also the detailed analysis of][where an age of 600 Myr is adopted]{Sanz-Forcada+19}. We adopt a spectral type of F9V and an age of 0.47 Gyr in this work. $\iota$ Hor hosts one confirmed exoplanet that was first detected via radial velocity (RV) measurements by \citet{Kurster+00}. It is a giant planet with \mpsini$\sim$2.3 \Mjup\ and an orbital period of $\sim$300 days \citep[corresponding to an orbital distance of $\sim$0.9 AU;][]{Stassun+17}.

$\iota$ Hor is one of the most thoroughly observed stars in our sample, with 32 publicly available \XMM\ observations taken between 2011 and 2018. A detailed study of the coronal activity and structure of the star is presented in \citet{Sanz-Forcada+19}, who combined the \XMM\ monitoring with observations from TESS and the STIS instrument on the {\em Hubble Space Telescope}. They corroborate a 1.6 year X-ray activity cycle for the star, although $\iota$ Hor does not exhibit the same type of dramatic changes in luminosity as seen in other stars in our sample (see Figure~\ref{fig:time_vs_Lx}). Our X-ray analysis of $\iota$ Hor is consistent with \citet{Sanz-Forcada+19}, although we do not undertake a more extensive study of the abundances of the star. Light curves of $\iota$ Hor are shown in Figure Set~\ref{fig:LightCurves-FigureSet}, and the best-fit spectrum is described in Table~\ref{table:spectral_modeling}.
\subsection{$\nu$ Phe}
$\nu$ Phe is characterized either as a late F-type \citep[F9V;][]{Gray+06} or early G-type \citep[G0V;][]{Turnbull15} star. There are no confirmed or candidate exoplanets in the system, and the star was undetected in a 16 ks \XMM\ observation.
\subsection{$\gamma$ Pav}
$\gamma$ Pav is a known F-type star, with the specific spectral type ranging from F9V \citep{Gray+06} to F7V \citep{Turnbull15}. There are significant discrepancies in the estimated age of $\gamma$ Pav: \citet{Holmberg+09} quote an age of $\sim$1 Gyr based on photometric observations and comparison to stellar evolution models, while and asteroseismic analysis by \citet{Mosser+08} yields an age of 7.25 Gyr. There are no known or candidate exoplanets orbiting $\gamma$ Pav. The star was only marginally detected (2.8$\sigma$) in a 25.9 ks \XMM\ observation. The low X-ray luminosity of $\gamma$ Pav is consistent with the older age estimate of \citet{Mosser+08}.
\subsection{$\beta$ Vir}
There are no significant discrepancies in the stellar physical parameters for $\beta$ Vir, and the star does not host any currently known or candidate exoplanets. $\beta$ Vir was observed once with \XMM\ (observation ID 0044740201) and was detected with $\sim$18,450 net counts on the PN camera (thick filter). The X-ray light curve does not show evidence of significant variability (see Figure Set~\ref{fig:LightCurves-FigureSet}). The spectrum is well described by a two-temperature thermal plasma model (see Table~\ref{table:spectral_modeling}).
\subsection{GJ 1095}
The spectral type of GJ 1095 is given as F9V by \citet{Gray+03} or G0V by \citet{Turnbull15}. There are no known or candidate exoplanets orbiting GJ 1095. The star was only marginally detected (2.9$\sigma$) in a 96.2 ks \Chandra\ observation.
\subsection{LHS 237}
There are no significant discrepancies in the stellar physical parameters of LHS 237, and no age estimate is currently available in the literature. LHS 237 is a triple-star system: the inner spectroscopic binary has an orbital period of $\sim$10 years \citep[the angular separation of $\sim$0.3$^{\prime\prime}$ is unresolvable by \XMM;][]{Tokovinin+12}, and the third component is one of the coolest known white dwarfs (NLTT 18141 = GJ 288B) located $\sim$14.5$^{\prime}$ from the central binary \citep{Holberg+13}. There are no known or candidate exoplanets in the system.

LHS 237 was serendipitously imaged by \XMM\ for 15.6 ks in observation 0840210501 (the intended target for the observation was a Wolf-Rayet star). The inner spectroscopic binary is not detected, and the white dwarf is outside the \XMM\ field of view.
\subsection{$\iota$ Per}
There are no significant discrepancies in the stellar physical parameters for $\iota$ Per. The spectral type is either F9.5V \citep{Gray+03} or G0V \citep{Turnbull15}. There are no known or candidate exoplanets in the system. $\iota$ Per was observed for 4.9 ks with \Chandra\ in ACIS-S 1/4 subarray mode, but the star was not detected.
\subsection{$\beta$ Hyi}
There are no significant disagreements in the stellar physical parameters of $\beta$ Hyi, and the star does not host any currently known or candidate exoplanets. $\beta$ Hyi was observed once with \Chandra/ACIS-S in 1/4 subarray mode and once with \XMM. The star was detected in both observations with sufficient counts to extract light curves, but the star is too faint for accurate spectral modeling to be performed. There is no strong evidence for X-ray variability in either the \Chandra\ or \XMM\ light curves of $\beta$ Hyi (see Figure Set~\ref{fig:LightCurves-FigureSet}).
\subsection{$\beta$ Com}
There are only minor differences in the reported age of $\beta$ Com: \citet{Takeda+07} report an age $<$1.12 Gyr, while \citet{Turnbull15} report an age of 1.7 Gyr. All other spectral parameters are taken from \citet{Stassun+17}. There are currently no known or candidate exoplanets orbiting $\beta$ Com.

$\beta$ Com was robustly detected (with $\sim$38,000 net counts) in a 57.2 ks observation by \XMM. We extracted a light curve from this observation and found marginal evidence for count rate variability (see Figure Set~\ref{fig:LightCurves-FigureSet}). Given the large number of counts available, we split the observation into 5 ks-long sub-exposures and extracted spectra from each sub-exposure to search for evidence of underlying spectroscopic variability. Throughout the observation, a two-temperature thermal plasma model was sufficient to describe the X-ray spectrum of $\beta$ Com, with the best fit temperatures and normalizations remaining constant within the uncertainties. All best-fit sub-exposure spectral parameters for $\beta$ Com are provided in Table~\ref{table:all_spectral_fits}, and the average best-fit spectral parameters are listed in Table~\ref{table:spectral_modeling}. The best-fit spectral models are shown in Figure Set~\ref{fig:Spectra-FigureSet}.

\subsection{LHS 208 ($\pi$ Men)}
All spectral parameters for LHS 208 are taken from \citet{Huang+18}. LHS 208 hosts three confirmed exoplanets. HD 390931 b is a giant planet that was detected via RV measurements \citep{Jones+02}; it has \mpsini$\sim$12 \Mjup\ and an orbital period of $\sim$2090 days \citep[corresponding to an orbital distance of $\sim$3.3 AU;][]{Feng+22}. $\pi$ Men c is a super-Earth that was discovered via transits \citep{Gandolfi+18}; it has an \mpsini$\sim$3.5 \Mearth\ with an orbital period of $\sim$6 days \citep[$\sim$0.07 AU orbital distance;][]{Feng+22}. $\pi$ Men d was detected via RV measurements with \mpsini$\sim$13 \Mearth\ and an orbital period of $\sim$125 days \citep{Hatzes+22}. LHS 208 not detected in a 36.9 ks observation with \XMM\ or a 19.6 ks observation with \Chandra.
\subsection{$\rho$ CrB}
Some minor discrepancies, mass ranges from 0.95 \citep{Brewer+23} to 1.05 \Msun\ \citep{Stassun+17} and spectral type from G0V \citep{Gray+06} to G2V \citep{Turnbull15}. $\rho$ CrB hosts three confirmed exoplanets ($\rho$ CrB c, d, and e), with the existence of a fourth exoplanet controversial \citep[$\rho$ CrB b,][which may simply be the result of correlated stellar activity in the RV measurements;]{Brewer+23}. The masses of $\rho$ CrB c, d, and e are \mpsini\ $\sim$28 \Mearth, $\sim$22 \Mearth, and 3.8 \Mearth, respectively, with orbital periods of $\sim$102 days ($\sim$0.4 AU), $\sim$282 days ($\sim$0.8 AU), and $\sim$13 days ($\sim$0.11 AU), respectively. $\rho$ CrB is also approaching the end of its main sequence lifetime, with a prognosis of planetary engulfment for most of the known planets in the system \citep{kane2023f}. $\rho$ CrB was observed once with \Chandra/ACIS-S for 9.8 ks, but was not detected.
\subsection{GL 672}
There are no significant discrepancies in the stellar physical parameters for GL 672, and there are no known or candidate exoplanets in the system. GL 672 was observed for 9.9 ks by \Chandra/ACIS-S but was not detected.
\subsection{$\chi^1$ Ori}
$\chi^1$ Ori is a G0V-type star in a $\sim$14 year elliptical ($e\sim0.45$) orbit with a low-mass companion. RV measurements compiled by \citet{Han+02} were used to infer a secondary mass of $\sim$0.15 \Msun\ with a likely spectral type of M4V or M5V. The estimated angular semimajor axis of the orbit is $\sim$0.668$^{\prime\prime}$, unresolvable by \XMM\ but potentially spatially resolvable by \Chandra\ (see Section~\ref{sec:data}). The inferred spectral parameters may be biased due to the presence of this low-mass companion. There is significant disagreement in the literature about the age of the system. \citet{Turnbull15} and \citet{Takeda+07} estimate an age of 4.32 Gyr, while \citet{Mamajek+08} use activity-rotation diagnostics to estimate an age of 300-400 Myr. There are no confirmed or candidate exoplanets in the system.

Since only one \XMM\ observation is available for $\chi^1$ Ori, we are unable to identify the source of the X-ray emission -- whether it be from the G0V star, the low-mass companion, or both. There is one potential, minor flaring event during the observation, starting $\sim$21 ks after the beginning of the observation and lasting for $\sim$4 ks, where the peak count rate exceeds the median count rate by $\sim$40\% (see Figure Set~\ref{fig:LightCurves-FigureSet}). While this may not be a bona fide flare event, we classify this period as flaring so that its spectrum can be compared to the remainder of the observation. The light curve otherwise shows a count rate that is either constant within the uncertainties (which we classify as quiescent) or somewhat elevated above the median count rate level by $\sim$20\% (which we classify as elevated).

Figure Set~\ref{fig:Spectra-FigureSet} shows the quiescent, elevated, and potential flare spectra for the $\chi^1$ Ori system. The elevated and flare spectra are nearly indistinguishable, and both show excess emission over the quiescent spectrum at wavelengths $\lesssim$10\AA. This difference in short-wavelength emission is driven primarily by the hottest thermal plasma component in our model, which has a temperature of $\sim$0.8 keV for the elevated and flaring spectra but $\sim$0.7 keV in the quiescent spectrum (see also Table~\ref{table:spectral_modeling}). The integrated X-ray luminosity of $\chi^1$ Ori is constant within the uncertainties across the three spectra.

\subsection{GL 788}
There are no significant discrepancies in reported stellar parameters of GL 788. The system does not host any known or candidate exoplanets. GL 788 was detected in a 9.6 ks \XMM\ observation with $\sim$4470 net counts. There was no evidence of count rate variability apparent in the X-ray light curve (see Figure Set~\ref{fig:LightCurves-FigureSet}). The spectrum was well-described with a single, cool thermal plasma ($kT\sim$0.2 keV).
\subsection{47 UMa}
There are no significant discrepancies in the stellar physical parameters reported in the literature for 47 UMa. The star hosts three confirmed giant exoplanets that were detected via radial velocity measurements. 47 UMa b, c, and d have masses of \mpsini$\sim$2.4 $M_{\rm Jup}$, $\sim$0.5 $M_{\rm Jup}$, and 1.5 $\sim$2.4 $M_{\rm Jup}$, respectively, and have orbital periods (distances) of $\sim$1077 days ($\sim$2 AU), $\sim$2287 days ($\sim$7.8 AU), and $\sim$19000 days ($\sim$12.4 AU), respectively. All planetary parameters were taken from \citet{Rosenthal+21}.

47 UMa was detected at $\sim$12$\sigma$ significance in a 6.2 ks \XMM\ observation. We extracted a light curve from the observation but found no evidence of variability (see Figure Set~\ref{fig:LightCurves-FigureSet}), and the star was too faint to enable spectroscopic modeling.
\subsection{GL 620.1A}
GL 620.1A is either co-moving with \citep{Mugrauer19} or in a wide binary with \citep{Holberg+02} the DA-type white dwarf WD 1620-391. The angular distance between the two components is $\sim$5.8$^{\prime}$, which would correspond to a physical separation of $\sim$4500 AU at the distance of GL 620.1A. GL 620.1A is classified as a barium star, and its over-abundance in $s$-process elements has been attributed to pollution due mass transfer during the white dwarf progenitor's AGB phase \citep{PortodeMello+97}. There is some disagreement in the literature over the spectral type and age of GL 620.1A. Spectral types range from G1V \citep{Gray+06} to a later G3/5V \citep{Turnbull15}; we adopt a spectral type G1V. While \citet{Turnbull15} quote an age of 2 Gyr, other estimates of the age of GL 620.1A are lower: \citet{Ghezzi+10} and \citet{Takeda+07} derive upper limits of $<$1 Gyr and $<$0.68 Gyr, respectively, while \citet{Sanz-Forcada+10} adopt an age of 0.4 Gyr. One giant exoplanet is confirmed via RV measurements of GL 620.1A \citep{Mayor+04}; the planet has \mpsini$\sim$1.2 \Mjup\ and an orbital period of $\sim$528 days (orbital distance of $\sim$1.3 AU).

GL 620.1A is detected in both publicly available \XMM\ observations (Figure~\ref{fig:Xray_imaging_binaries}) with sufficient counts to enable spectral modeling (see Figure Set~\ref{fig:Spectra-FigureSet} and Table~\ref{table:spectral_modeling}). There is no significant variability present in the count rate light curves of GL 620.1A (Table~\ref{tab:variability_metrics}, see Figure Set~\ref{fig:LightCurves-FigureSet}), so we do not attempt to divide the observations into sub-exposures.

\subsubsection{The White Dwarf WD 1620-391}\label{sec:WD1620-391}
While GL 620.1A is robustly detected in both publicly available \XMM\ observations, WD 1620-391 is not (see Figure~\ref{fig:Xray_imaging_binaries}a). The 3$\sigma$ upper limits on the count rates of 0.014 ct s$^{-1}$ (observation ID 0822070201) and 0.0236 ct s$^{-1}$ (observation ID 0822070301). We use WebPIMMs\footnote{See \url{https://heasarc.gsfc.nasa.gov/cgi-bin/Tools/w3pimms/w3pimms.pl}} to convert the observed \XMM/PN count rates (both observations were taken with the thick filter) to a flux assuming an APEC thermal plasma model with $kT\approx0.5$ keV \citep[the approximate best-fit model for the accreting white dwarf G 29-38, which is detected in X-rays by both \XMM\ and \Chandra;][]{Cunningham+22,Estrada-Dorado+23}. We find flux (luminosity) upper limits of 2.0$\times10^{-14}$ \flux\ (4$\times10^{26}$ \lum) and 3.4$\times10^{-14}$ \flux\ (6.8$\times10^{26}$ \lum) for observations 0822070201 and 0822070301, respectively. These luminosity limits are a factor of $\sim$5-7 higher than the observed luminosity of G 29-38; targeted observations with a more sensitive X-ray instrument are needed to determine if WD 1620-391 is emitting X-rays at a levels comparable to other X-ray-detected white dwarfs.

\subsection{GL 311 ($\pi^1$ UMa)}
There are no significant discrepancies in the stellar physical parameters reported in the literature for GL 311. The system does not host any currently known or candidate exoplanets.

The variability metrics for GL 311 (see Table~\ref{tab:variability_metrics}) suggest that count rate variability is observed in the available \XMM\ observation (observation number 0111400101; see Figure Set~\ref{fig:LightCurves-FigureSet}). There are two minor flare-like events in the light curve. The first begins at $\sim$21 ks after the beginning of the observation, and reaches a peak count rate that is a factor of $\sim$2 higher than the median count rate. This event lasts for approximately 8 ks and it immediately followed by a second count rate increase, which exceeds the median count rate by a factor of $\sim$1.6 and lasts for approximately 5 ks. The post-flare count rate is elevated by $\sim$15\% relative to the pre-flare count rate. Spectroscopic modeling before, during, and after the flaring event show that the temperatures of all three thermal components increased in temperature during the flaring event, although a modest decrease in the normalization (related to the VEM) during the flare led to a constant (within the uncertainties) luminosity (Figure Set~\ref{fig:Spectra-FigureSet}; see also Table~\ref{table:spectral_modeling}). 

Thus, although GL 311 clearly exhibits a degree of rapid X-ray variability (on a timescale of a few hours), this activity does not have a significant affect on the overall X-ray luminosity output of the star. The relative stability of the star's X-ray luminosity is also apparent in Figure~\ref{fig:time_vs_Lx}, where the luminosity curve for GL 311 is near-vertical.
\subsection{$\alpha$ Cen}
$\alpha$ Centauri A (G2V) and B (K1V) are members of the extensively studied triple system that also includes Proxima Centauri (M5.5V; see Section~\ref{sec:ProxCen}). $\alpha$ Cen A and B are in a $\sim$80 year elliptical orbit \citep[e$\sim$0.5;][]{Akeson+21}. One candidate exoplanet, $\alpha$ Cen B b, was reported \citep{Dumusque+12} but was later found to be a false positive signal \citep{Rajpaul+16}. There are currently no other known or candidate exoplanets in the $\alpha$ Cen AB system. 

The semi-major axis of the $\alpha$ Cen system is $\sim$18$^{\prime\prime}$, but due to the brightness of the stars there is considerable blending in the \XMM\ images (see Figure~\ref{fig:Xray_imaging_binaries}). We therefore do not attempt to extract X-ray products (light curves, spectra) for each star individually. The $\alpha$ Cen system exhibits significant X-ray variability (see Figure Set~\ref{fig:LightCurves-FigureSet}). The system is well-described by a three-component thermal plasma model, and increases in the observed count rates are driven largely by increases in the hottest plasma temperature (from $\sim$0.6 keV during quiescent periods to $\sim$0.8-0.9 keV during times associated with flaring events) and normalization (which increases by a factor of $\sim$12). The spectra of $\alpha$ Cen are shown in Figure Set~\ref{fig:Spectra-FigureSet} (see also Table~\ref{table:spectral_modeling}). Due to the significant blending of the two components, we do not use the quiescent $\alpha$ Cen spectrum as a baseline for inferring other stellar luminosities.
\subsection{44 Boo}
44 Boo (also called $\iota$ Boo) is a triple star system with a G0V primary (44 Boo A) in $\sim$210 year orbit with a spectroscopic binary composed of a K0V (44 Boo Ba) dwarf and a K4V (44 Boo Bb) dwarf \citep{Zasche+09,Zirm11}. 44 Boo A is separated from the spectroscopic binary by $\sim$0.2$^{\prime\prime}$. The 44 Boo B binary has an orbital period of only $\sim$6.4 hours and is a possible contact binary \citep{Lu+01}. There are no known or candidate exoplanets in the 44 Boo system. 

44 Boo is robustly detected with \XMM\ in an 18.8 ks observation, with count rate variations clearly seen in the light curve (see Figure Set~\ref{fig:LightCurves-FigureSet}). We do not include the final $\sim$5 ks in our spectral analysis due to a significant background flaring event during the observation. Our spectral modeling revealed no significant differences between the minor flaring event from $\sim$6-10 ks and the relatively constant count rate period from 15-20 ks. The spectrum of 44 Boo did change during periods of declining count rate (the first 6 ks and from 10-15 ks) as the result changes in the hottest thermal plasma component: $kT_3$ decreased from $\sim$2.1-2.2 keV during quiescent/flaring times to $\sim$1.5 keV during periods of secular decline, while the normalization of this component (related to the VEM) more than doubled. These changes to APEC \#3 result in a decrease in X-ray flux at the shortest wavelengths ($\lesssim$10 \AA); see Table~\ref{table:spectral_modeling} and Figure Set~\ref{fig:Spectra-FigureSet}.
\subsection{18 Sco}
There are minor differences in the reported spectral type of 18 Sco: \citet{Turnbull15} quote a spectral type of G5V, but \citet{Gray+06} classifies the star as a G2V. We adopt a spectral type of G2V. 18 Sco hosts one candidate super-Earth with \mpsini\ $\sim$6.8 $M_{\oplus}$ and an orbital period of $\sim$20 days \citep{Laliotis2023}.  The star is detected in both a 20.1 ks \XMM\ observation and a 62.2 ks observation with \Chandra/ACIS-S. However, the 18 Sco is not bright enough in either observation for timing or spectroscopic analysis to be performed. 
\subsection{51 Peg}
There are no significant discrepancies reported for the stellar parameters of 51 Peg. The star is famous for hosting the first discovered exoplanet \citep{Mayor+95}, the hot Jupiter 51 Peg b with \mpsini\ $\sim$0.46 $M_{\rm Jup}$ and an orbital period of $\sim$4.2 days \citep[$\sim$0.05 AU][]{Rosenthal+21}. 51 Peg was observed with both \Chandra\ and \XMM, but was not detected in either observation.
\subsection{$\zeta$ Ret}
The wide binary $\zeta$ Ret contains $\zeta^1$ Ret, which is listed as a G3V/G5V by \citet{Turnbull15} but G2V as \citet{Gray+06}, and $\zeta^2$ Ret, a G2V star \citep{delPeloso+00}. The two stars are separated by $\sim$5.2$^{\prime}$. The system does not hot any known or candidate exoplanets.

Although this separation would be easily resolvable by \XMM, only $\zeta^1$ Ret is contained with the fields of view of the \XMM\ cameras. The star is robustly detected (with $\sim$7700 net counts) in a 20.9 ks \XMM\ observation (ID 0404920101). No significant variability is observed in the X-ray light curve of $\zeta^1$ Ret (see Figure Set~\ref{fig:LightCurves-FigureSet}), and the spectrum is well-described by a two-component thermal plasma model (with temperatures of $\sim$0.1 keV and $\sim$0.4 keV; see Figure Set~\ref{fig:Spectra-FigureSet} and Table~\ref{table:spectral_modeling}).
\subsection{HD 136352}
There are no significant discrepancies in the stellar physical parameters reported in the literature for HD 136352. The star hosts three confirmed exoplanets, all of which have masses in the super-Earth/sub-Neptune regime. HD 136352 b, c, and d have \mpsini\ of $\sim$4.7 \Mearth, $\sim$11 \Mearth, and $\sim$8.8 \Mearth, respectively, with orbital periods (distances) of $\sim$11.6 days ($\sim$0.1 AU), $\sim$28 days ($\sim$0.17 AU), and $\sim$107 days ($\sim$0.43 AU), respectively. The inner two planets were found to transit the host star by \citet{kane2020c}, providing bulk densities for the planets. Subsequent observations by \citet{Delrez+21} revealed that the third, outer planet also transits the host star. HD~136352 is marginally detected ($\sim$2.7$\sigma$) in a 16.9 ks \XMM\ observation with a count rate of 0.0041$\pm$0.0015 ct s$^{-1}$ with the PN camera and medium filter (Table~\ref{tab:match_stars}).
\subsection{GL 327}
There are no significant discrepancies in reported stellar parameters of GL 327 in the literature. The system does not host any currently known or candidate exoplanets. GL 327 was robustly detected with $\sim$10,500 net counts in a 23.5 ks exposure with \XMM\ (observaton ID 0404920201). Count rate variability was observed in the X-ray spectrum as shown in Figure Set~\ref{fig:LightCurves-FigureSet}, so we split the observation into three sub-exposures and modeled the three spectra independently. In all three sub-exposures, a single-temperature thermal plasma component was sufficient to describe the observations. During the quiescent period from $\sim$6-21 ks, the plasma temperature was $\sim$0.4 keV, which increased to $\sim$0.5 keV during the descending and flaring periods. The normalization during periods of increased X-ray count rate additionally increased by $\sim$20\%. The spectra of GL 327 are shown in Figure Set~\ref{fig:Spectra-FigureSet} (see also Table~\ref{table:spectral_modeling}.
\subsection{$\mu$ Ara}
No significant discrepancies in stellar parameters are present in the literature for $\mu$ Ara, although there is some uncertainty about the evolutionary stage; we adopt a spectral type of G3IV/V \citep{Turnbull15}. $\mu$ Ara hosts four known (giant) exoplanets, all detected via RV measurements. HD 160691 b, c, d, and e have masses of  \mpsini$\sim$1.7 \Mjup, $\sim$2.4 \Mjup, $\sim$10.2 \Mearth, and $\sim$0.5 \Mjup, respectively, and the planets have orbital periods (orbital distances) of $\sim$646 days ($\sim$1.5 AU), $\sim$4470 days ($\sim$5.5 AU), $\sim$10 days ($\sim$0.09 AU), and $\sim$307 days ($\sim$0.9 AU), respectively. All planetary data were taken from \citet{Gozdziewski+07}. $\mu$ Ara is not detected in either of the two publicly available \XMM\ observations.
\subsection{$\kappa^1$ Cet}
There is some discrepancy in the reported age estimates of $\kappa^1$ Cet: \citet{Turnbull15} lists an age of 2.2 Gyr and \citet{Takeda+07} derive an upper limit of $<$2.76 Gyr, but many studies report significantly younger ages. For example, \citet{Mamajek+08} derive an age of 300-400 Myr, while \citet{Gudel+97} estimated an age of 750 Myr and and \citet{Dorren+94} estimate an age of 650 Myr. There are no currently known or candidate exoplanets orbiting $\kappa^1$ Cet.

$\kappa^1$ Cet is relatively well-observed in X-rays, with five \XMM\ observations and two \Chandra\ observations available in the archives, and the star is easily detected in all exposures. It was also observed twice with \Chandra\ using ACIS-S7 in 1/8 subarray mode with a 0.44 s readout time in both observations. While there are clear indications of variability in the light curves of $\kappa^1$ Cet the maximum observed count rate only ever exceeds the median count rate by a factor of $\sim$2 (see Figure Set~\ref{fig:LightCurves-FigureSet}). However, despite the only modest changes in count rate, the spectra during these minor flare-like events is quite distinct from the lower count rate quiescent periods, with significant excess emission a wavelengths shortward of 10 \AA\ compared to quiescent times. The change in count rate is being driven by a significant increase in the temperature (by $\sim$0.2 keV, or $\sim$2.3 MK) and normalization (a factor of $\sim$2.2 increase) of the hottest component in a three-thermal plasma component model (Figure Set~\ref{fig:Spectra-FigureSet}; see also Table~\ref{table:spectral_modeling}).
\subsection{HD 140901}
There are no significant discrepancies in the published stellar physical parameters of HD 140901. Two planets have been identified via RV measurements. HD 140901 c has \mpsini\ $\sim$1.8 $M_{\rm Jup}$ with an orbital period of $\sim$14390 days \citep[orbital distance of $\sim$11.8 AU;][]{Philipot+23}. The existence of HD 140901 b is controversial; if the planet exists, it has a mass of \mpsini\ of $\sim$16$M_{\oplus}$ with an orbital period of $\sim$9 days (orbital distance 0.085 AU), but the RV data of HD 140901 can be equally well described by a retrograde orbital inclination of $\sim$138$^{\circ}$ \citep{Feng+22,Philipot+23}. 

HD 140901 was observed once by \Chandra/ACIS-I for 24.7 ks. The star was robustly detected with $\sim$220 net counts. The light curve shows clear evidence for variability (see Figure Set~\ref{fig:LightCurves-FigureSet}), but the star is not bright enough to enable spectral modeling.
\subsection{GJ 777}
GJ 777 is a binary star system composed of a G6IV primary (GJ 777A) and an M4.5V secondary (LHS 2509), which are separated by $\sim$3000 AU ($\sim$3$^{\prime}$).
GJ 777A hosts two currently known exoplanets, detected via RV measurements: HD 190360 b has \mpsini\ $\sim$1.8 $M_{\rm Jup}$ and an orbital period of $\sim$2854 days \citep[$\sim$3.9 AU;][]{Feng+21}, and HD 190360 c has \mpsini\ $\sim$19 $M_{\oplus}$ and an orbital period of $\sim$17 days \citep[$\sim$0.13 AU;][]{Rosenthal+21}. No X-ray emission is detected coincident with the GJ 777 binary system in either of two snapshot \XMM\ observations.
\subsection{$\xi$ Boo}
$\xi$ Boo is a visual binary \citep[$\xi$ Boo A has a spectral type G8V and $\xi$ Boo B has a spectral type K4V;][]{Turnbull15} with an angular separation of $\sim$4.9$^{\prime\prime}$ \citep[orbital period $\sim$150 years;][]{Wielen62}, the stars are unresolvable by \XMM\ (see Figure~\ref{fig:Xray_imaging_binaries}). No discrepancies in stellar physical parameters in the literature. No known or candidate exoplanets. $\xi$ Boo A is a known BY Draconis variable \citep{Finley+19}.

The $\xi$ Boo binary is robustly detected by \XMM, although the two components cannot be resolved. The system exhibits dramatic X-ray variability, as shown in the light curve in Figure Set~\ref{fig:LightCurves-FigureSet}. The X-ray spectra of $\xi$ Boo are described by three-component thermal plasma models in all variability states. The flaring spectrum requires hotter temperatures for all three plasma components than the quiescent spectrum, with the normalization of the hottest component increasing by a factor of $\sim$3.9 during flaring times compared to quiescent times. Figure Set~\ref{fig:Spectra-FigureSet} shows the best fit spectral models for $\xi$ Boo (see also Table~\ref{table:spectral_modeling}).
\subsection{$\delta$ Pav}
There are no significant discrepancies in the stellar physical parameters reported in the literature for $\delta$ Pav. The star does not host any currently known or candidate exoplanets.

$\delta$ Pav was detected in a 32.6 ks \XMM\ observation (ID number 0780510401) $\sim$1000 net counts. The $A^2$ statistic provides evidence that the star may be variable on short timescales, however there are not a sufficient number of counts to split the observation into sub-exposures (see Figure Set~\ref{fig:LightCurves-FigureSet}). We attempted to model the spectrum of $\delta$ Pav despite the low number of counts; the spectrum required two thermal plasma components to achieve a statistically acceptable fit (see Table~\ref{table:spectral_modeling}). The best-fit spectrum in shown in Figure Set~\ref{fig:Spectra-FigureSet}. 

\subsection{LHS 2156}
There are no significant discrepancies in the stellar physical parameters reported in the literature for LHS 2156. There are no known or candidate exoplanets in the system.

LHS 2156 was detected with $\sim$60,000 net counts in a 68.4 ks \XMM\ observation with the PN camera (thick filter). Figure Set~\ref{fig:LightCurves-FigureSet} shows the X-ray light curve, which exhibits count rate variability over the duration of the observation. The spectra of LHS 2156 are well described by three-component thermal plasma models. The best-fit temperatures and normalizations of two cooler plasma components remain constant within the uncertainties in all sub-exposures. The hottest thermal plasma component increases in temperature by $\sim$0.1 keV during flaring periods compared to quiescent periods, an the normalization increases by $\sim$25\%. The best-fit spectra for LHS 2156 are shown in Figure Set~\ref{fig:Spectra-FigureSet} (see also Table~\ref{table:spectral_modeling}).
\subsection{82 Eri}
There is some discrepancy in the literature about the age of 82 Eri: \citet{Turnbull15} and \citet{Takeda+07} quote ages $>$12 Gyr, but a much younger age of 5.76$\pm$0.66 Gyr was found by \citep{Pepe+11}. 82 Eri hosts four confirmed exoplanets, all of which were discovered via RV measurements \citep{Pepe+11}. HD 20794 b, c, d, and e have \mpsini\ values of $\sim$2.8 \Mearth, $\sim$2.5 \Mearth, $\sim$3.5 \Mearth, and $\sim$4.8 \Mearth, respectively, with orbital periods (distances) of $\sim$18 days ($\sim$0.13 AU), $\sim$43 days ($\sim$0.23 AU), $\sim$89 days ($\sim$0.26 AU), and $\sim$147 days ($\sim$0.5 AU), respectively. All planetary parameters were taken from \citet{Feng+17}. 82 Eri has been observed by both \XMM\ and \Chandra\ (see Table~\ref{tab:observation_log}), but was not detected in either observation. 
\subsection{$\tau$ Ceti}
There are no significant discrepancies in the stellar physical parameters of $\tau$ Ceti reported in the literature. $\tau$ Ceti hosts four exoplanets, all confirmed via radial velocity measurements \citep{Feng+17tauCet}, with some evidence of additional planets in or near the $\tau$ Cet HZ \citep{Dietrich+21}. $\tau$ Ceti e and f both have masses \mpsini\ $\sim$3.9 $M_{\oplus}$ with orbital periods of $\sim$163 days ($\sim$0.55 AU) and $\sim$636 days ($\sim$1.3 AU), respectively, while $\tau$ Ceti g and h both have masses of \mpsini\ $\sim$1.8 $M_{\oplus}$ and orbital periods of $\sim$20 days ($\sim$0.13 AU) and $\sim$49 days ($\sim$0.24 AU), respectively.

$\tau$ Ceti was observed by \Chandra\ ACIS-S for 45.1 ks in 1/8 subarray mode and detected at $\sim$7$\sigma$ significance with $\sim$110 net counts. We extracted a light curve from this observation and find an $A^2$ statistic that is strongly suggestive of variability (see Table~\ref{tab:variability_metrics} and Figure Set~\ref{fig:LightCurves-FigureSet}), but due to the relatively low number of counts we did not attempt to extract a spectrum. $\tau$ Ceti was marginally detected in an 11.9 ks \XMM\ observation, but with insufficient counts for light curve extraction or spectral modeling to be performed. The luminosity estimates for $\tau$ Ceti suggest that the star's X-ray luminosity changes by a factor of $\sim$2.
\subsection{55 Cnc}
55 Cnc is a binary star system containing a G8V star (55 Cnc A) and a M4.5V dwarf (55 Cnc B). No significant discrepancies in stellar physical parameters. The two components are separated by $\sim$1,065 AU \citep[$\sim$1.4$^{\prime}$;][]{Raghavan+06}. Five exoplanets are known to orbit 55 Cnc A. 55 Cnc b was the first to be discovered \citep{Butler+97} with a mass of \mpsini\ $\sim$0.8 $M_{\rm Jup}$ and an orbital period $\sim$14.7 days \citep[corresponding to an orbital distance of $\sim$0.11 AU;][]{Bourrier+18,Nelson+14}. 55 Cnc c, d, e, and f has masses of \mpsini\ $\sim$0.16 $M_{\rm Jup}$, $\sim$3.9 $M_{\rm Jup}$, $\sim$8 $M_{\oplus}$, and $\sim$0.15 $M_{\rm Jup}$, respectively, with orbital periods of $\sim$44 days ($\sim$0.24 AU), $\sim$4870 days ($\sim$5.5 AU), $\sim$0.7 days ($\sim$0.02 AU), and $\sim$260 days ($\sim$0.8 AU), respectively.

Cnc 55 was observed twice by \Chandra/ACIS-S in 1/8-subarry mode (for 10.7 ks and 18 ks) and once for 10.4 ks by \XMM. 55 Cnc A was not detected by \Chandra, but was detected (at $\sim$5.6$\sigma$) significance by \XMM. No evidence for strong variability was observed in the \XMM\ light curve of 55 Cnc A (Table~\ref{tab:variability_metrics}, see also Figure Set~\ref{fig:LightCurves-FigureSet}). Despite the relatively low number of net counts, we attempted to model the X-ray spectrum of 55 Cnc A. We find that a single-component thermal plasma model is sufficient to describe the X-ray spectrum (see Table~\ref{table:spectral_modeling} and Figure Set~\ref{fig:Spectra-FigureSet}). 55 Cnc B was not detected in the \XMM\ observation or one of the \Chandra\ observations (the star was not within the field of view of the second \Chandra\ observation).
\subsection{70 Oph}
70 Oph is a visual and spectroscopic binary containing a K0V (70 Oph A) star and a K5V (70 Oph B) star with an orbital period of $\sim$88 years \citep{Pourbaix00}. The orbit is highly elliptical, with an eccentricity $e=0.50$ that causes the orbital separation of the stars to vary from $\sim$11.4 AU to $\sim$34.8 AU \citep{Eggenberger+08}. These orbital separations correspond to angular separations of $\sim$2.3-6.8$^{\prime\prime}$, unresolvable by \XMM\ but potentially spatially resolvable by \Chandra\ (see Section~\ref{sec:data}). \citet{Eggenberger+08} find a best-fit age of 6.2$\pm$1.0 Gyr, in broad agreement with the 6.8-7.0 Gyr found by \citet{Tang+08}. These ages (which where derived using astroseismic observations) are in contrast to the significantly younger age ($\sim$680 Myr) predicted by \citet{Mamajek+08}. There are no currently known or candidate exoplanets in the system.

There are three \XMM\ observations available for the 70 Oph system. The individual stars are not resolvable in the \XMM\ observations (see Figure~\ref{fig:Xray_imaging_binaries}); we therefore cannot separate the X-ray emission from the K0V star (70 Oph A) from the K5V star (70 Oph B) with the currently available observations. 

Despite the changes in X-ray count rate observed in the light curves (see Figure Set~\ref{fig:LightCurves-FigureSet}), the best-fit spectral models of 70 Oph across different VarFlags are nearly indistinguishable (Figure Set~\ref{fig:Spectra-FigureSet}; see also Table~\ref{table:spectral_modeling}), and the X-ray luminosity of the two stars changes only by $\sim$40\% between the flaring and quiescent periods. Since we cannot separate the X-ray emission of the binary components individually, we do not use the best-fit spectrum of the 70 Oph system as template for inferring luminosities of other stars in our sample.
\subsection{40 Eri}
The 40 Eri triple system contains a K0V star (40 Eri A) that is either in a wide orbit or co-moving with a white dwarf (40 Eri B)-M4.5V dwarf (40 Eri C) pair. The projected angular distance between 40 Eri A and 40 Eri BC is $\sim$1.3$^{\prime}$. While the orbital period of the wider A-BC pair is very long, the orbital period of the B-C pair is measured to be $\sim$230 years \citep{Mason+17,Bond+17}. The angular separation between 40 Eri B and 40 Eri C is $\sim$6.9$^{\prime\prime}$, easily resolvable by \Chandra\ (see Figure~\ref{fig:XMM_vs_Chandra_imaging}). 40 Eri C has a visual magnitude of $\sim$11, which we use to estimate \Lbol, and \citet{Johnson+83} estimate \Teff\ to be $\sim$3,100 K. There is one confirmed planet orbiting 40 Eri A, which was discovered via RV measurements. HD 26965 b has \mpsini\ of $\sim$8.5 \Mearth\ with an orbital period of $\sim$42 days \citep{Ma+18}. No currently known or candidate planets are associated with the 40 Eri BC pair.

The first soft (0.1-2.0 keV) X-ray detection of the 40 Eri system was obtained by HEAO-1, which measured a luminosity of (9.6$\pm$3.2)$\times10^{28}$ \lum\ \citep[corrected for different distance assumed in that work;][]{Cash+79}. Due to the poor spatial resolution of HEAO-1, the binary components could not be resolved. Different scenarios for X-ray production are explored in \citet{Cash+79}, and the X-ray emission was attributed to 40 Eri C. 

The \Chandra\ image of the system clearly shows X-ray sources coincident with both 40 Eri A and C (although not the white dwarf, 40 Eri B; see discussion below). 40 Eri C is the brighter of the two sources, with roughly an order of magnitude more net counts than observed for 40 Eri A. There is no evidence for strong variability in either star (see Figure Set~\ref{fig:LightCurves-FigureSet}). While the coolest thermal plasma component in the 40 Eri C spectrum is similar in temperature to other late-type stars in our sample, the hotter two thermal plasmas are considerably hotter (see Table~\ref{table:spectral_modeling} and Figure Set~\ref{fig:Spectra-FigureSet}). Despite the lack of obvious variability in 40 Eri C, the best-fit X-ray spectral parameters most closely resemble those of WX UMa in its flaring and post-flaring states. It is possible this star indeed exhibits significant coronal variability, but it was not detectable in the short snapshot \Chandra\ observation available in the archive.

\subsubsection{The White Dwarf 40 Eri B}
The white dwarf 40 Eri B is not detected in the 5 ks \Chandra/ACIS-S observation of the system. The 3$\sigma$ upper limit on the count rate is 0.0033 ct s$^{-1}$. We again use WebPIMMs to convert the observed count rate upper limit to a flux (and luminosity) upper limit, assuming the same spectral model as for WD 1620-391 (Section~\ref{sec:WD1620-391}). We find a flux upper limit of 1.6$\times10^{-13}$ \flux, which corresponds to a luminosity of $\sim$5$\times10^{25}$ \lum. This luminosity upper limit is below the measured X-ray luminosity of G 29-38 \citep[$\sim$8$\times10^{25}$ \lum;][]{Cunningham+22,Estrada-Dorado+23}, strongly suggesting a lack of a cold accretion disk around this white dwarf.

\subsection{$\delta$ Eri}
There are no significant discrepancies in the reported physical parameters for $\delta$ Eri, and the star does not host any known or candidate exoplanets. $\delta$ Eri was observed once with \XMM\ (observation observation) for 56.2 ks. There was marginal evidence for significant variability in the X-ray light curve of $\delta$ Eri (see Figure Set~\ref{fig:LightCurves-FigureSet}), but the star was too faint for spectral modeling.
\subsection{GL 451A (Groombridge 1830)}
There are no significant discrepancies in the stellar physical parameters reported in the literature for GL 451A. The system does not host any currently known or candidate exoplanets. GL 451A was observed once with \Chandra/ACIS-S (observation ID 9931) for 32.8 ks in 1/8-subarray mode across ACIS detectors 5-7 with a readout time of 0.54 s. No X-ray source is found coincident with the star's position.
\subsection{GL 117}
\citet{Turnbull15} report an age of 1.5 Gyr for GL 117 and \citet{Takeda+07} derive an upper limit of $<$1.2 Gyr. These ages are in contrast with the very young age ($\sim$100 Myr) predicted by \citet{Mamajek+08}. There are no currently known or candidate exoplanets orbiting GL 117.

GL 117 is robustly detected in \XMM\ observation 0203060501, and there are low-level count rate variations observable in the light curve (see Figure Set~\ref{fig:LightCurves-FigureSet}). The largest of these variations reaches a count rate level that is $\sim$30\% above the median count rate. Given the ample number of X-ray counts recorded by \XMM, we divide the light curve into sub-exposures but we find no significant differences in the best fit spectral parameters nor the predicted luminosities (see Figure Set~\ref{fig:Spectra-FigureSet} and Table~\ref{table:spectral_modeling}). The bright X-ray luminosity of the star, particularly when compared to other K dwarfs in our sample, are consistent with a young age for GL 117.
\subsection{GL 783}
GL 783 is a binary system containing a K3V star (A) and an M4.5V star \citep[B;][]{Turnbull+03}. The two stars were stated to be separated by $\sim$8$^{\prime\prime}$ by \citet{Allen+00} but Gaia DR3 positions indicate a separation of $\sim$2.6$^{\prime\prime}$ \citep{GaiaDR3}, both of which are unresolvable by \XMM\ but potentially spatially resolvable by \Chandra\ (see Section~\ref{sec:data}). GL 783B has an apparent visual magnitude of $\sim$12.5 \citep{Allen+00}, which we use to estimate \Lbol. At the time of this writing there were no published values of \Teff\ for GL 783B. The system does not host any currently known or candidate exoplanets.

There is one \XMM\ observation available for the GL 783 system, and the individual stars are not resolvable (see Figure~\ref{fig:Xray_imaging_binaries}). The therefore cannot separate the X-ray emission from the K3V star (GL 783A) from the M4.5V star (GL 783B) with the currently available observations. The $A^2$ statistic provides evidence that one or both stars may be variable on short timescales (see Figure Set~\ref{fig:LightCurves-FigureSet}), however there are not a sufficient number of counts to split the observation into sub-exposures. We attempted to model the spectrum of the GL 783 system despite the low number of counts; the spectrum required three thermal plasma components to achieve a statistically acceptable fit (see Figure Set~\ref{fig:Spectra-FigureSet} and Table~\ref{table:spectral_modeling}). The coolest thermal plasma component ($\sim$0.1 keV) is similar to the coolest components of the other FGK stars in our sample, while the temperatures of the hottest ($\sim$1 keV) and intermediate ($\sim$0.3 keV) components more closely resemble components found in the later-K and M dwarfs in our sample. 

Since we cannot separate the X-ray emission of the binary components individually, we do not use the best-fit spectrum of the GL 783 system as template for inferring luminosities of other stars in our sample. Follow-up observations with higher spatial resolution are needed to associate the observed X-ray emission with one or both binary components.

\subsection{GL 183}
There is some discrepancy in the reported spectral type of GL 183: \citet{Gray+06} report a spectral type K3V, while K4III is reported by \citet{Turnbull15}. We adopt a spectral type K3V. There is additional tension in the age estimates, with a younger age ($\sim$2 Gyr) reported by \citet{Mamajek+08} and older ages reported by \citet[][5.3 Gyr]{Turnbull15} and \citet[][$<$5.45 Gyr]{Takeda+07}. There are no currently known or candidate exoplanets orbiting GL 183.

GL 183 was detected in a 18.4 ks \XMM\ observation (ID number 0780510301) $\sim$650 net counts. The $A^2$ statistic provides evidence that the star may be variable on short timescales (see Figure Set~\ref{fig:LightCurves-FigureSet}), however there are not a sufficient number of counts to split the observation into sub-exposures. We attempted to model the spectrum of GL 183 despite the low number of counts and found the spectrum required two thermal plasma components to achieve a statistically acceptable fit (Figure Set~\ref{fig:Spectra-FigureSet}; see also Table~\ref{table:spectral_modeling}), and the fit parameters were reasonably well-constrained.
\subsection{GJ 667}
The GJ 667 is a triple star system wherein the two more massive components, A and B (both of which are early/mid K dwarfs) orbit one another in a highly elliptical orbit ($e\approx0.58$). The average angular separation of the two stars is $\sim$1.8$^{\prime\prime}$, possibly resolvable by \Chandra. The third star in the system, the M1.5V dwarf GJ 667 C, is separated from the AB pair by $\sim$0.5$^{\prime}$ (which would be easily resolvable with both \Chandra\ and \XMM). There is no published age estimate for the system. GJ 667C has been studied previously in X-rays by \citet{Brown+23}, and is known to host two confirmed super-Earth exoplanets: GJ 667C b, with \mpsini\ $\sim$5.6 \Mearth\ and an orbital period of $\sim$7.2 days \citep[$\sim$0.05 AU;][]{Robertson+14,Bonfils+13}, and GJ 667C c, with with \mpsini\ $\sim$3.8 \Mearth\ and an orbital period of $\sim$28 days \citep[$\sim$0.125 AU;][]{Anglada-Escude+13}. An additional three planets in wider orbits were reported by \citet{Anglada-Escude+13} but were unable to be confirmed by \citet{Robertson+14}; their existence remains controversial.

GJ 667 was observed twice with \Chandra/ACIS-S in 1/8-subarray mode. GJ 667 A and B were not within the field of view of either observation. GJ 667 C was detected in both observations. We extracted an X-ray light curve from \Chandra\ ObsID 17317 (with an exposure time of 18.2 ks, see Figure Set~\ref{fig:LightCurves-FigureSet}), but the star was too faint for spectral modeling to be performed. There were insufficient counts in the second \Chandra\ observation (ObsID 17318, with an exposure time of 9.1 ks) for light curve or spectral analysis.
\subsection{LHS 1875}
There is some discrepancy in the literature about the age of LHS 1875: \citet{Turnbull15} quote an age of $\sim$3.1 Gyr, while \citet{Takeda+07} provide a significantly younger upper limit of $<$0.48 Gyr. There are no known or candidate exoplanets orbiting LHS 1875. LHS 1875 was detected with $\sim$9260 net counts in a 21.4 ks \XMM/PN observation. No evidence for significant X-ray variability was observed (see Figure Set~\ref{fig:LightCurves-FigureSet}), and the spectrum was well-described by a three-component thermal plasma model (see Table~\ref{table:spectral_modeling}).
\subsection{GL 570}
There is a significant age discrepancy in the literature for the GL 570 quadruple-star system: \citet{Takeda+07} report a very young age of $<$0.6 Gyr, while \citet{Turnbull15} adopt an ade of $\sim$3 Gyr. There are no other significant discrepancies in the stellar physical parameters for the stars in this system. There are no currently known or candidate exoplanets in the system.

GL 570A (K4V) orbits a binary red dwarf pair (GL 570B, an M1.5V dwarf, and GL 570C, an M3V dwarf) with a semi-major axis of $\sim$190 AU (corresponding to an angular separation of $\sim$0.5$^{\prime}$). GL 570A and the GL 570BC pair are both detected in X-rays and are easily resolvable with \Chandra\ (see Figure~\ref{fig:Xray_imaging_binaries}). The red dwarfs the make up the BC pair are separated by $\sim$0.2$^{\prime\prime}$ \citep{Mariotti+90,Forveille+99} and cannot be resolved. The fourth component, GL 570D, is a brown dwarf in a very wide orbit ($\sim$4$^{\prime}$) from the GL 570 ABC triple system \citep{Burgasser+00}. GL 570D is not detected in X-rays.

We extract X-ray light curves for both GL 570A and the GL 570BC pair. There is clear evidence for variability in both light curves (see Figure Set~\ref{fig:LightCurves-FigureSet}), although there are not enough net counts from GL 570BC to enable spectroscopic modeling on sub-exposures. We therefore divide observation of GL 570A into three distinct sub-exposures (the first $\sim$20 ks, 20-30 ks, and 30-40 ks) and independently model the spectra, but only extract one spectrum of GL 570BC from the entire observation. All spectra are well described by a two-component thermal plasma model. The flaring event seen from 20-30 ks in the GL 570A light curve is characterized by changes in plasma temperatures ($\sim$0.1 keV for the cooler component and $\sim$0.2 keV for the hotter component) and an increase in the hot-component normalization by $\sim$170\%. The different best-fit spectral models for GL 570A, as well as a comparison of the flaring-to-quiescent spectra, are shown in Figure Set~\ref{fig:Spectra-FigureSet}. The GL 570A model is overall hotter than the GL 570BC model (see Table~\ref{table:spectral_modeling}).

\subsection{61 Cyg}
The 61 Cygni AB system is a wide binary containing a K5V star (A) and a K7V star (B), with an angular separation of $\sim$12$^{\prime\prime}$ (the orbital period of the system is $\sim$659 years). Most age estimates suggest the system is old \citep[6$\pm$1 Gyr; e.g.,][]{Kervella+08}. The system is resolved by \XMM\ \citep{Robrade+12}. The system does not host any currently known or candidate exoplanets.

61 Cyg has been observed 38 times with \XMM, making it the best X-ray-observed star system in our sample. The observations are spaced $\sim$6 months apart and span $\sim$18 years, and the two stars are well-resolved in all available images (see Figure~\ref{fig:Xray_imaging_binaries}). Both stars are observed to flare (see Figure Set~\ref{fig:LightCurves-FigureSet}), and all the spectra extracted from both stars are well describe by a three-component thermal plasma model. The three best-fit temperatures in the models of 61 Cyg A remain constant within the uncertainties during different variability epochs. The changes in count rate observed during flaring events in 61 Cyg A are driven by changes in the normalizations of the two hotter thermal plasma components (and hence the VEMs): the hottest component normalization increases by an order of magnitude, and the intermediate temperature component normalization increases by a factor of $\sim$2.5. The spectra of 61 Cyg B show similar changes in normalization during flaring events, but the two {\it cooler} plasma components also increase in temperature during flaring events (by $\sim$0.1 keV). The best-fit spectra for 61 Cyg A and B are shown in Figure Set~\ref{fig:Spectra-FigureSet} (see also Table~\ref{table:spectral_modeling}). 
\subsection{GL 412}
The GL 412 AB system contains a M1V primary (A) and a UV Ceti-type flaring M6.6V secondary \citep[B;][also referred to as WX UMa]{Mann+15}. X-ray flares were previously observed from the system with ROSAT \citep{Schmitt+95} and attributed to GL 412B; GL 412A was not believed to be a significant source of X-rays. The system does not host any currently known or candidate exoplanets.

The two stars have an angular separation of $\sim$31$^{\prime\prime}$ and are resolved in the \XMM\ image (observation number 0742230101; see Figure~\ref{fig:XMM_vs_Chandra_imaging}). The image clearly suggests that the observation caught a significant WX UMa flaring event and that the M1V primary, while fainter, emits a detectable quantity of X-ray radiation. A third X-ray source is detected $\sim$37$^{\prime\prime}$ from WX UMa, but is not obviously associated with the GL 412 binary. We measure a count rate of 0.057$\pm$0.002 ct s$^{-1}$ in the PN image (medium filter) for GL 412A. 

The light curve of GL 412B is shown in Figure Set~\ref{fig:LightCurves-FigureSet}. Spectral modeling shows that the X-ray luminosity increases by a factor of $\sim$30 between quiescent and flaring times. The quiescent spectrum (the first $\sim$2 ks of the observation) is well-described with only two thermal plasma components. The steep, $\sim$0.5 ks flare, the elevated count rate (from $\sim$3-7 ks), and subsequent declining period all require three components to adequately describe the X-ray spectra. The flaring spectrum exhibits a very hot thermal plasma component, with a temperature of $\sim$3.5 keV ($\sim$40 MK), which cools as the flare subsides (it has a temperature of $\sim$2.2 keV during the ``elevated'' period and $\sim$1.4 keV during the ``descending'' count rate period; see Figure Set~\ref{fig:Spectra-FigureSet} and Table~\ref{table:spectral_modeling}).
\subsection{GJ 832}
No significant discrepancies in stellar physical parameters. One confirmed exoplanet, GJ 832 b with \mpsini\ $\sim$0.6 $M_{\rm Jup}$ and an orbital period of 3853 days \citep[corresponding to an orbital semi-major axis of $\sim$3.7 AU,][]{Bailey+09,Philipot+23}.

GJ 832 was observed twice by \XMM, once during a 8.9 ks (observation 0748010201) exposure during which the star was detected but too faint to enable timing or spectroscopic modeling, and once during a 28.6 ks exposure (observation 0860303301) when the star was overall brighter and underwent a short flaring event (see Figure Set~\ref{fig:LightCurves-FigureSet}). The quiescent spectrum is well-described by two thermal plasma components, while the lower signal-to-noise during the short flaring period allows the flare spectrum to be modeled with a single thermal plasma component. Overall, the luminosity of GJ 832 increases by a factor of $\sim$2.5 during the flare. The best-fit spectral models and a comparison of the flaring-to-quiescent spectrum are shown in Figure Set~\ref{fig:Spectra-FigureSet} (see also Table~\ref{table:spectral_modeling}). The spectrum extracted during the short flaring time during the observation shows an excess of X-ray flux at $\sim$2-30 \AA, but shows weaker X-ray emission (compared to the quiescent spectrum) at wavelengths $\lesssim$2 \AA.

\subsection{Kapteyn's Star}
There are no significant discrepancies in the stellar physical parameters reported in the literature for Kapteyn's Star. Although the detection of sub-Neptune mass planet was reported by \citet{Anglada-Escude+14}, this was later refuted by \citet{Bortle+21} who argued that the observed RV variations were artifacts of stellar activity and rotation. There are not currently any additional candidate exoplanets around Kapteyn's Star.

Kapteyn's Star has been observed numerous times by \XMM, \Chandra, and \Swift\ (see Table~\ref{tab:observation_log}), but is not detected in any of archival X-ray observations. We explored whether Kapteyn's Star would be detected in a stacked \Chandra\ X-ray image (the \Chandra\ background is lower and significantly less affected by spurious flaring events than \XMM). The stacked \Chandra\ image has an effective exposure time of 72.5 ks. Kapteyn's Star is still not detected in the stacked image, with a count rate upper limit of $<$1.24$\times10^{-4}$ ct s$^{-1}$. Assuming the best fit quiescent spectrum of Wolf 1055, this count rate corresponds to a luminosity upper limit of $<$1.1$\times10^{24}$ \lum.
\subsection{Wolf 1055}
There are no significant discrepancies in the stellar physical parameters of Wolf 1055 or its binary companion VB 10 \citep[GL 572B, a M8V-type BY Draconis flare star][]{Burt+21}. There is one confirmed exoplanet orbiting Wolf 1055: HD 180617 b (also identified as GL 572 Ab), with \mpsini\ $\sim$12 $M_{\oplus}$ and an orbital period of $\sim$106 days ($\sim$0.34 AU). 

Wolf 1055 and VB 10 are separated by $\sim$1.2$^{\prime}$ and are easily resolvable with both \XMM\ and \Chandra. Both Wolf 1055 was observed once with \XMM\ (for 24.2 ks) and twice with \Chandra\ (with exposure times of 12.2 ks and 29.2 ks). Both Wolf 1055 and VB 10 are detected in all three observations. A comparison of the \XMM\ observation and the longer \Chandra\ observation (ObsID 7428) is shown in Figure~\ref{fig:Xray_imaging_binaries}. Wolf 1055 is bright enough in two out of these three observations to enable spectral modeling. Although VB 10 is bright enough to be significantly detected in all three observations and we are able to extract light curves from all three observations. The light curves of both stars obtained during \XMM\ observation 0504010101 are shown in Figure Set~\ref{fig:LightCurves-FigureSet}.

The quiescent spectrum of Wolf 1055 is well-described by a three-component thermal plasma model, while the lower signal-to-noise during the flaring period in the second half of the \XMM\ observation 0504010101 allows the flare spectrum to be modeled with a single thermal plasma component. Overall, the luminosity of Wolf 1055 increases by $\sim$80\% during the flare. The best-fit spectral models and a comparison of the flaring-to-quiescent spectrum are shown in Figure Set~\ref{fig:Spectra-FigureSet} (see also Table~\ref{table:spectral_modeling}). The spectrum extracted during the flaring period during the second half of observation shows an excess of X-ray flux at $\sim$2 \AA, and shows weaker X-ray emission (compared to the quiescent spectrum) at wavelengths $\lesssim$2 \AA.
\subsection{AD Leo}
There are no significant discrepancies in the stellar physical parameters reported for AD Leo; all stellar parameters are taken from \citet{Kossakowski+22} and references therein. There are no confirmed or candidate exoplanets orbiting AD Leo.

AD Leo is a known flaring star, and exhibits strong X-ray flaring both times the star was observed by \XMM. Both light curves are shown in Figure Set~\ref{fig:LightCurves-FigureSet}. High count rates enable spectroscopic modeling over multiple sub-exposures. The X-ray spectra during quiescent periods are generally well-described with a two-component thermal plasma model, while periods of flaring and enhanced variability require three thermal plasma components to adequately describe the observed spectra. Increases in X-ray count rate are driven almost entirely by the appearance of this third, hot ($\sim$1.5 keV) thermal plasma component, which drives a significant increase (by three- to five-orders of magnitude) in X-ray emission below $\sim$2 \AA\ (see Figure Set~\ref{fig:Spectra-FigureSet}).
\subsection{Wolf 1061}
There are no significant discrepancies in stellar physical parameters of Wolf 1061. No age estimate is available for the star. Wolf 1061 hosts three currently known exoplanets \citep{Wright+16}: Wolf 1061 b, c, and d have masses of \mpsini\ $\sim$1.9 \Mearth, $\sim$3.4 \Mearth, and $\sim$7.7 \Mearth, respectively, with orbital periods of $\sim$4.9 days ($\sim$0.04 AU), $\sim$17.9 days ($\sim$0.09 AU), and $\sim$217 days ($\sim$0.47 AU), respectively. All planetary data taken from \citet{Astudillo-Defru+17}.

Wolf 1061 was detected in one 38.3 ks observation with Chandra/ACIS-S with $\sim$240 net counts. The light curve shows evidence for variability (see Figure Set~\ref{fig:LightCurves-FigureSet}), but the star is not bright enough to enable spectroscopic modeling.
\subsection{Luyten's Star}
There are no significant discrepancies in the reported stellar parameters for Luyten's Star in the literature. No age estimate is available. There are two confirmed exoplanets orbiting Luyten's Star: GJ 273b and GJ 273c, which were detected via RV measurements by \citet{Astudillo-Defru+17}. GJ 273b and GJ 273c have masses of $\sim$2.9 $M_{\oplus}$ and $\sim$1.2 $M_{\oplus}$, respectively, and the planets have orbial periods of $\sim$18.6 days ($\sim$0.09 AU) and $\sim$4.7 days ($\sim$0.03 AU), respectively.

Luyten's star is detected at $\sim$5$\sigma$ significance in a 28.4 ks observation with \Chandra/ACIS-S. The light curve shows evidence of X-ray variability (see Figure Set~\ref{fig:LightCurves-FigureSet}), but are are unable to extract X-ray spectra due to the low number of net counts.
\subsection{Proxima Centauri}\label{sec:ProxCen}
Proxima Centauri is a member of an extensively studied triple system along with $\alpha$ Cen A and $\alpha$ Cen B. There is currently one confirmed Earth-like exoplanet, Proxima Centauri b, which has \mpsini\ $\sim$1 \Mearth\ and an orbital period of $\sim$11 days \citep[orbital distance $\sim$0.05 AU;][]{Faria+22,Damasso+20,Anglada-Escude+16}. Two candidate exoplanets, Proxima Centauri c and d, are awaiting follow-up observations. If present, Prox Cen c \citep{Damasso+20} and d \citep{Faria+22} are estimated to have \mpsini\ values of $\sim$6 \Mearth\ and $\sim$0.26 \Mearth, respectively, with orbital periods (distances) of $\sim$1900 days ($\sim$1.5 AU) and $\sim$5 days ($\sim$0.029 AU).

Proxima Centauri exhibits significant flaring events, such as those illustrated in Figure~\ref{fig:ProxCen_lc} (see also Figure Set~\ref{fig:LightCurves-FigureSet}). The X-ray spectrum of Prox Cen is well-described by a two-component thermal plasma model (with temperatures of $\sim$0.25 keV and $\sim$0.85 keV) during quiescent times and periods of elevated or rising count rates. A third thermal plasma component is required to describe the spectra during flaring periods and during periods of descending count rates following a flare. During the flare, this third component is hot ($\sim$1.5 keV), and cools to $\sim$0.6 keV as the count rate decreases (see Figure Set~\ref{fig:Spectra-FigureSet} and Table~\ref{table:spectral_modeling}). The flaring spectrum drives dramatic increases in both X-ray luminosity. Although the flaring spectrum of Prox Cen exhibits a significant increase in the X-ray flux at wavelengths $<$10 \AA, the spectrum remains elevated (by an order of magnitude) compared to the quiescent spectrum out to long wavelengths (see Figure~\ref{fig:FQ_extend}). This is in contrast to earlier-type stars, which generally show a similar X-ray emission between quiescent and flaring periods at wavelengths $>$100 \AA.

\centerwidetable
\begin{deluxetable*}{cccccccccc}
\tablecaption{Stars with Available \Swift/XRT Imaging}
\label{tab:Swift_stars}
    \tablehead{
    \colhead{Star} & \colhead{Alternate} & \colhead{R.A.} & \colhead{Dec.} & \colhead{Spectral} & \colhead{Distance} & \colhead{Mass} & \colhead{$T_{\rm eff}$} & \colhead{\Lbol} & \colhead{log\Lx} \\ \cline{3-4}
    Name & Name(s)   &  \multicolumn{2}{c}{(J2000)}     &  Type$^*$   & (pc)    & (\Msun) & (K)      & (\Lsun) & ([\lum]) \\
    }
    \startdata
    71 Ori & HIP 29650 & 06:14:50.77 & +19:09:20.29 & F6V & 20.87 & 1.35 & 6533 & 2.94 & $<$27.59 \\
    $\gamma$ Lep A & GL 216A & 05:44:27.45 & -22:27:00.08 & F6.5V & 8.88 & 1.22 & 6258 & 2.34 & 27.53 \\
    $\phi^2$ Cet & GL 37 & 00:50:07.34 & -10:38:43.26 & F7V & 15.77 & 1.17 & 6250 & 1.85 & $<$26.94 \\
    $\eta$ CrB & GL 584A & 15:23:12.23 & +30:17:17.7 & G2V & 18.82 & 1.14 & 6029 & 1.64 & $<$27.75 \\
    61 Vir & GL 506 & 13:18:24.97 & -18:18:31.0 & G7V & 8.50 & 0.91 & 5585 & 0.84 & $<$26.78 \\
    GL 567 & LHS 5279 & 14:53:23.27 & +19:09:13.54 & K0V & 11.54 & 0.93 & 5258 & 0.54 & 28.28 \\
    GL 68 & LHS 1287 & 01:42:29.41 & +20:15:55.87 & K1V & 7.60 & 0.88 & 5190 & 0.45 & $<$26.29 \\
    $\mu$ Cas & GL 53A & 01:08:22.74 & +54:54:47.53 & K1V & 7.55 & 0.82 & 5316 & 0.46 & $<$26.77 \\
    GL 785 & LHS 488 & 20:15:18.88 & -27:02:01.61 & K2V & 8.79 & 0.85 & 5071 & 0.4 & $<$26.50 \\
    AK Lep & GL 216B & 05:44:26.19 & -22:25:24.26 & K2V & 8.89 & 0.80 & 4869 & 0.3 & 28.00 \\
    GL 688 & HD 160346 & 17:39:16.72 & +03:33:17.32 & K2V & 10.71 & 0.76 & 4982 & 0.34 & $<$26.47 \\
    Lacaille 8760 & AX Mic & 21:17:10.80 & -38:52:20.84 & K9V & 3.97 & 0.56 & 3599 & 0.1 & 26.89 \\
    Wolf 1453 & GL 205 & 05:31:28.21 & -03:41:11.50 & M1V & 5.70 & 0.56 & 3690 & 0.07 & 27.53 \\
    Lacaille 9352 & GL 887 & 23:06:00.94 & -35:50:49.79 & M1V & 3.29 & 0.54 & 3676 & 0.04 & 26.71 \\
    BR Pis & GL 908 & 23:49:13.59 & +02:23:48.90 & M1V & 5.90 & 0.41 & 3685 & 0.029 & $<$26.80 \\
    GL 229A\tablenotemark{b} & LHS 1827 & 06:10:34.46 & -21:52:04.16 & M1.5V & 5.76 & 0.54 & 3912 & 0.06 & 27.55 \\
    GL 1 & HD 225213 & 00:05:31.99 & -37:22:03.88 & M1.5 & 4.36 & 0.38 & 3696 & 0.02 & 26.75 \\
    Kruger 60 A & GJ 860A & 22:27:58.11 & +57:41:38.52 & M3V & 4.01 & 0.34 & 3344 & 0.01 & 27.45 \\
    Struve 2398B\tablenotemark{a} & GL 725B & 18:42:43.94 & +59:38:18.09 & M3V & 3.52 & 0.25 & 3345 & 0.021 & 27.57 \\
    Struve 2398A\tablenotemark{a} & GL 725A & 18:42:43.94 & +59:38:06.52 & M3.5V & 3.52 & 0.33 & 3401 & 0.015 & 27.57 \\
    \enddata
    \tablerefs{$^a$Unresolved binary. $^b$Unresolved binary with a T7 brown dwarf.}
\end{deluxetable*}

\begin{figure}
\figurenum{15}
    \centering
    \includegraphics[width=1\linewidth]{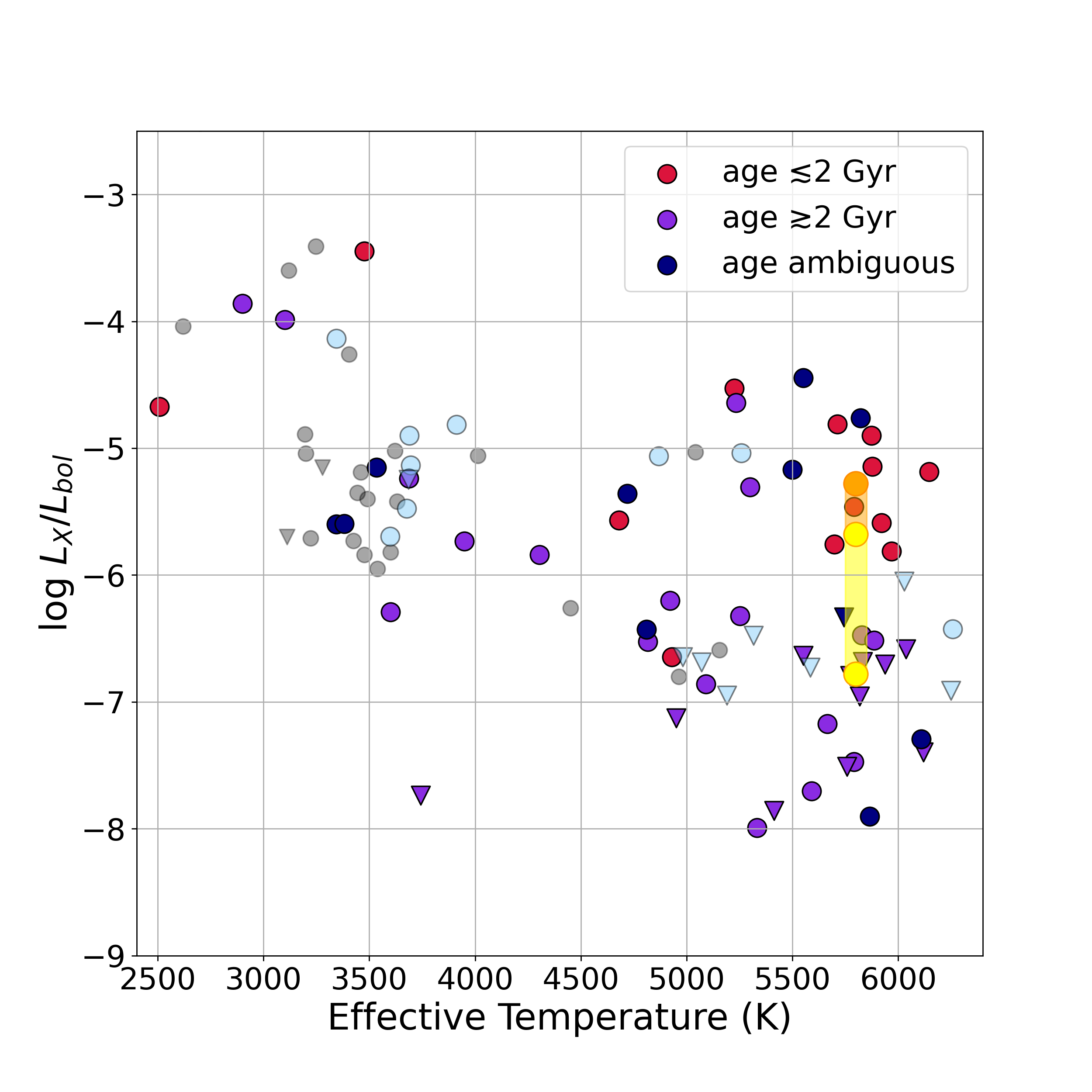}
    \caption{Same as Figure~\ref{fig:MUSCLES_compare} with \Swift/XRT sample in light blue.}
    \label{fig:MUSCLES_2Gyr_Swift}
\end{figure}

\section{Updated X-ray Luminosities for \Swift/XRT-Imaged Stars}\label{appendix:swift}
We searched the \Swift/XRT archive for all nearby stars that did not have available \XMM\ or \Chandra\ imaging. Nineteen of these stars (mostly K and M dwarfs; one system, Struve 2398, is a binary that is unresolved in the \Swift\ imaging) have fluxes or flux upper limits available from \Swift, which we summarize in Table~\ref{tab:Swift_stars}. The R.A. and Decl. coordinates for each star is taken from \citet{GaiaDR3} and the stellar physical parameters are from \citet{TESSinput}. We use the quiescent spectrum from Table~\ref{table:spectral_modeling} of the star that mostly closely matches each \Swift\ star's stellar physical parameters and the \Swift\ response files\footnote{See \url{https://swift.gsfc.nasa.gov/proposals/swift_responses.html}} to convert \Swift\ 0.3-10 keV count rates to luminosities. Figure~\ref{fig:MUSCLES_2Gyr_Swift} shows the \Lx/\Lbol\ ratio as a function of $T_{\rm eff}$ (as in Figure~\ref{fig:MUSCLES_compare}) with the nineteen \Swift\ stars added.

\bibliography{sample631}{}
\bibliographystyle{aasjournal}

\end{document}